\RequirePackage{fix-cm}
\documentclass[twocolumn,epjc3,nopacs]{svjour3}  
\smartqed  
\RequirePackage{graphicx}
\RequirePackage{enumitem}
\journalname{}

\usepackage{xcolor}
\usepackage{hyperref}
\usepackage[utf8]{inputenc}
\usepackage[T1]{fontenc}
\usepackage[mathlines]{lineno}
\usepackage{cite}
\usepackage[toc,acronym]{glossaries}

\makeglossaries

\newglossaryentry{gw170817}
{name=GW170817,
    description={The first binary neutron star merger event that was detected by its gravitational wave signal by the LIGO-Virgo collaboration on 17.08.2017 \cite{TheLIGOScientific:2017qsa}}}
    
\newglossaryentry{na61}{
name=NA61/SHINE,
description={experiment in the CERN northern area, devoted to a two-dimensional scan in collision energy and system size}
}

\newglossaryentry{brahms}{
name=BRAHMS,
description={Broad RAnge Hadron Magnetic Spectrometers experiment at RHIC}}

\newglossaryentry{phobos}{
name=PHOBOS,
description={experiment at RHIC}}

\newglossaryentry{phenix}{
name=PHENIX,
description={Pioneering High Energy Nuclear Interaction eXperiment at RHIC}}

\newglossaryentry{star}{
name=STAR,
description={Solenoid Tracker At Rhic experiment at RHIC}}

\newglossaryentry{bman}{
name=BM@N,
description={Baryonic Matter at the Nuclotron experiment at the NICA accelerator facility of JINR Dubna}}

\newglossaryentry{vhlle}{
name=vHLLE,
description={Harten-Lax-van Leer-Einfeldt approximate Riemann solver for 3+1 dimensional relativistic viscous hydrodynamic simulations of quark-gluon/hadron matter expansion in ultra-relativistic heavy-ion collisions}}

\newacronym{acquisition}{AcQuisition}{}
\newacronym{acs}{ACS}{}
\newacronym{adc}{ADC}{Analog-Digital Converter}
\newacronym{afe}{AFE}{Analog Front-End module}
\newacronym{afi}{AFI}{}
\newacronym{ags}{AGS}{Alternating Gradient Synchrotron}
\newacronym{ak}{AK}{}
\newacronym{aleph}{ALEPH}{}
\newacronym{alice}{ALICE}{}
\newacronym{alxi}{ALXi}{}
\newacronym{ampt}{AMPT}{}
\newacronym{aps}{APS}{}
\newacronym{asg}{ASG}{}
\newacronym{auau}{AuAu}{}
\newacronym{bc}{BC}{}
\newacronym{bebe}{BeBe}{}
\newacronym{bes}{BES}{Beam Energy Scan}
\newacronym{bibi}{BiBi}{}
\newacronym{bm}{BM}{}
\newacronym{boost}{BOOST}{}
\newacronym{bw}{BW}{Blast-Wave}
\newacronym{caen}{CAEN}{}
\newacronym{cc}{CC}{}
\newacronym{cep}{CEP}{Critical End Point}
\newacronym{cern}{CERN}{}
\newacronym{cm}{CM}{}
\newacronym{cometa}{COMETA}{}
\newacronym{csr}{CSR}{Chiral Symmetry Restoration}
\newacronym{daq}{DAQ}{Data AcQuisition}
\newacronym{dca}{DCA}{Distance of Closest Approach}
\newacronym{dcm}{DCM}{}
\newacronym{dcs}{DCS}{}
\newacronym{dds}{DDS}{}
\newacronym{delphi}{DELPHI}{}
\newacronym{dirac}{DIRAC}{}
\newacronym{dst}{DST}{Data Summary Tape}
\newacronym{eas}{EAS}{Extended Air Shower}
\newacronym{ecal}{ECal}{Electromagnetic Calorimeter}
\newacronym{eleff}{ElEff}{}
\newacronym{elpur}{ElPur}{}
\newacronym{em}{EM}{}
\newacronym{emc}{EMC}{}
\newacronym{eos}{EoS}{Equation of State}
\newacronym{ep}{EP}{}
\newacronym{eps}{EPS}{}
\newacronym{epsf}{EPSF}{}
\newacronym{evb}{EVB}{}
\newacronym{fair}{FAIR}{}
\newacronym{fairsoft}{FAIRsoft}{}
\newacronym{fchal}{FCHal}{Forward Hadronic Calorimeter}
\newacronym{fd}{FD}{}
\newacronym{fe}{FE}{}
\newacronym{fee}{FEE}{Front-End Electronics}
\newacronym{ffd}{FFD}{Fast Forward Detector}
\newacronym{ffde}{FFDE}{}
\newacronym{ffdw}{FFDW}{}
\newacronym{ffe}{FFE}{}
\newacronym{fh}{FH}{}
\newacronym{fhcal}{FHCal}{}
\newacronym{flp}{FLP}{First Level Processor}
\newacronym{flps}{FLPs}{}
\newacronym{fpga}{FPGA}{Field Programmable Gate Arrays}
\newacronym{ftp}{FTP}{}
\newacronym{fwhm}{FWHM}{}
\newacronym{gb}{GB}{}
\newacronym{gbm}{GBM}{}
\newacronym{geant}{GEANT}{}
\newacronym{geco}{GECO}{}
\newacronym{gnu}{GNU}{}
\newacronym{gsi}{GSI}{}
\newacronym{gui}{GUI}{Graphical User Interface}
\newacronym{gw}{GW}{}
\newacronym{hcal}{HCAL}{}
\newacronym{hdmi}{HDMI}{}
\newacronym{hlle}{HLLE}{}
\newacronym{hmc}{HMC}{}
\newacronym{hrgm}{HRGM}{Hadron Resonance Gas Model}
\newacronym{ht}{HT}{}
\newacronym{hub}{HUB}{}
\newacronym{hv}{HV}{}
\newacronym{ipd}{IPD}{Intelligent Power Distributor}
\newacronym{its}{ITS}{Inner Tracker System}
\newacronym{jam}{JAM}{}
\newacronym{jinr}{JINR}{Joint Institute for Nuclear Research}
\newacronym{jk}{JK}{}
\newacronym{la}{LA}{}
\newacronym{lgpl}{LGPL}{}
\newacronym{lhc}{LHC}{Large Hadron Collider}
\newacronym{lhep}{VBLHEP}{Veksler and Baldin Laboratory of High Energy Physics}
\newacronym{ligoscientific}{LIGOScientific}{}
\newacronym{lit}{MLIT}{Meshcheryakov Laboratory of Information Technology}
\newacronym{lqcd}{LQCD}{Lattice QCD}
\newacronym{lv}{LV}{}
\newacronym{lvds}{LVDS}{}
\newacronym{lvn}{LVN}{}
\newacronym{maps}{MAPS}{Monolithic Active Pixel Sensors}
\newacronym{mbb}{MBB}{}
\newacronym{mc}{MC}{}
\newacronym{mcord}{MCORD}{}
\newacronym{mcppmt}{MCP-PMT}{Micro Channel Plate Photomultiplier}
\newacronym{mdc}{MDC}{Mini Data Center}
\newacronym{mev}{MeV}{}
\newacronym{mg}{MG}{}
\newacronym{micc}{MICC}{Multifunctional Information and Computing Complex}
\newacronym{mpd}{MPD}{Multi-Purpose Detector}
\newacronym{mpddst}{MPDDst}{}
\newacronym{mpdroot}{MPDRoot}{}
\newacronym{mppc}{MPPC}{Multi-Pixel Photon Counter}
\newacronym{mq}{MQ}{}
\newacronym{mrpc}{MRPC}{Multigap Resistive Plate Chamber}
\newacronym{mrpcs}{MRPCs}{}
\newacronym{ms}{MS}{}
\newacronym{msc}{MSC}{}
\newacronym{mtca}{MicroTCA}{MicroTCA}
\newacronym{na}{NA}{}
\newacronym{nbd}{NBD}{}
\newacronym{nbti}{NbTi}{}
\newacronym{ncbj}{NCBJ}{}
\newacronym{neutral}{NEUTRAL}{}
\newacronym{nica}{NICA}{Nuclotron-based Ion Collider fAcility}
\newacronym{nl}{NL}{}
\newacronym{nn}{NN}{}
\newacronym{nupecc}{NUPECC}{}
\newacronym{pacs}{PACS}{}
\newacronym{pca}{PCA}{Point of Closest Approach}
\newacronym{pcb}{PCB}{}
\newacronym{pec}{PEC}{Protective Earth Cable}
\newacronym{phqmd}{PHQMD}{}
\newacronym{phsd}{PHSD}{}
\newacronym{pid}{PID}{Particle Identification}
\newacronym{pilas}{PiLas}{}
\newacronym{platform}{PLATFORM}{}
\newacronym{pmt}{PMT}{}
\newacronym{pmts}{PMTs}{}
\newacronym{prc}{PCR}{Primary Cosmic Ray}
\newacronym{ps}{PS}{}
\newacronym{pwg}{PWG}{}
\newacronym{qcd}{QCD}{Quantum Chromodynamics}
\newacronym{qgm}{QGM}{}
\newacronym{qgp}{QGP}{}
\newacronym{qgsm}{QGSM}{}
\newacronym{rack}{RACK}{}
\newacronym{rhic}{RHIC}{Relativistic Heavy-Ion Collider}
\newacronym{rf}{RF}{}
\newacronym{rms}{RMS}{}
\newacronym{roc}{ROC}{Read-Out Chamber}
\newacronym{root}{ROOT}{}
\newacronym{rp}{RP}{}
\newacronym{run}{RUN}{}
\newacronym{sb}{SB}{}
\newacronym{sc}{SC}{}
\newacronym{scada}{SCADA}{}
\newacronym{service}{SERVICE}{}
\newacronym{shine}{SHINE}{}
\newacronym{sipm}{SiPM}{Silicon Photomultiplier}
\newacronym{sipms}{SiPMs}{}
\newacronym{sis}{SIS}{Super Ion Synchrotron}
\newacronym{smash}{SMASH}{Simulating Many Accelerated Strongly-interacting Hadrons}
\newacronym{smm}{SMM}{}
\newacronym{spd}{SPD}{}
\newacronym{sps}{SPS}{Super Proton Synchrotron}
\newacronym{sqgp}{sQGP}{strongly-coupled Quark-Gluon Plasma}
\newacronym{svjour}{SVJour}{}
\newacronym{tca}{TCA}{}
\newacronym{tdc}{TDC}{}
\newacronym{tdr}{TDR}{Technical Design Report}
\newacronym{tdrs}{TDRs}{}
\newacronym{te}{TE}{}
\newacronym{tev}{TeV}{}
\newacronym{tex}{TeX}{}
\newacronym{theseus}{THESEUS}{}
\newacronym{tm}{TM}{}
\newacronym{tof}{TOF}{Time-of-Flight}
\newacronym{tom}{TOm}{}
\newacronym{tp}{TP}{}
\newacronym{tpc}{TPC}{Time Projection Chamber}
\newacronym{trb}{TRB}{}
\newacronym{tw}{TW}{}
\newacronym{tyvek}{TYVEK}{}
\newacronym{urqmd}{UrQMD}{Ultrarelativistic Quantum Molecular Dynamics}
\newacronym{uv}{UV}{}
\newacronym{vblhep}{VBLHEP}{}
\newacronym{vhl}{VHL}{}
\newacronym{vme}{VME}{}
\newacronym{wiener}{WIENER}{}
\newacronym{wls}{WLS}{Wave Length Shifter}
\newacronym{wut}{WUT}{}
\newacronym{xp}{XP}{}

\begin{document}

\title{Status and initial physics performance studies of the MPD experiment at NICA
}

\author{The MPD Collaboration\thanksref{e1}}

\authorrunning{V.~Abgaryan {\it et al.} (MPD Collaboration)}
\titlerunning{First Physics in MPD}

\institute{The full list of Collaboration Members is provided at the end of the manuscript\label{e1}}

\date{Received: \today / Accepted: date}

\maketitle

\begin{abstract}
The \acrlong{nica} (\acrshort{nica}) is under construction at the \acrlong{jinr} (\acrshort{jinr}), with commissioning of the facility expected in late 2022. 
The \acrlong{mpd} (\acrshort{mpd}) has been designed to operate at NICA and its components are currently in production. 
The detector is expected to be ready for data taking with the first beams from NICA.
This document provides an overview of the landscape of the investigation of the QCD phase diagram in the region of maximum baryonic density, where NICA and MPD will be able to provide significant and unique input. 
It also provides a detailed description of the MPD set-up, including its various subsystems as well as its support and computing infrastructures.
Selected performance studies for particular physics measurements at MPD are presented and discussed in the context of existing data and theoretical expectations.

\keywords{NICA \and MPD \and QCD}
\end{abstract}

\tableofcontents

\section{\label{sec:nicaintro}Introduction}

The Multi-Purpose Detector (MPD) is one of the two dedicated heavy-ion collision experiments of the Nuclotron-based Ion Collider fAcility (NICA), one of the flagship projects, planned to come into operation at the Joint Institute for Nuclear Research (JINR) in 2022. Its main scientific purpose is to search for novel phenomena in the baryon-rich region of the QCD phase diagram by means of colliding heavy nuclei in the energy range of 4 GeV $\leq\sqrt{s_{\rm NN}}\leq 11$ GeV.

A wealth of results, obtained by colliding heavy ions at different beam energies, has been gathered by experiments carried out at several facilities such as the Super Ion Synchrotron (\acrshort{sis}), the \acrlong{ags} (\acrshort{ags}), the \acrlong{sps} (\acrshort{sps}), the \acrlong{rhic} (\acrshort{rhic}) and the \acrlong{lhc} (\acrshort{lhc}). 
The new experimental program at the NICA-MPD will fill a niche in the energy scale, which is not yet fully explored, and the results will bring about a deeper insight into hadron dynamics and multiparticle production in the high baryon density domain.

Recent Lattice Quantum Chromo Dynamics (\acrshort{lqcd}) calculations have shown that for vanishing baryon chemical potential, $\mu_B$, a crossover transition, from the confined/broken chiral symmetry phase to the deconfined/partially restored chirally symmetry phase, happens at a pseudocritical temperature $T_c(\mu_B=0)\simeq 156.5\pm 1.5$ MeV~\cite{Bazavov:2018mes,Aoki:2006we}. \acrshort{lqcd} calculations also show that this crossover transition happens at energy densities higher than 0.5 GeV/fm$^3$. Microscopic model calculations indicate that such densities can be achieved in the center of the fireball created in head-on collisions of heavy-ions at energies above 
$\sqrt{s_{\rm NN}}=3-5$~GeV~\cite{Mendenhall:2020fil}. Different effective model calculations~\cite{Ayala:2019skg, Asakawa:1989bq,Ayala:2014jla} suggest that for low temperatures and high baryon chemical potentials, the transition from the ordinary hadron matter phase to a phase where chiral symmetry is restored is of first order. If the temperature is increased, the end of this first order phase transition line in the $T$ vs.~$\mu_B$ plane should happen at a \acrlong{cep} (\acrshort{cep}). It should be emphasized that the existence of such a CEP has not been established~\cite{Ayala:2017ucc}. Furthermore, its location in models that predict its existence is widely spread over the phase diagram~\cite{Stephanov:2007fk}. \acrshort{lqcd} calculations cannot be used to directly determine the position of the CEP, due to the severe sign problem~\cite{Ding:2015ona}. Approximations are used to gain some insight. For example, recent results employing the Taylor series expansion around $\mu_B=0$ or the extrapolation from imaginary to real $\mu_B$  values, suggest that the CEP cannot be located at $\mu_B/T\leq 2$ and $145\leq T\leq 155$ MeV~\cite{Sharma:2017jwb}. 
More recently, \acrshort{lqcd} calculations for two light quarks and a physical strange quark allowed to extract the chiral phase transition temperature $T_c^0=132^{+3}_{-6}$ MeV \cite{Ding:2019prx} (see also Ref.~\cite{Kotov:2021rah}). Using $T_{CEP}<T_c^0$~\cite{Schmidt:2021pey} together with the systematic energy dependence of the chemical freeze out temperature~\cite{Cleymans:2005xv,Andronic:2017pug} 
one can deduce that the range $\sqrt{s_{\rm NN}}< 6$~GeV is the most appropriate one to search for the CEP~\cite{Senger:2021dot}. 

It is expected that quark matter at relatively low temperatures is in a color superconducting phase with large pairing gaps~\cite{Alford:1997zt,Rapp:1997zu,Berges:1998rc}, which corresponds 
to sufficiently high critical temperatures. Therefore, this 
phase can also be discovered in the NICA energy range~\cite{Blaschke:2010ka}.
The emergence of color superconductivity could change the character of both, the chiral and the superconducting phase transition at low temperatures into a crossover, which would entail the existence of a second CEP or even the absence of a CEP~\cite{Hatsuda:2006ps}.

The investigation of the properties of nuclear matter inside neutron stars is one of the goals of modern astrophysics. The recent observation of a neutron star merger, both by direct detection of gravitational
waves~\cite{TheLIGOScientific:2017qsa} as well as in the electromagnetic spectrum~\cite{GBM:2017lvd} initiated a new era of multi-messenger astronomy. Recent model calculations reveal that in a neutron star merger, nuclear matter reaches densities and temperatures similar to those occurring in heavy-ion collisions in the NICA energy range~\cite{Blacker:2020nlq,Most:2019onn}, that are therefore relevant to investigate the onset
of deconfinement, albeit at higher isospin density. In other words, heavy-ion collisions at NICA and neutron star mergers probe similar regions of the QCD phase diagram. Therefore, the MPD offers a unique opportunity to complement the study of neutron star mergers by obtaining data from a terrestrial laboratory experiment~\cite{Klahn:2012uq}. Furthermore, if the observations of neutron stars and their mergers could reveal the existence of a first-order phase transition, such finding would necessarily imply the existence of a CEP in the QCD phase diagram~\cite{Blaschke:2013ana,Bauswein:2018bma}.

Collisions of heavy nuclei at moderate energies (in particular those involving exotic-beams) produce isospin imbalanced matter, due to their rich neutron content~\cite{Napolitani:2009pc}. This imbalance can be characterized by a finite isospin chemical potential $\mu_I$. Unlike the case of a finite $\mu_B$, \acrshort{lqcd} simulations for finite $\mu_I$ are not affected by the sign problem and can be computed using Monte Carlo (MC) techniques~\cite{Brandt:2017zck}, thus providing reliable benchmarks. It is found that for a temperature dependent threshold value of $\mu_I$  charged pions can be created, leading to charged pion condensation~\cite{Migdal:1978az,Ruck:1976zt}, which may play an important role in the description of neutron stars and can be searched for by means of a systematic analysis of heavy-ion collisions in the high baryon density domain~\cite{Khunjua:2019nnv}. 

\begin{figure}
    \centering
    \includegraphics[width=0.85\linewidth]{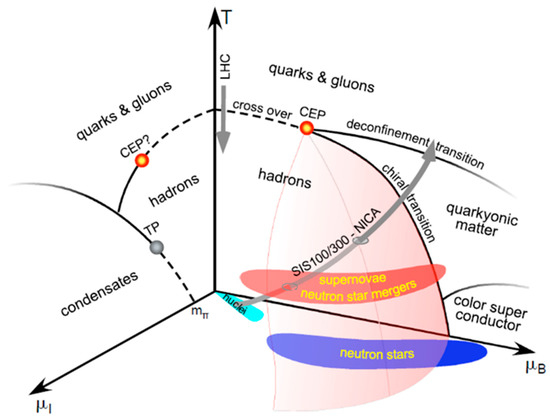}
    \caption{Schematic representation of the hypothetical QCD phase diagram in the $T$, $\mu_B$ and $\mu_I$ directions. The landmarks of the critical endpoints (CEP) and the tri-critical points (TP) are indicated together with the regions to be accessed by the NICA and FAIR experiments which neighbor the region that can be populated by supernova explosions and neutron star merger events. Figure taken from Ref.~\cite{NUPECC:2017lrp}.}.
    \label{fig:phd-nupecc}
\end{figure}

The conjectured rich structure of the phase diagram is illustrated in Fig.~\ref{fig:phd-nupecc},
taken from Ref.~\cite{NUPECC:2017lrp}, which indicates lines of first-order transitions (solid lines) as well as crossover transitions (dashed lines) between different phases of low-energy QCD. 
The CEPs occur where solid and dashed lines meet. 
A tricritical point (TP) occurs where three phases meet and this may be realised, {\it e.g.}, when between the hadronic and the quark-gluon plasma transition lines a third phase like quarkyonic matter or color-superconductivity can be realised. 
As discussed in Ref.~\cite{Andronic:2009gj}, it may turn out that the CEP is a TP because chiral symmetry restoration and deconfinement may not coincide at high baryon densities. 
Calculations within the thermal model indicate that the highest baryon density is achieved in the NICA energy range~\cite{Randrup:2006nr} which makes the MPD particularly well suited to experimentally search for the existence of the CEP.

When fluctuations of conserved charges are well described by hadron degrees of freedom in equilibrium, their cumulants should be consistent
with models such as the \acrlong{hrgm} (\acrshort{hrgm})~\cite{BraunMunzinger:1994xr,Andronic:2005yp}. 
On the other hand, when fluctuations deviate from those in the HRGM,
they can be used as experimental signals of non-hadron and/or non-equilibrium physics. Near the CEP, higher order cumulants of conserved charges behave anomalously. 
In particular, they change sign in the vicinity of the critical point~\cite{Asakawa:2015ybt}. 
They are also sensitive to the increase of correlation lengths~\cite{Stephanov:2008qz}.
The MPD will search for the location of the CEP by means of the analysis of fluctuations using ratios of cumulants of conserved charges, as a function of collision energy and system size. 

Other novel phenomena that the MPD is suited to study include signatures of vortical motion~\cite{Rogachevsky:2010ys,Ivanov:2018eej,STAR:2017ckg,Ayala:2020soy, Ayala:2021xrn} and magnetic fields~\cite{Skokov:2009qp,Tawfik:2016cot} produced in non-central heavy-ion collisions, the search for exotic hadrons made of tetra- and penta-quark configurations~\cite{Ma:2006hs} and the search for light nuclei formation to study its influence on the \acrlong{eos} (\acrshort{eos}) at high baryon densities~\cite{Blaschke:2020gqr,Mohs:2020awg}, among others.

The civil construction of both the NICA complex and the hall to host the MPD has been completed. Assembly of the MPD started in the middle of 2020 with the installation of the iron yoke on its support legs and the delivery at the end of the same year of the superconducting solenoid to JINR from Italy. The assembly of the detector in the Stage 1 configuration (see Sec. 3) shall be completed in time for the commissioning of the NICA collider ring expected in late 2022. The initial luminosity is planned to be at least $10^{24}$~cm$^{-2}$s$^{-1}$ with a relatively quick increase to at least $10^{25}$~cm$^{-2}$s$^{-1}$. The design luminosity goal for NICA with all components, such as an Electron Cooling System and the full set of RF cavities, is $10^{27}$~cm$^{-2}$s$^{-1}$. Symmetric collisions of heavy ions will be performed in the initial stages of the NICA operation. Several types of ions are under consideration. These include $^{197}$Au ions, which were used in previous and ongoing experiments at RHIC; $^{208}$Pb ions, which were used for extensive data runs at SPS; and $^{209}$Bi ions, which are very similar to Pb ions, but provide more reliable operation of the NICA injection and acceleration complexes during the commissioning and first running phases. 
For heavy ions, such as Au and Bi, the kinetic energy of the beam provided by the Nuclotron will be in the range from 2.5 to 3.8 GeV per nucleon.
In the first year of operation, additional acceleration of the beams in the NICA collider is not foreseen. Therefore the initial collision energy $\sqrt{s_{\rm{NN}}}$ may vary from $7$ up to $9.46$~GeV, with the collision energy of $9.2$~GeV being preferred, so that results can be compared with those of RHIC-STAR that collected data at the same energy. Delivering Au+Au collisions at $\sqrt{s_{\rm{NN}}}$ up to 11~GeV remains the key goal of the NICA project that will be accomplished after the initial commissioning stage of operation.

The MPD is designed as a 4$\pi$ spectrometer capable of detecting charged hadrons, electrons and photons in heavy-ion collisions at high luminosity. It will provide precise 3-D tracking and a high-performance particle identification (PID) system based on a large-volume gaseous Time Projection Chamber (\acrshort{tpc}), \acrlong{tof} (\acrshort{tof}) measurements and calorimetry. It is expected that the MPD will produce event-by-event information on charged particle tracks coming from the primary interaction vertices, together with identification of those particles, and information on the collision centrality.

This paper is organized as follows: In Sec.~\ref{sec:goals} we describe in detail the physics motivation and the physics goals of the MPD experimental program. We emphasize the measurements that can be performed during the first stage of MPD operation, positioning them within the landscape of existing results from previous and current heavy-ion experiments. In Sec.~\ref{sec:detector} we describe the different components of the MPD and in Sec.~\ref{sec:itresources} the software developments and the computing software requirements. In Sec.~\ref{sec:observables} we present feasibility and performance studies for selected physics measurements that can be carried out by the MPD Collaboration. A summary is provided in Sec.~\ref{sec:conclusions}.

\section{\label{sec:goals}Brief survey of the MPD physics goals}

The diversity of the data in the field of relativistic heavy-ion collisions, obtained by
experiments at the SIS, AGS, SPS, RHIC and LHC, is already quite large and impressive.
In recent years, the STAR-BES program has produced a wealth of results to describe the bulk properties of the medium created in Au+Au reactions for $\sqrt{s_{\rm NN}}$=7.7, 11.5, 14.6, 19.6, 27, 39, 62.4 and 200 GeV~\cite{Adamczyk:2017iwn} by measuring several observables at mid rapidity.  
The new experimental MPD program planned with the high intensity NICA beams promises to provide deeper knowledge of the dynamics of hadronic interactions and multiparticle production mechanisms at high baryon density complementing the energy range covered by the STAR-BES program.
 
The properties of the matter created in nucleus-nucleus (A+A) collisions at NICA energies will be characterized on an event-by-event basis, using a variety of observables. The global quantities, such as multiplicity and transverse energy ($E_\mathrm{T}$) are considered as the main tools to reveal the energy density achieved in the collision. The {\it first-day} key global observables that can be measured with the initial data sample, are to be selected to check the reliability of the new experimental MPD setup and to place the results among the landscape of the available world data. These observables are expected to provide the basic information for more focused physics studies relevant to the onset of quark confinement, chiral symmetry restoration and for the search of the CEP on the phase-diagram  of strongly interacting matter.
\begin{figure}[t]
\includegraphics[width=8cm]{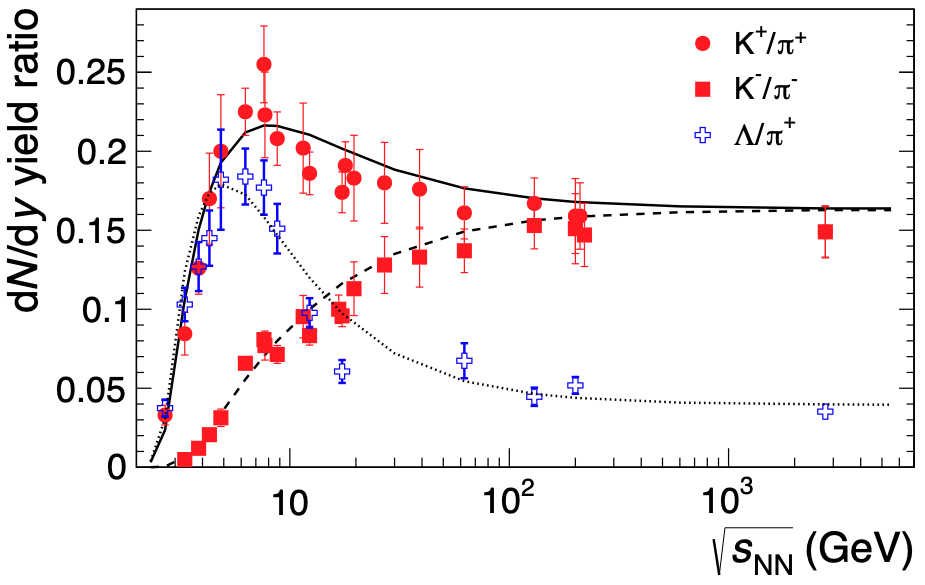}
\caption{\label{fig:K+pi+} 
Dependence of the $K^+/\pi^+$, $K^-/\pi^-$ and $\Lambda/\pi^+$ ratios on the collision energy. The data points are from different experiments~\cite{Andronic:2014zha,Afanasiev:2002mx} whereas the lines represent the results of the thermal statistical model of chemical freezeout.
Figure taken from Ref.~\cite{Andronic:2017pug}.
}
\end{figure}


\subsection{Hadrochemistry}\label{sec2-1}

Existing data on hadroproduction from SPS and RHIC, obtained from single-particle spectra and yields, suggest that the QCD  transitions (deconfinement and chiral symmetry restoration) occur in the NICA energy range.  Moreover, this energy range is appropriate to study in detail the interplay between the hadron and parton phases. The structure of the QCD phase diagram  may be tested by measuring abundances of hadron species, while different regions of the diagram are accessible by varying the collision energy.  Experimental results on hadron abundances produced in heavy-ion collisions in the range from AGS to LHC energies indicate that the final state of such collisions is close to chemical equilibrium. Thus, the yields can be fitted by the thermal statistical  model using
two free parameters, namely, $T$ and $\mu_B$. Assuming the measured multiplicities to be preserved throughout the final hadron-resonance cascade expansion, the analysis reveals the hadronization point along the QCD parton-hadron boundary line in terms of the extracted $T$ and $\mu_B$ values. 
Within the thermal statistical model, one can show that the NICA energy range covers the region where the matter created in nuclear collisions transitions from net baryon- to meson-dominated matter.
\begin{figure}[t]
\begin{center}
\includegraphics[width=8cm]{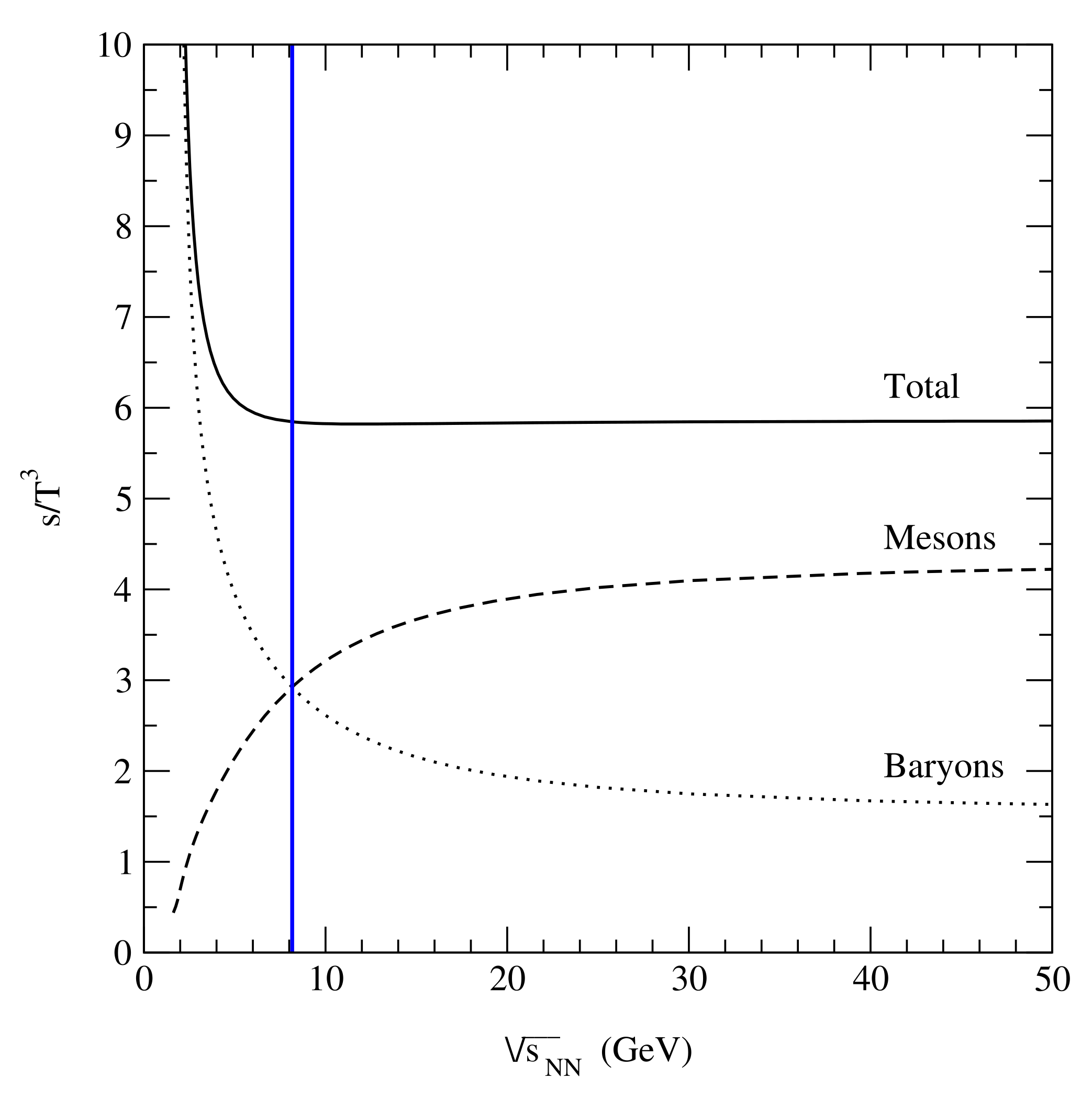}
\end{center}
\caption{\label{fig:meson-baryon} Entropy fraction carried by mesons and baryons as a function of the collision energy along the chemical freeze-out line in the QCD phase diagram. Figure taken from Ref.~\cite{Cleymans:2004hj}.
}
\end{figure}

An example of a relevant observable to study the onset of deconfinement~\cite{Gazdzicki:2020iix} is the kaon-to-pion ratio. Figure~\ref{fig:K+pi+} shows the excitation functions for $K^+/\pi^+$, $K^-/\pi^-$ and $\Lambda/\pi^+$ ratios, from AGS up to LHC energies, including results from several experiments (see Refs.~\cite{Andronic:2014zha,Afanasiev:2002mx} for a compilation of the plotted data). The $K^+/\pi^+$ and $\Lambda/\pi^+$ ratios show a peak structure, whereas the $K^-/\pi^-$ ratio exhibits a monotonic rise. This feature is very well described by the thermal statistical model of chemical freezeout (represented by the lines in Fig.~\ref{fig:K+pi+}) where it appears to be due to the drop of the baryochemical potential with increasing collision energy and to the transition from baryon dominance to meson dominance, as illustrated in Fig.~\ref{fig:meson-baryon}. A similar behaviour is also observed in other model calculations, some of which even show a sharper peak, followed by a plateau in the $K^+/\pi^+$ ratio as a function of $T/\mu_B$~\cite{Afanasiev:2002mx,Alt:2007aa}. It should be pointed out that the experimental results at energies below $\sqrt{s_{\rm NN}}\sim 10$~GeV, shown in Fig.~\ref{fig:K+pi+}, were obtained by different experiments and exhibit relatively large uncertainties. The MPD will offer the capability to cover this energy range using a single experimental set up and to provide results of higher precision. 
Furthermore, the analysis of hadron abundances may allow us to address the much debated {\it onset of deconfinement} problem. If nucleus-nucleus dynamics within the NICA energy range crosses the phase transition or crossover line starting from a particular (threshold) energy, then, below this value, the actual hadron multiplicity would not stem from the QCD hadronization phase transition (responsible for chemical equilibration among the species) and we would expect a sizeable change in the observed hadron freeze-out pattern, such as a sequential chemical freeze-out in inverse order of the inelastic cross-section. Important aspects of the MPD operation, relevant for these studies, such as tracking, particle identification, hyperon reconstruction and more, are discussed in detail in Secs.~\ref{sec:detector} and~\ref{sec:observables}. 

\subsection{Anisotropic flow measurements}
Anisotropic flow measurements in relativistic-heavy ion collisions at RHIC and the LHC have provided compelling evidence
for the formation of a \acrlong{sqgp} (\acrshort{sqgp}), a state of matter with partonic degrees of freedom
and low specific shear viscosity $\eta/s$~\cite{Busza:2018rrf,Snellings:2014kwa}. 
The latter term is the ratio of shear viscosity $\eta$ to entropy density $s$ of the formed matter.
The anisotropic collective flow,
as manifested by the anisotropic emission of particles in the plane transverse to the beam 
direction, is one of the promising  observables in this context due to its sensitivity to the transport properties of the strongly interacting matter,
namely, the EoS, the speed of sound ($c_s$) and the specific shear ($\eta/s$)  and bulk ($\zeta/s$)
viscosities \cite{Snellings:2014kwa,Lacey:2005qq}. 
The azimuthal anisotropy of produced particles is quantified
by the Fourier coefficients $v_n$ of the expansion of the particle azimuthal distribution as~\cite{Voloshin:1994mz} \begin{eqnarray}dN/d\varphi \propto 1 + \sum_{n\geq 1} 2 v_{n} \cos (n(\varphi-\Psi_{n})),
\end{eqnarray}
where $n$ is the order of the harmonic, $\varphi$ is the azimuthal angle of a given particle, and $\Psi_n$ is the azimuthal angle of the $n^{\mathrm{th}}$-order event plane.
The $n^{\mathrm{th}}$-order flow coefficients $v_n$ can be calculated as $v_{n} = \langle{\cos[n(\varphi - \Psi_n)]}\rangle$,
where the angle brackets denote an average over particles and events.

Relativistic viscous hydrodynamic models have been successful in describing the observed anisotropy $v_n$ for the produced particles
in the collisions of heavy ions at RHIC and the LHC~\cite{Gale:2012rq,Heinz:2013th,Bernhard:2019bmu}.
In this framework, the values of the coefficients $v_{n}$ have been attributed to an eccentricity-driven
hydrodynamic expansion of the plasma produced in the collision zone. This means that a finite eccentricity
moment $\varepsilon_n$ drives uneven pressure gradients in and out of the event plane $\Psi_n$, and the resulting
expansion leads to the anisotropic flow of particles about this plane. The event-by-event  geometric fluctuations
in its initial density distribution are found to be responsible for a finite elliptic flow signal $v_2$ in the collisions
with almost zero impact parameter, and for the presence of odd harmonic moments in the initial geometry $\varepsilon_n$
and the final momentum anisotropy $v_n$~\cite{Snellings:2014kwa}. The proportionality constant between
$v_n$ and  $\varepsilon_n$ is found to be sensitive to the transport properties of the strongly
interacting matter. Bayesian parameter estimation methods can be used to extract estimates of the temperature-dependent
specific shear $\eta/s(T)$ and bulk viscosity $\zeta/s(T)$ simultaneously from the experimental data \cite{Bernhard:2019bmu}. 

\begin{figure}[t]
\begin{center}
\includegraphics[width=7.5cm]{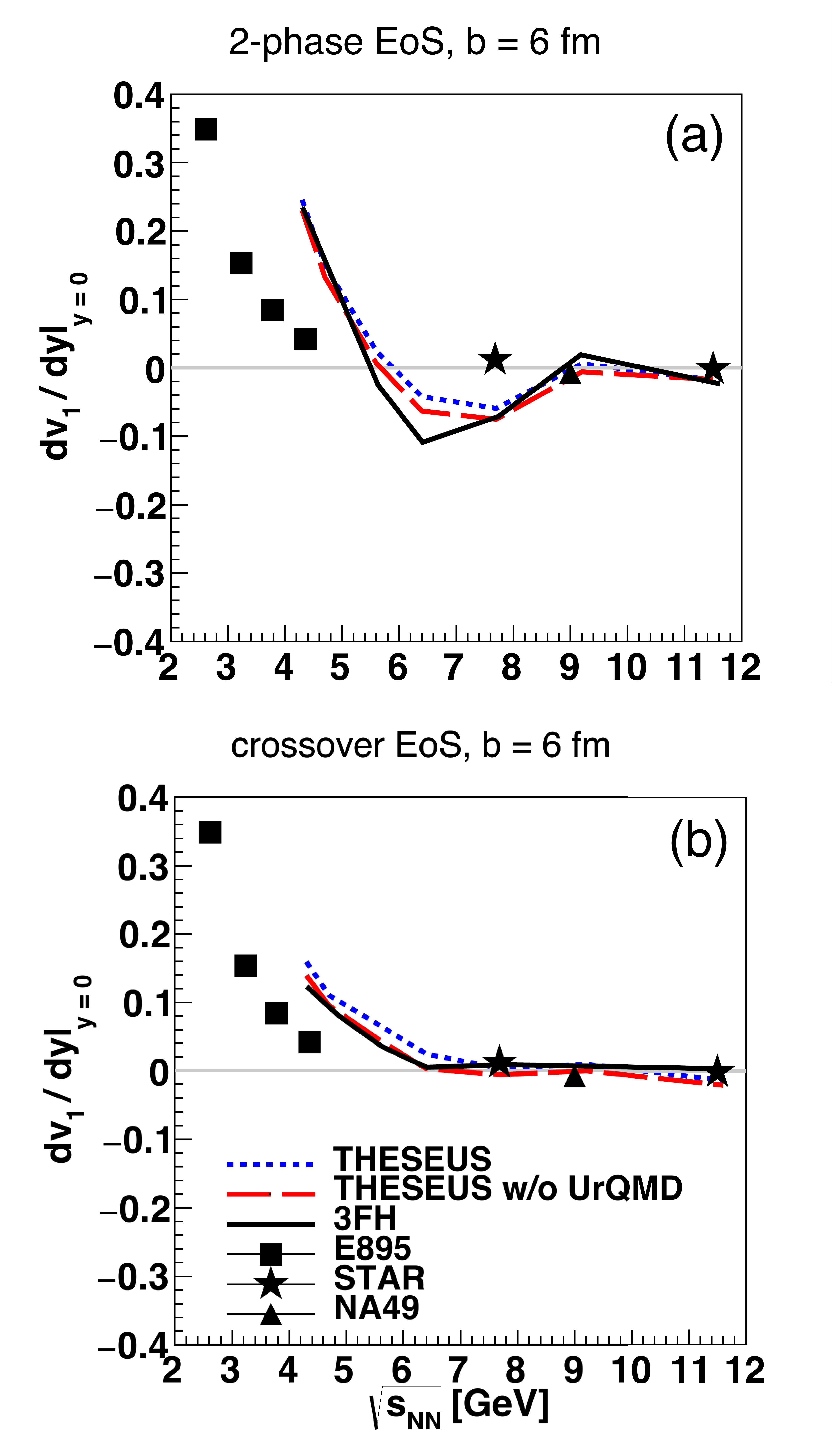}
\caption{\label{fig:p-antiflow} Antiflow of protons in the MPD energy range $\sqrt{s_{\rm NN}}=5-9$ GeV for the first order phase transition (upper panel) and its absence for the crossover scenario (lower panel).
The results are obtained by calculations with the Frankfurt-Wroclaw-Dubna three-fluid-hydrodynamics model (THESEUS)~\cite{Batyuk:2016qmb}.}
\end{center}
\end{figure}

The predicted first-order phase transition  between hadronic and sQGP phases can also be characterized by a dramatic drop in
the pressure, or a softening of the EoS~\cite{Stoecker:2004qu}. Signals such as anisotropic flow are very promising
in this context
due to their sensitivity to the EoS. The rapidity-odd component of the directed flow ($v_1$) can probe
the very early stages of the collision as it is generated  during the passage time of the two colliding nuclei
$t_{\rm pass}=2R/(\gamma_s \beta_s)$, where $R$ is the radius of the nucleus
at rest, $\beta_s$ is the spectator velocity in the center-of-mass and $\gamma_s$ is the corresponding
Lorentz factor. Both hydrodynamic  and transport model calculations indicate that the directed
flow of charged particles, especially baryons at midrapidity, is very sensitive
to the EoS~\cite{Stoecker:2004qu,Rischke:1996nq}. 
The slope of the
rapidity dependence $dv_1/dy$ close to mid-rapidity is a convenient way to characterize the
overall magnitude of the rapidity-odd component of the directed flow signal~\cite{Batyuk:2016qmb,Singha:2016mna}.
A minimum in  $dv_1/dy$ in the  midrapidity region ($y\sim 0$)  could be related to the softening of the EoS
due to a first order phase transition between hadronic matter
and sQGP~\cite{Batyuk:2016qmb,Singha:2016mna,Keane:2017kdq,Stoecker:2004qu,Rischke:1996nq}, see Fig.~\ref{fig:p-antiflow}. 

\begin{figure}[t]
	\centering
	\includegraphics[width=0.44\textwidth] {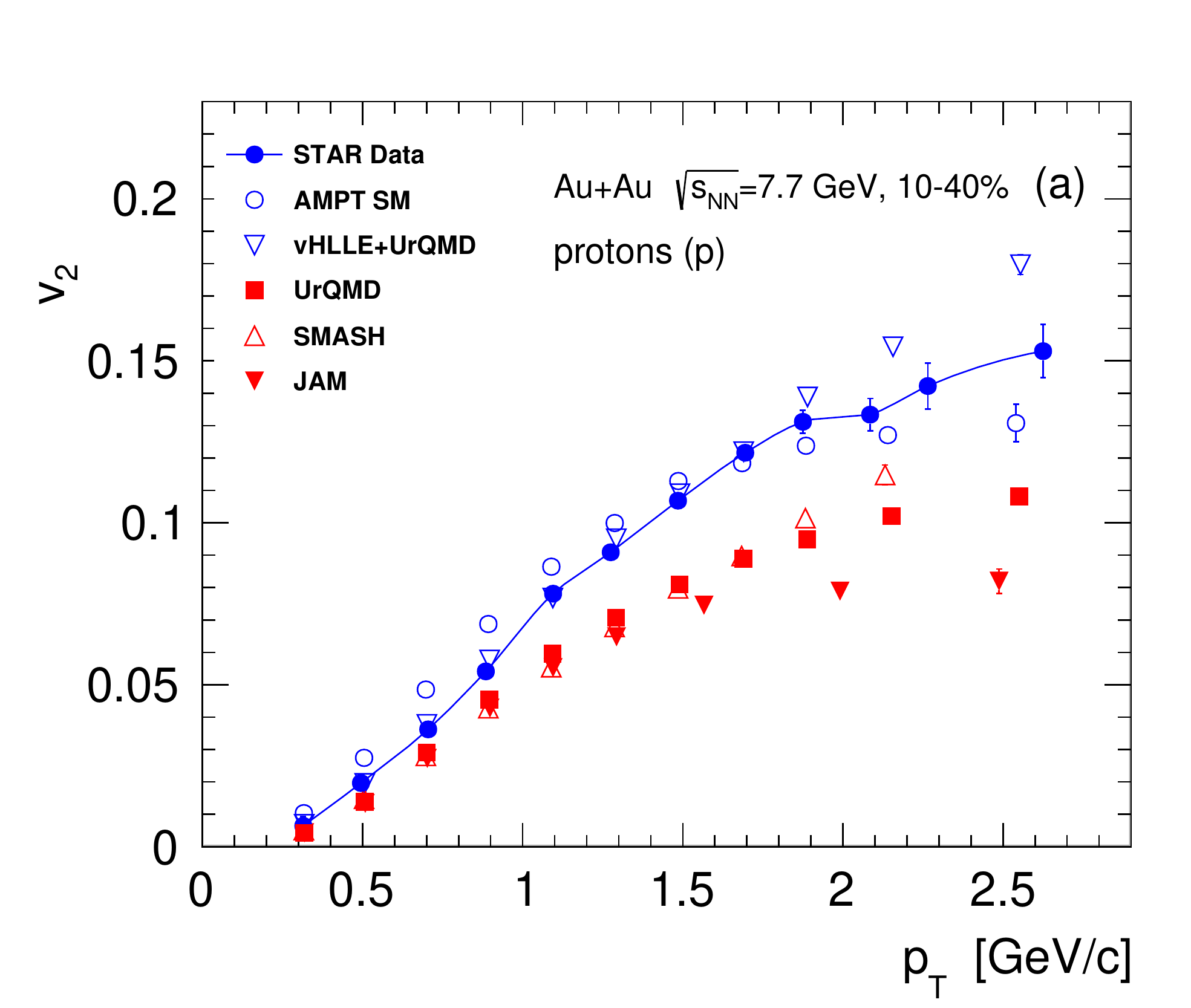}
	\includegraphics[width=0.44\textwidth] {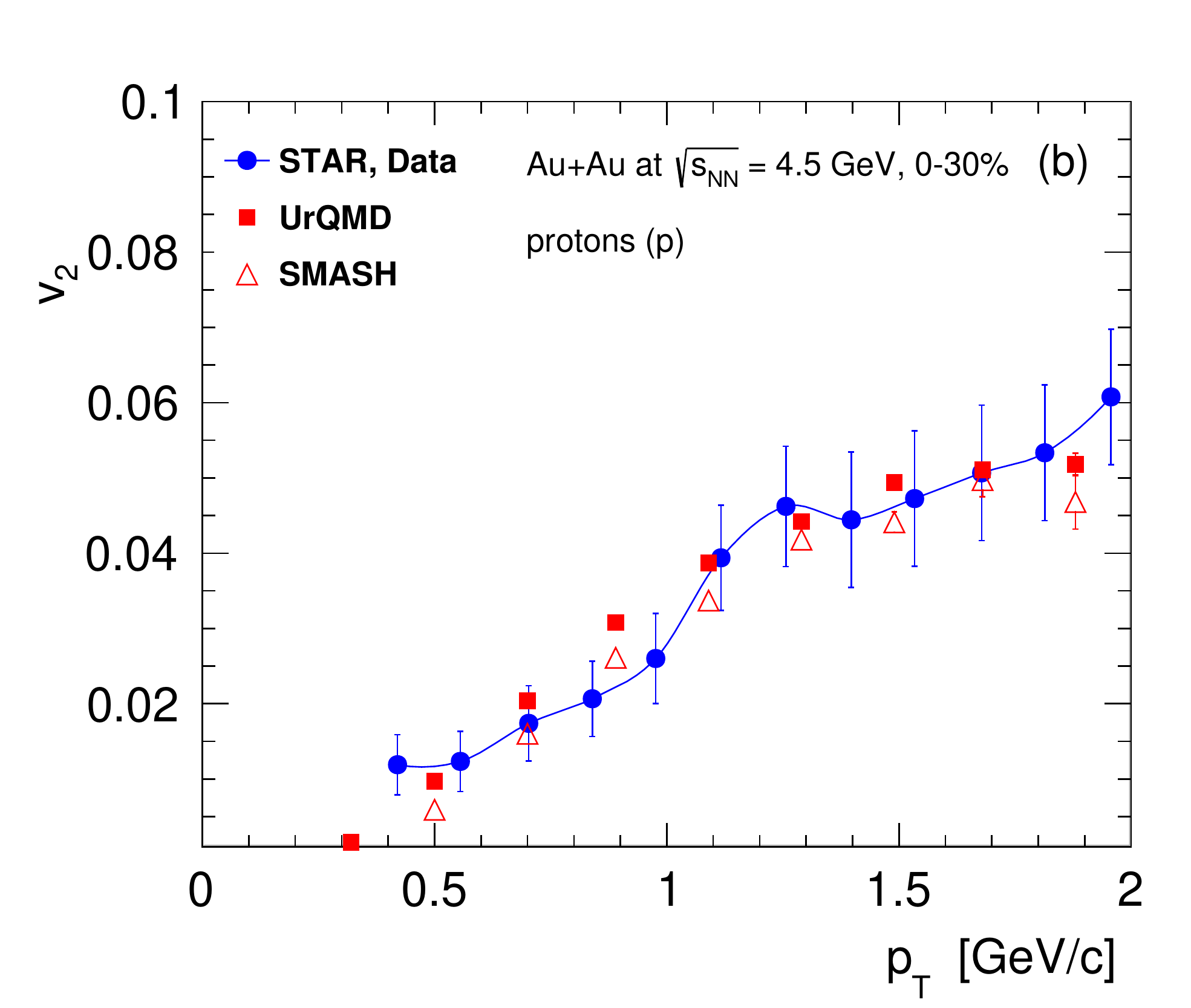}
	\caption{ $p_{\rm T}$ dependence of $v_2$ of protons from (a) 10-40\% midcentral Au+Au collisions at
          $\sqrt{s_{\rm NN}}$ = 7.7~GeV and (b) 0-30\% central Au+Au collisions at
          $\sqrt{s_{\rm NN}}$ = 4.5~GeV. Blue closed circles correspond to
          the experimental data from STAR experiment \cite{Adamczyk:2015fum,Adam:2020pla}, other symbols
          to the results from different models as indicated.}
	\label{fig:v2bar7745}
\end{figure}

The elliptic flow $v_2$ is one of the most extensively studied observables in
relativistic nucleus-nucleus collisions and was measured in
different experiments in  the last three  decades~\cite{Lacey:2005qq,Taranenko:2019uyv}.
However, high-statistics differential measurements of $v_2$ as a function
of centrality, $p_{\rm T}$ and rapidity,
for different particle species, are available only at two beam energy domains:  \acrshort{rhic}/\acrshort{lhc} ($\sqrt{s_{\rm NN}}$ = 7.7 - 5200 GeV) and
\acrshort{sis} ($\sqrt{s_{\rm NN}}$ = 1-2 GeV)~\cite{Taranenko:2019uyv}.
The collision energy dependence of the
elliptic flow ($v_2$)  for inclusive and identified hadrons at mid-rapidity
in Au+Au collisions, has been studied very extensively by the STAR experiment 
at $\sqrt{s_{\rm NN}}$ = 7.7 - 200 GeV during the \acrshort{rhic} \acrshort{bes}-I program \cite{Adamczyk:2012ku,Adamczyk:2015fum}.
The elliptic flow signal $v_2(p_{\rm T})$  for  charged hadrons
shows a very small change over such a wide range of collision energies \cite{Adamczyk:2015fum}.
According to the hybrid transport+viscous hydrodynamics approach, this $v_2(p_{\rm T})$ behavior
may result from the interplay between the hydrodynamic and hadronic transport phases~\cite{Auvinen:2013sba,Karpenko:2013wva}.
Calculations show
that the  transport
dynamics of hadrons  become more important at lower energies and are
able to compensate for the reduction of the hydrodynamically
produced $v_2$ flow \cite{Auvinen:2013sba,Karpenko:2013wva}.
In order to describe  the existing $v_2(p_{\rm T})$ results at NICA energies,
calculations with
state-of-the-art models were performed \cite{Taranenko:2020tzn,Parfenov:2020msw}. The following models
were used:  hybrid vHLLE+\acrshort{urqmd} \cite{Karpenko:2013wva},
cascade version of \acrshort{urqmd}~\cite{Bleicher:1999xi,Bass:1998ca}, \acrshort{smash} \cite{Weil:2016zrk}, JAM \cite{Nara:2019crj} and
the string melting version of AMPT \cite{Lin:2004en}.  Hybrid models with QGP formation, {\it e.g.} 
the viscous hydro + hadronic cascade  \gls{vhlle}+\acrshort{urqmd} model \cite{Karpenko:2013wva}
or the string melting version of AMPT \cite{Lin:2004en}
provide a relatively good description of $v_2(p_{\rm T})$ of protons in Au+Au collisions
at $\sqrt{s_{\rm NN}}$ = 7.7~GeV and above, see Fig.~\ref{fig:v2bar7745} (a). 
Pure hadronic transport models
(\acrshort{urqmd}, SMASH, JAM) generally underestimate the measured $v_2$ values. The situation
is different for Au+Au collisions at $\sqrt{s_{\rm NN}}$ = 4.5~GeV, see Fig.~\ref{fig:v2bar7745} (b).
Here, the pure hadronic transport system (as described by the \acrshort{urqmd} and SMASH models)
seems to explain the new $v_2(p_{\rm T})$ measurements from the STAR fixed-targed program  \cite{Adam:2020pla}.

The results of model to data comparison for $v_2$ at  $\sqrt{s_{\rm NN}}=7.7$ GeV and 4.5 GeV may indicate that at
NICA energies a transition occurs from partonic to hadronic matter. The high-statistics differential measurements of $v_n$, that are
anticipated from the MPD experiment at NICA, are expected to
provide  valuable information about this parton-hadron transient energy domain.
The  performance of the MPD detector for differential anisotropic flow measurements of identified hadrons
at NICA energies will be discussed in Sec.~5.5.

\subsection{Intensity interferometry}

\begin{figure}[t]
\includegraphics[width=8cm]{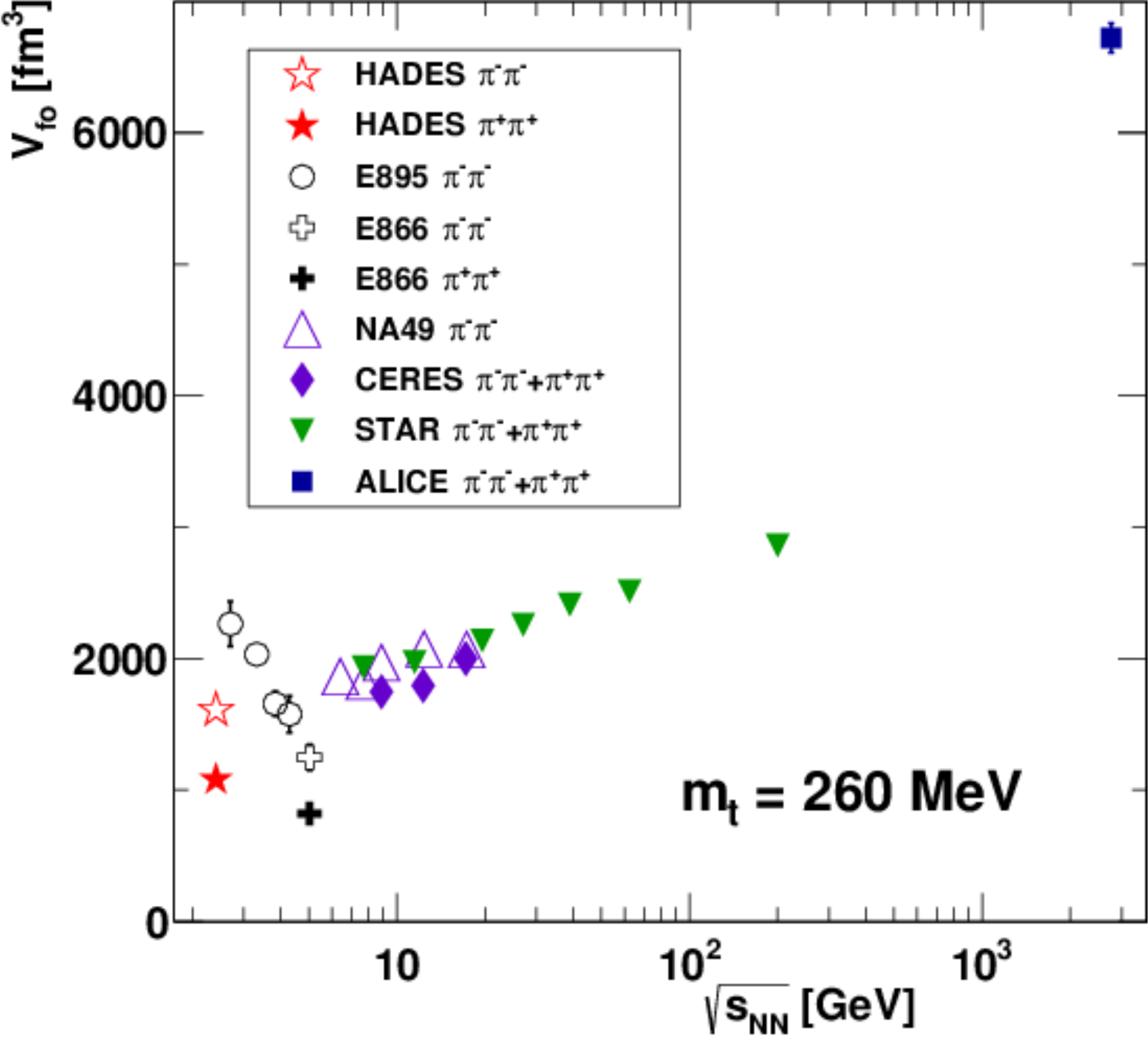}
\caption{\label{fig:Vfreezeout} Dependence of the freeze-out volume for pions on the collision energy. Compilation taken from~\cite{Adamczewski-Musch:2018owj}.}
\end{figure}

Intensity interferometry, usually referred to as {\it femtoscopy}, is used extensively in heavy-ion collision studies to determine the size of the particle-emitting produced system and consequently, details of the space-time dynamics of that system’s evolution~\cite{Kopylov:1972qw,Kopylov:1975rp,Podgoretsky:1989bp,Lednicky:2003mq,Pratt:1984su,Lisa:2005dd}. In particular, two-pion measurements are straightforward to perform due to the high statistics of pion production and the well understood methodology. The technique of the correlation function used in the measurements is, to a good approximation, insensitive to single particle acceptance effects. Thus, there are no strict requirements on the precision of the calibration process. At the same time it provides a critical and sensitive probe of the two-particle tracking and PID efficiency. As a result, measurements of the two-pion femtoscopic correlation functions are usually among the first ones performed at accelerator experiments immediately after their startup~\cite{Adler:2001zd,Aamodt:2010jj} and as such, are excellent candidates for first-day physics measurements. Femtoscopy measurements have been performed for several decades, as a function of collision energy, colliding system, collision centrality, pair transverse momentum, reaction plane orientation and more~\cite{Aamodt:2011mr,Aamodt:2011kd,Abelev:2013pqa,Abelev:2014pja,Adam:2015pya,Adam:2015vja,Adam:2015vna,Adamova:2017opl,Acharya:2018dpu,Adler:2001zd,Adams:2003vd,Adams:2004yc,Abelev:2009bw,Abelev:2009tp,Aggarwal:2010aa,Adamczyk:2014mxp,Adamczewski-Musch:2018owj}. The dependence on collision energy of the freeze-out volume, obtained from two-pion interferometry shown in Fig.~\ref{fig:Vfreezeout}, is of particular interest for the MPD. A rather striking non-monotonic behavior of the volume is observed in the NICA energy range. However, it is unclear whether this can be explained by the onset of multiparticle production in elementary reactions or by systematic uncertainties. The measurements at energies above $\sqrt{s_{\rm NN}}=7.7$ GeV were performed with detectors in a collider geometry, whereas the results obtained at lower energies suffer from limited statistics and were performed in fixed-target experiments. Some results are several decades old, and were not analyzed with modern femtoscopic techniques. More precise data, based on a large-statistics sample in a collider geometry experiment and analyzed with state-of-the-art techniques will be provided by the MPD.

It has been argued~\cite{Batyuk:2017smw,Wielanek:2016xzf} that a first-order phase transition will extend the lifetime of the system created in a heavy-ion collision. An expanding system living longer will naturally reach a larger size at freeze-out. The femtoscopic size of the system can be measured in three directions in the so-called Bertsch-Pratt decomposition: \lq\lq long'' along the beam axis, \lq\lq out'' along the transverse momentum of the pair, and \lq\lq side'' perpendicular to the other two. It is argued that the emission duration strongly affects the size of the system in the direction of collective flow (associated with the \lq\lq out'' direction), while the \lq\lq side'' direction is unaffected~\cite{Sinyukov:1994vg,Sinyukov:2011mw}. The analysis of $R_{\rm out}^{2}-R_{\rm side}^{2}$ is proposed as a sensitive probe of the duration of the particle emission stage~\cite{Lacey:2015yxg}. In particular the existence of the deconfined phase and the phase transition itself may strongly affect these radii. Their measurement can potentially provide critical signatures of such transition. Therefore, measuring the size of the colliding system, in the NICA collision energy range, is an important ingredient for the search of the existence and nature of the transition between deconfined and hadronic matter. 

\subsection{Fluctuations}
As previously emphasized, an important feature of the conjectured QCD phase structure is the existence of a CEP, followed by a first order phase transition line at higher values of $\mu_B$. If the phase trajectory of the system passes near this point, significant variations of the system’s thermodynamical parameters are expected~\cite{cep:stepanov}. Such variations can be found in the analyses of event-by-event fluctuations of conserved charges, for example, the baryon number or strangeness. QCD-based calculations indicate that the moments of event-by-event multiplicity distributions, as well as their combinations, are sensitive to the correlation length, a characteristic parameter of a phase transition, and to the susceptibilities of the conserved charges~\cite{cep:karsch}. Moreover, higher moments of the distributions of conserved quantities have stronger dependence on the correlation length. The promising higher moments are the skewness, $S=(\delta N)^3/\sigma^3$ and kurtosis $k=[(\delta N)^4/\sigma^4]-3$, where $\delta N=N-M$, $M$ is the mean and $\sigma$ is the standard deviation of the distribution.

\begin{figure}[t]
\includegraphics[width=8cm]{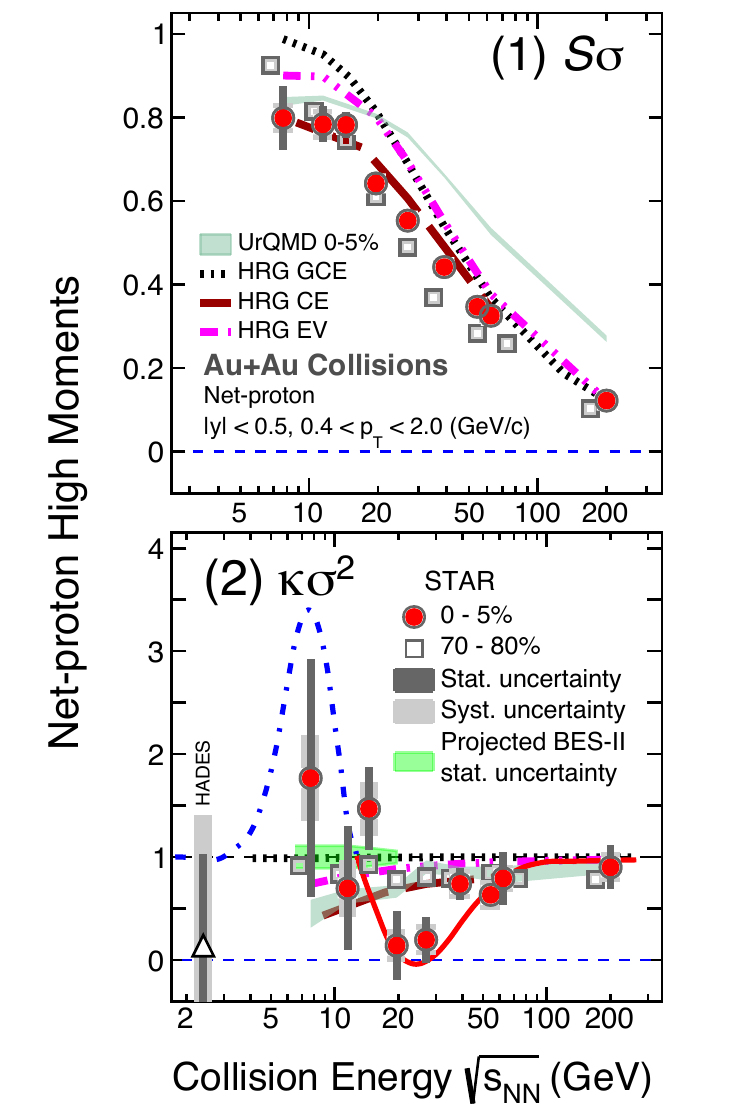}
\caption{\label{fig:skewness-kurtosis} 
Dependence of scaled skewness (1) and kurtosis (2) for 70-80\% peripheral (open squares) and 0-5\% central (filled circles) Au+Au collisions on the collision energy. 
The solid red and the dashed blue line in (2) are schematic representations of the expectation from a QCD based model calculation in the presence of a critical end point. Compilation taken from Ref.~\cite{Mohanty:2021uva}.}
\end{figure}

Measurements of event-by-event fluctuations have been performed by the NA49, PHENIX, NA61/SHINE and STAR Collaborations~\cite{cep:na49_1,cep:na49_2,cep:na49_3,cep:phenix_2,cep:star_1,Aduszkiewicz:2015jna}. Recent STAR measurements from the RHIC-\acrshort{bes} program~\cite{Adam:2020unf} indicate a non-monotonic behaviour of the excitation function for the net-proton moments in central Au+Au collisions in the region below $\sqrt{s_{\rm NN}}$=20 GeV, which can be a hint for the critical point in the range of finite baryon number density, see 
Fig.~\ref{fig:skewness-kurtosis}. 
The MPD experiment at NICA will be able to scan the region of the collision energies $\sqrt{s_{\rm NN}}$= 4-11 GeV with significantly higher precision.

\subsection{Short-lived resonances}
The study of short-lived hadronic resonances such as $\rho(770)$, $K^{*}(892)$, $\phi(1020)$, $\Sigma(1385)$, $\Lambda(1520)$ and $\Xi(1530)$ has played an important role in the physical program of many heavy-ion experiments. Resonances probe crucial aspects of hadronic and heavy-ion collisions. They carry information about the hadron chemistry and strangeness production, about the reaction dynamics and processes that shape the particle transverse momentum ($p_{\rm T}$) spectra, and about the density and lifetime of the hadronic phase.

Since the mid 80s, the enhanced strangeness production in heavy-ion collisions is considered as a signature of the Quark-Gluon Plasma formation. Such an enhancement has been experimentally observed in heavy-ion collisions at AGS, SPS, RHIC, and LHC energies \cite{Koch:1986ud,Adam:2015vsf,Agakishiev:2011ar,Aggarwal:2010ig,Abelev:2007xp}. The enhancement was observed for ground-state hadrons ($K$, $\Lambda$, $\Xi$, $\Omega$) as well as for resonances ($\phi(1020)$, $\Sigma(1385)$, $\Lambda(1520)$, $\Xi(1530)$). Canonical suppression models \cite{Vislavicius:2016rwi} reproduce most of these results except for measurements of the $\phi(1020)$ meson, which is predominantly made of $s\bar{s}$ pairs and thus has hidden strangeness. Since this meson is not sensitive to canonical suppression, it represents a key observable for the study of the mechanisms responsible for strangeness production. Previous measurements of the $\phi(1020)$ meson in heavy-ion collisions at energies $\sqrt{s_{\rm{NN}}}=17-5000$~GeV revealed that it behaves like a particle with open strangeness. Currently, there is no model that reproduces the experimentally measured enhancement of particles containing $s$-quarks in high-multiplicity p+p, p+A and A+A collisions.

Another well-known phenomenon is the increase of baryon-to-meson ratios at intermediate momentum in central heavy-ion collisions at RHIC and LHC energies \cite{Adam:2017zbf,Abelev:2014uua,Adare:2010pt,Adare:2014eyu}. The driving force for the observed enhancement is not well defined. It could be a particle mass effect due to collective radial flow or a quark content effect inherent to coalescence models. In this sense, the measurement of baryon-to-meson ratios for particles of similar mass could help to disentangle these different production mechanisms.  Resonances like $\phi(1020)$ and $K^{*}(892)$ are mesons with masses close to that of the proton (which is a baryon). Measurements of $p_{\rm T}$-differential p/$\phi(1020)$ and p/$K^{*}(892)$ ratios will help in the study of mechanisms that shape particle spectra at low-to-intermediate transverse momenta. Measurements at RHIC and the LHC showed a flattening of the p/$\phi(1020)$ ratio in central heavy-ion collisions at $p_{\rm T} < 4$~GeV/$c$, indicating that particle spectral shapes are driven by particle masses, as predicted by hydrodynamics. However, the measurements could not completely rule out the coalescence models.

In the net baryon-rich regime of heavy-ion collisions, the subthreshold production of kaons and other strange hadrons is a characteristic feature found, {\it e.g.}, by the KaoS Collaboration at the SIS. This is important to be confirmed by the MPD in the energy regime just above SIS since this effect is a sensitive indicator of modifications of hadron properties and kinematic conditions due to the presence of a dense nuclear medium. 

\begin{figure}[t]
\includegraphics[width=8cm]{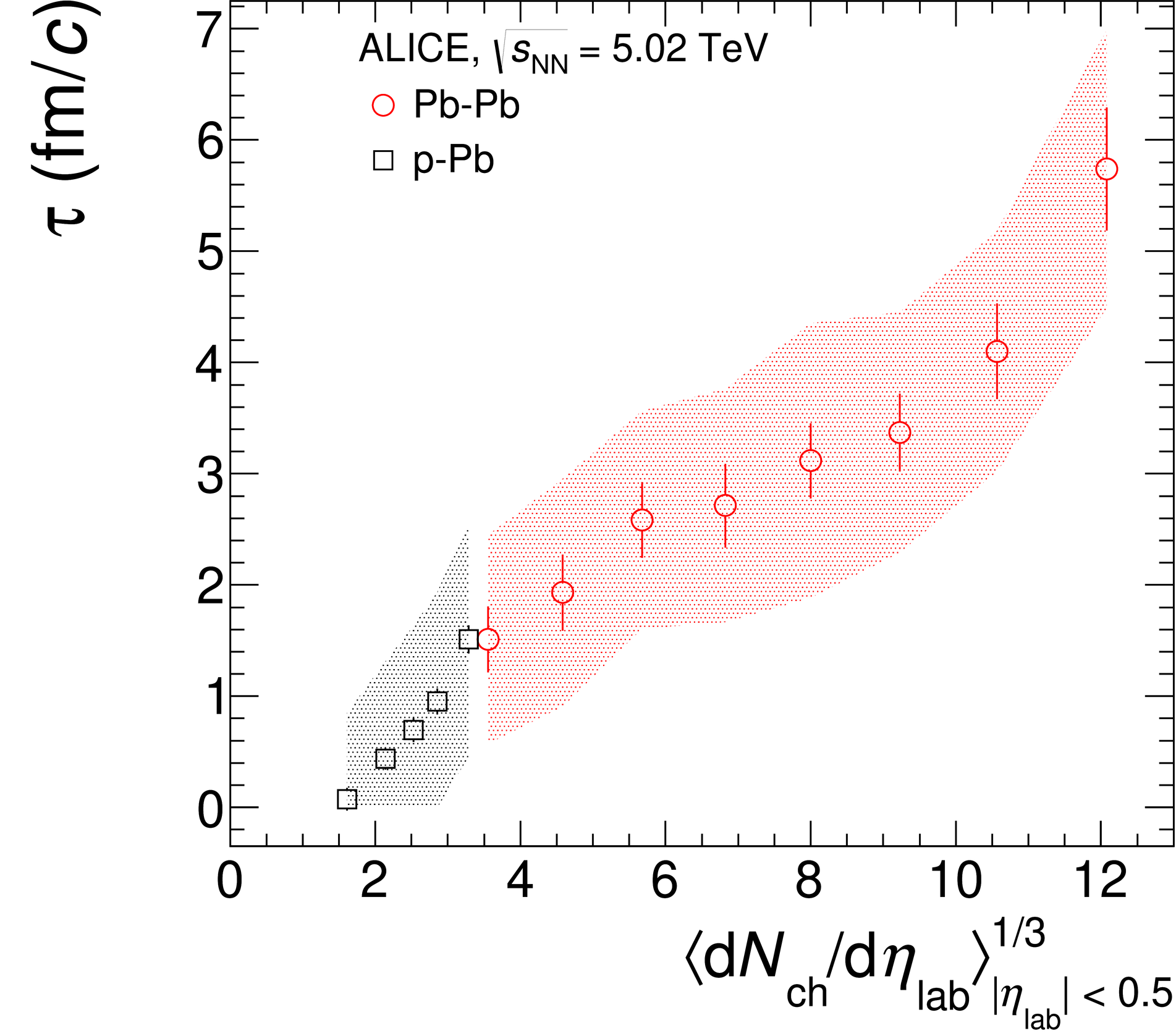}
\caption{\label{fig:hadphase} 
Lower limit on the hadronic phase lifetime between chemical and kinetic freezeout as a function of the charged particle multiplicity in p+Pb and Pb+Pb collisions at $\sqrt{s_{\rm{NN}}}=5$~TeV. Taken from \cite{Acharya:2019qge}.
}
\end{figure}

Resonances are characterized by short lifetimes. A fraction of these particles decay within the fireball. Decay daughters can scatter in the hadronic phase, change direction or magnitude of their momentum vector, thus preventing the reconstruction of the parent particles. At the same time, copiously produced hadrons in the gas can recombine and form new resonances. As a result, resonance yields in the final state are defined by the resonance yield at chemical freeze-out, the lifetime of the resonances, and the lifetime and density of the hadronic phase
as well as the rescattering cross-sections. Resonances cover a wide range of lifetimes from $\sim1$~fm/$c$ for $\rho(770)$ up to $\sim45$~fm/$c$ for $\phi(1020)$. This makes these particles well suited to study  the hadronic phase properties. Previous measurements at RHIC and LHC energies~\cite{Acharya:2018qnp,Adams:2004ep,Adams:2006yu,ALICE:2018ewo,Adams:2004ux,Adler:2004hv} showed that the production of resonances with lifetimes shorter than 20 fm/$c$ is suppressed in central heavy-ion collisions compared to measurements in p+p and peripheral heavy-ion collisions. This suppression was interpreted as a result of the rescattering of daughter particles in the hadronic phase. At the same time, the production of resonances with longer lifetimes is not affected when going from p+p to central heavy-ion collisions. Fig.~\ref{fig:hadphase} shows the obtained lifetime of the hadronic medium in A+A and p+A collisions at $\sqrt{s_{\rm{NN}}}=5$~TeV as a function of the multiplicity of produced charged hadrons. At high multiplicities, which are characteristic of events with the formation of the QGP, the lifetime of the hadronic medium reaches values as high as $\tau \sim 6$~fm/$c$, which is comparable with the QGP lifetime~\cite{Acharya:2019qge}.

Heavy ions at NICA will collide at energies of $\sqrt{s_{\rm NN}} = 4-11$~GeV corresponding to final state charged-particle multiplicities of $dN_{ch}/d\eta \sim 100-200$ at mid-rapidity~\cite{Adare:2015bua}. At such multiplicities, the production of resonances at RHIC and LHC energies is significantly modified. Study of the resonance properties in Au+Au collisions at $\sqrt{s_{\rm NN}} = 4-11$~GeV using the UrQMD, PHSD and AMPT event generators confirmed the expectations~\cite{Ivanishchev:2019nqn,Ivanishchev:2019jgh}. The event generators confirm modifications of the resonance yields similar to those observed at RHIC and the LHC. It means that the lifetime and density of the hadronic phase in heavy-ion collisions at NICA are expected to be large -- of the same order as the total system lifetime. The effect of the hadronic phase should be taken into account when model predictions for different observables are compared to experimental measurements.

\subsection{Electromagnetic probes}
Electromagnetic probes, virtual (e$^+$e$^-$ or $\mu^+\mu^-$ pairs) and real photons are unique probes of the QGP.  They are sensitive to its two main characteristic properties, the deconfinement of quarks and gluons and the chiral symmetry restoration. They are penetrating probes in the sense that once produced, they propagate through the strongly interacting matter created in ultra-relativistic heavy-ion collisions, experiencing only the electromagnetic force and carrying information about the medium at the time of production. The main topic of interest is the identification of the thermal radiation emitted by the system
both, at the early stage of the collision where QGP formation and
deconfinement occur, as well as during the subsequent hadronization phase,
where chiral symmetry restoration effects take place.
For a theoretical review see Ref.~\cite{Rapp:2009yu} and for an experimental one see Ref.~\cite{Tserruya:2009zt}.

The MPD is very well suited for the study of electron pairs, covering the entire invariant mass range of interest from the low-masses of the $\pi^0$ Dalitz decay up to the J/$\psi$, as well as real photons. It benefits from a large acceptance, excellent particle ID and the NICA high integral luminosity. 
Electrons will be identified via their $\langle{\rm d}E/{\rm d}x\rangle$ in the \acrshort{tpc}, time of flight in the \acrshort{tof} and $E/p=1$ in the \acrshort{ecal}, whereas real photons will be identified in the \acrshort{ecal}. The MPD will focus on an energy range that has not been covered in the study of EM probes, but is crucial for understanding the thermal radiation. 

\begin{figure}[t]
\includegraphics[width=8cm]{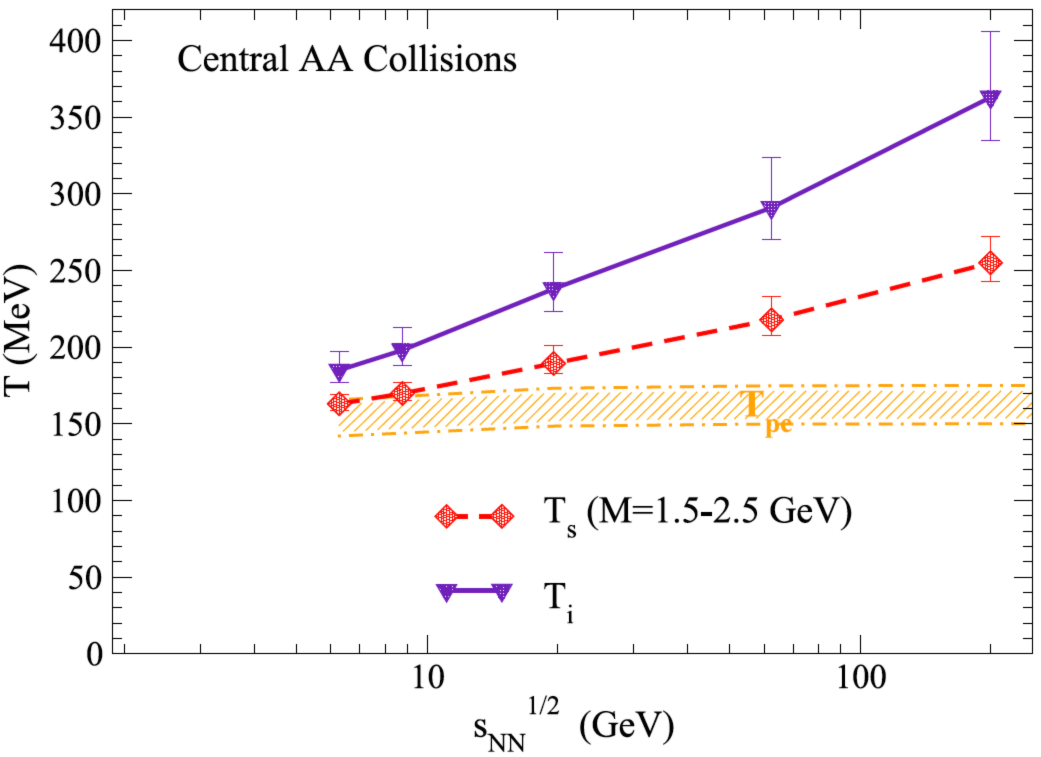}
\caption{\label{fig:thermometer} Excitation function of the inverse-slope parameter, $T_s$, from intermediate-mass dilepton spectra ($M=1.5-2.5$ GeV, diamonds connected with dashed line) as \lq\lq thermometer'' and initial temperature $T_0$ (triangles connected with solid line) in central heavy-ion collisions (A$\simeq 200$). The hatched area schematically indicates the pseudo-critical temperature regime at vanishing (and small) chemical potential as extracted from various quantities computed in lattice QCD \cite{Borsanyi:2010bp}.
Taken from Ref.~\cite{Rapp:2014hha}.
}
\end{figure}

The dilepton yield in the intermediate mass region (IMR), 
$m_{l^+l^-} = 1 - 3$ GeV/c$^2$,  has been singled out as the most appropriate region to observe the thermal radiation from the QGP. However, so far this radiation has been identified only in In+In collisions at the SPS energy of 160 GeV \cite{Arnaldi:2008fw}. 
At higher energies, like at RHIC, the measurements are very challenging due to the contribution of dilepton pairs form the semi-leptonic decays of charm and bottom mesons. At NICA energies, this contribution is negligible allowing a relatively clean measurement of the thermal radiation in the IMR (with only an additional contribution from Drell-Yan pairs that can be independently measured in p+p and 
p+A collisions).  The inverse-slope parameter, $T_s$, of the invariant mass spectra in this mass range is closely related to the initial temperature $T_i$ of the fireball and thus can be regarded as a~\lq\lq thermometer" for the heavy-ion collision, see Fig.~\ref{fig:thermometer} and Ref.~\cite{Rapp:2014hha}. 

A very interesting result of these calculations is the prediction of QGP thermal radiation down to collision energies as low as 
$\sqrt{s_{\rm NN}} = 6-8$ GeV setting the onset of QGP formation well within the energy range covered by NICA. It should be emphasized that the model calculations of Ref.~\cite{Rapp:2014hha} are very successful in reproducing all dilepton measurements including the benchmark results from the SPS~\cite{Borsanyi:2010bp,CERES:2006wcq} and the RHIC dilepton program~\cite{Adamczyk:2015lme,STAR:2018xaj}, thus lending credibility to this prediction. 

\begin{figure}[t]
\includegraphics[width=8cm]{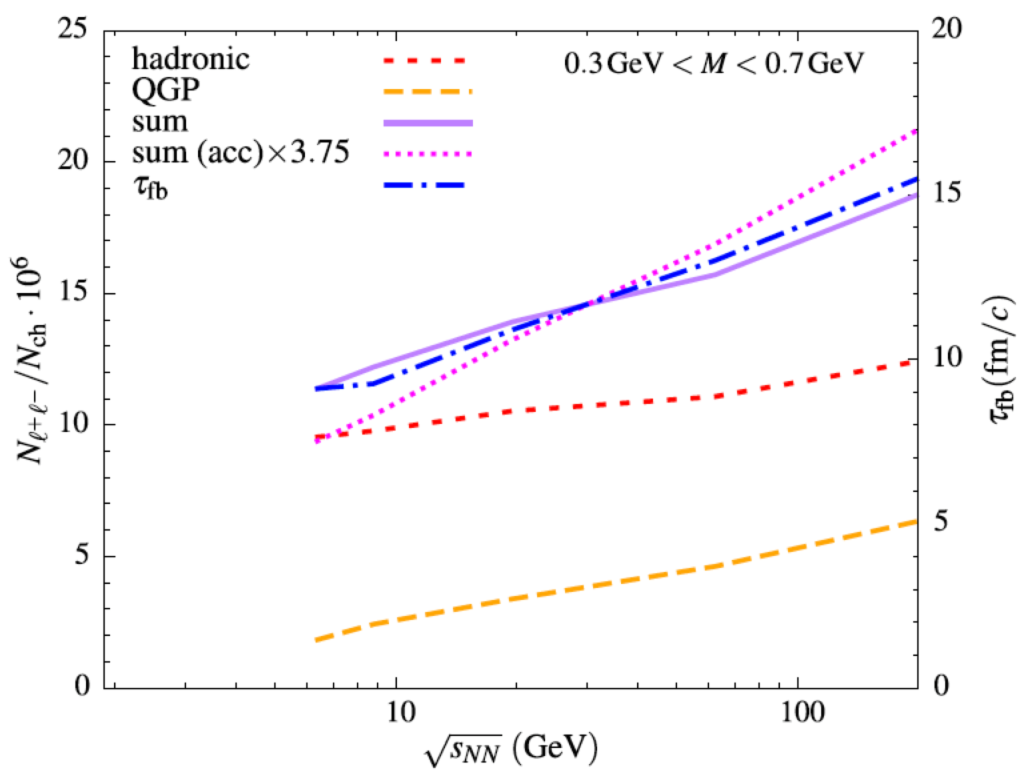}
\caption{\label{fig:chronometer} 
Excitation function of low-mass thermal dilepton radiation (\lq\lq excess spectra'') in $0-10\%$ central A+A
collisions (A$\simeq 200$), integrated over the mass range $M=0.3$~${\rm GeV}-0.7$~${\rm GeV}$, for QGP (dashed line) and in-medium hadronic (short-dashed line) emission and their sum (solid line). The underlying fireball lifetime (dot-dashed line) is given by the right vertical scale which therefore qualifies the corresponding experimental data as a \lq\lq chronometer''. Taken from Ref.~\cite{Rapp:2014hha}.
}
\end{figure}

The dilepton yield in the low-mass region (LMR), $m_{l^+l^-} < 1 $
GeV/c$^2$, has been studied extensively both theoretically and
experimentally. An enhancement has been observed in all heavy-ion
collision systems and at all energies measured
\cite{Arnaldi:2008fw,CERES:1995vll,CERES:2006wcq,STAR:2015tnn,STAR:2018xaj,PHENIX:2015vek,Adamczewski-Musch:2019byl}. 
At the SPS and RHIC energies, the excess, dominated by the thermal radiation from the hadron gas, is linked to the restoration of chiral symmetry that manifests itself in a broadening of the $\rho$  meson spectral function \cite{Rapp:2009yu,Tserruya:2009zt}. 
The same interpretation holds also for the dilepton excess recently observed by the HADES experiment in Au+Au collisions at  $\sqrt{s_{\rm NN}} = 2.42$ GeV~\cite{Adamczewski-Musch:2019byl}. 
The dilepton excess, integrated over an appropriate mass window below the $\rho$ mass, $m_{l^+l^-} = 0.3$–$0.7$~GeV/c$^2$, was found to be proportional to the fireball lifetime. This is illustrated in 
Fig.~\ref{fig:chronometer}, that shows the QGP (dashed line) and the hadronic (short-dashed line) contributions and their sum (solid line) together with the calculated fireball lifetime \cite{Arnaldi:2008fw}. The dilepton excess in the LMR can therefore be considered as a \lq\lq chronometer'' of the fireball \cite{Rapp:2014hha}. 
Dilepton results from the STAR experiment beam energy scan are in good agreement with this prediction \cite{STAR:2018xaj}. Deviations from the proportionality of the dilepton excess and the fireball lifetime could then signal lifetime variations associated with the conjectured critical phenomena and onset of a first order phase transition.   

Real photons are expected to carry the same physics information as virtual photons and are thus an important part of the physics program of relativistic heavy-ion experiments. The inclusive photon yield includes \lq\lq direct'' photons on top of an overwhelming background originating from hadronic decays, mostly from $\pi^0$ 
($\sim~90\%$ of all photons). The \lq\lq direct'' photons are produced either in hard scattering processes or emitted as thermal radiation from the produced medium. The background is orders of magnitude larger compared to the corresponding one of the dilepton measurement, making the real photon measurement much less sensitive to any new source, and in particular to the thermal photons from the QGP \cite{Tserruya:1995mh}.  Indeed, at the high energies of RHIC and LHC, a signal of direct photons, on top of the hadronic background, has been identified \cite{PHENIX:2008uif,STAR:2016use,ALICE:2015xmh}. 
The signal is well reproduced by pQCD calculations. On the other hand, at the lower SPS energy, only an upper limit \cite{WA80:1995xza} or a small effect at the 1-2 $\sigma$ level, could be established \cite{WA98:2000vxl}. 
The MPD will make measurements of real photons by detecting them in the \acrshort{ecal} or using the conversion method, but in both cases the measurements will be quite challenging. 

Another important goal of the MPD EM program is the measurement of flow for virtual as well as real direct photons. Of particular interest is the flow measurement of dileptons in the LMR and the IMR. Being produced at different stages of the collision one should expect large differences in the flow values of dileptons in the  LMR and the IMR.  Such a measurement has not been done so far and could provide a model independent confirmation of the origin of these two dilepton groups. The flow measurement of real photons is also of great interest. Results at RHIC and the LHC revealed large values of the $v_2$  and $v_3$ flow coefficients for  direct photons comparable to those for the light hadrons \cite{PHENIX:2011oxq,PHENIX:2015igl,ALICE:2018dti}. 
Theoretical calculations cannot simultaneously describe the direct photon yield and the flow values, leading to the so-called “direct photon puzzle”. The MPD has the potential to provide additional results or constraints to this puzzle.

\section{\label{sec:detector}MPD apparatus}

\begin{figure*}[th]
  \centering
  \includegraphics[width=0.8\linewidth]{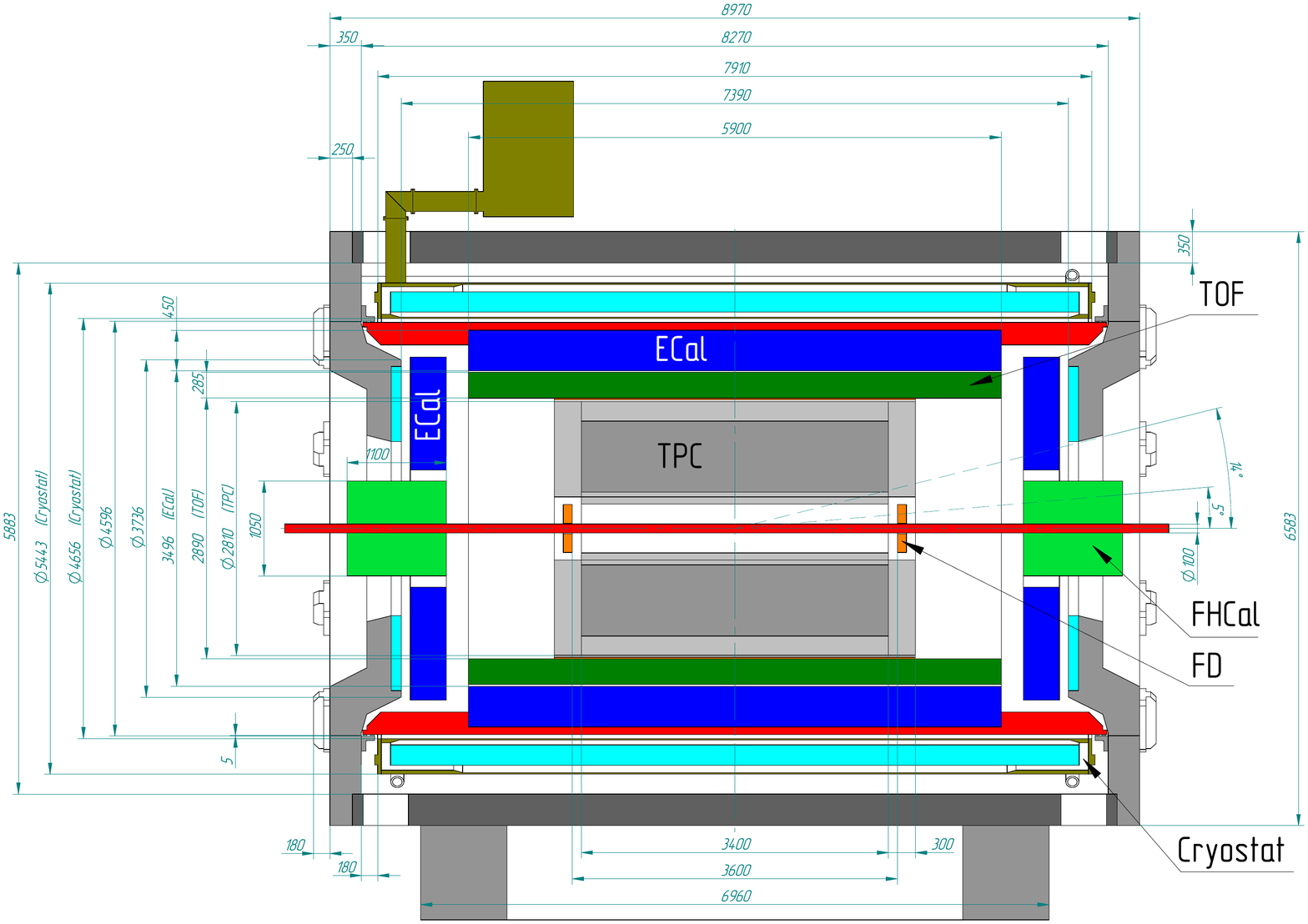}
  \caption{The overall schematic of the MPD subsystems in the first stage of operation (Stage 1) - cross-section by the vertical plane.}
  \label{fig:overallmpdcross}
\end{figure*}

\begin{figure}[th]
  \centering
  \includegraphics[width=0.8\linewidth]{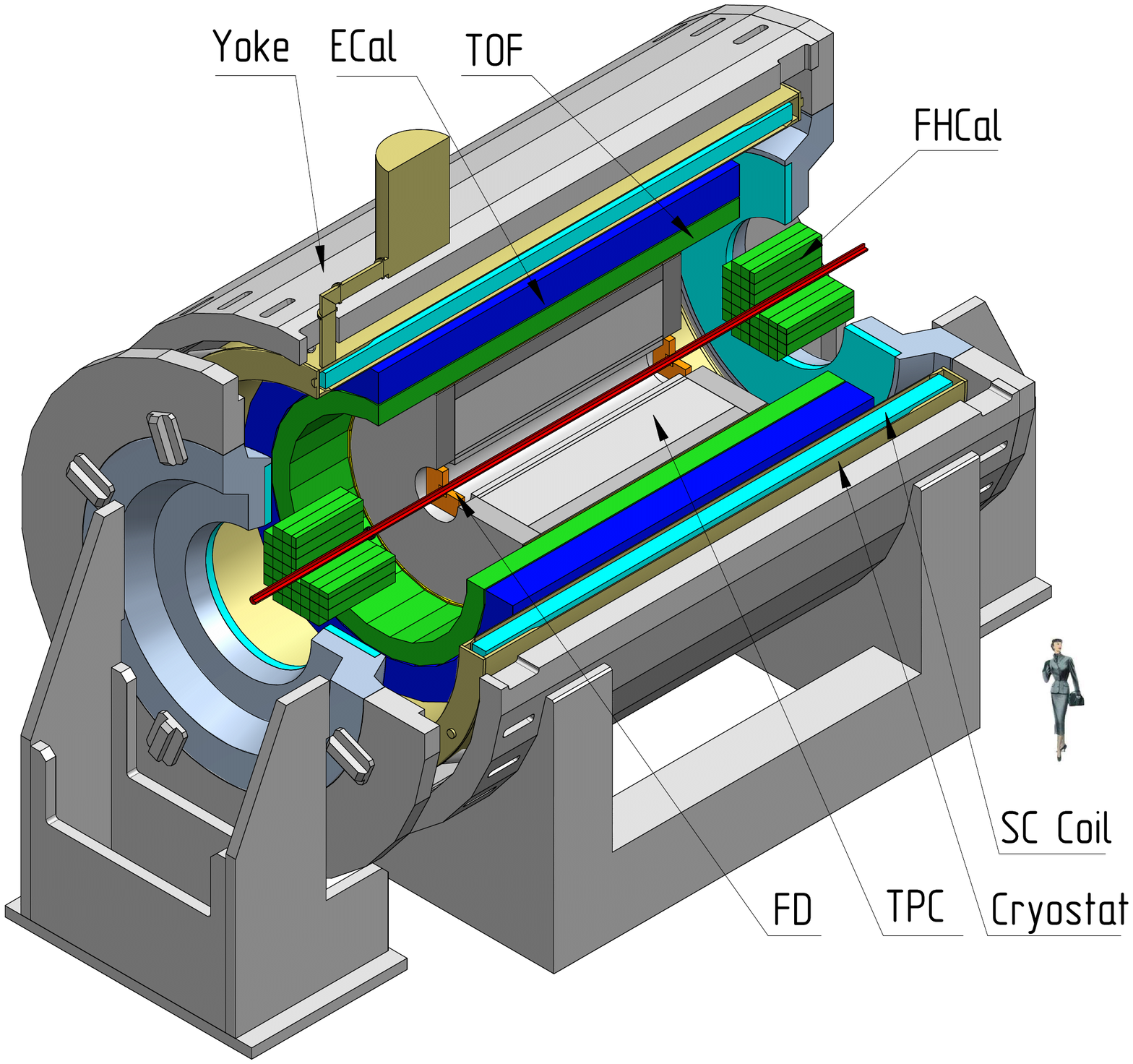}
  \caption{The overall schematic of the MPD subsystems in the first stage of operation (Stage 1) - three-dimensional view.}
  \label{fig:overallmpd}
\end{figure}

Work aimed to complete the various subsystems that make up the MPD has been systematically carried out for the last few years.
It is foreseen that the MPD will be installed in two stages. 
The first stage of the detector configuration is planned to be ready for commissioning with a beam at the end of 2022.
The overall set-up of the MPD and the spatial arrangement of detector subsystems in the first stage are shown in Figs.~\ref{fig:overallmpdcross}~and~\ref{fig:overallmpd}. The \lq\lq central barrel'' components have an approximate cylindrical symmetry. The beam line is surrounded by the large gaseous Time Projection Chamber (\acrshort{tpc}) which is enclosed by the \acrshort{tof} barrel. The \acrshort{tpc} is the main tracker, and in conjunction with the \acrshort{tof} they will provide precise momentum measurements and particle identification. The Electromagnetic Calorimeter (\acrshort{ecal}) is placed in between the \acrshort{tof} and the MPD Magnet. It will be used for detection of electromagnetic showers, and will play the central role in photon and electron measurements. 
In the forward direction, the \acrlong{ffd} (\acrshort{ffd}) is located still within the \acrshort{tpc} barrel. 
It will play the role of a wake-up trigger.
The Forward Hadronic Calorimeter (\acrshort{fhcal}) is located near
the Magnet end-caps. It will serve
for determination of the collision centrality and the orientation of the reaction plane for collective flow studies. Technical Design Reports (\acrshort{tdr}s) of all the first stage detectors are available in~\cite{MPD_TDR}.

Additional detectors are proposed in the later stages. 
The silicon-based \acrlong{its} (\acrshort{its}) will be installed close to the interaction point in the second stage of the MPD construction. It will greatly enhance tracking and secondary vertex reconstruction capabilities.
The miniBeBe detector, placed between the beam pipe and the \acrshort{tpc}, close to the beam, is designed to aid in triggering and start time determination for the \acrshort{tof}. 
The MCORD, installed on the outside of the MPD Magnet Yoke, will measure muons, also from the cosmic showers. 

In this section, we describe the characteristics of the components that the MPD will consist of
at the beginning of its operation, and briefly show plans for the additional detector components.

\subsection{\label{sec:magnet}Magnet}

\begin{figure}[th]
  \centering
  \includegraphics[width=0.95\columnwidth]{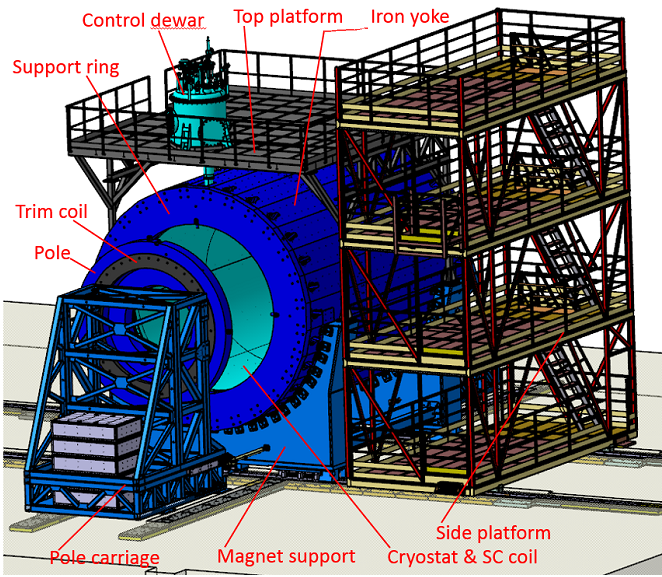}
  \caption{Schematic view of the MPD Magnet subsystem and support structures.}
  \label{fig:mpdmagnet}
\end{figure}

An essential component of the MPD is a solenoid magnet with a superconducting NbTi coil and a steel flux return yoke. It is designed to provide a highly homogeneous magnetic field of up to  $0.57$~T (with a default operational setting of $0.5$~T), uniform along the beam direction, in an aperture of 4596~mm in diameter to ensure appropriate transverse momentum resolution for reconstructed particles within the range of momenta of $0.1$–$3$~GeV/$c$.  
The MPD magnet consists of (Fig.~\ref{fig:mpdmagnet}):
\begin{itemize}
    \item A cryostat with a superconducting coil and a control Dewar;
    \item A flux return yoke with two support rings, 28 bars, and two poles with trim coils;
    \item The magnet support cradles;
    \item The auxiliary platforms for moving the poles;
    \item The roller skates for the movement of the magnet and its poles.
\end{itemize}
In addition, there are power supplies for the superconducting coil and for the trim coils in the poles, a SC coil quenching protection system, a cryogenic system with a cryogenic pipeline, a vacuum system, a helium refrigerator and a magnet control system.

The full magnet yoke has been assembled in its designated location in the MPD Pit, reaching the design mechanical precision of 200~$\mu$m or better.

\subsection{\label{sec:tpc}Time Projection Chamber}


\begin{figure}[th]
  \centering
  \includegraphics[width=0.9\columnwidth]{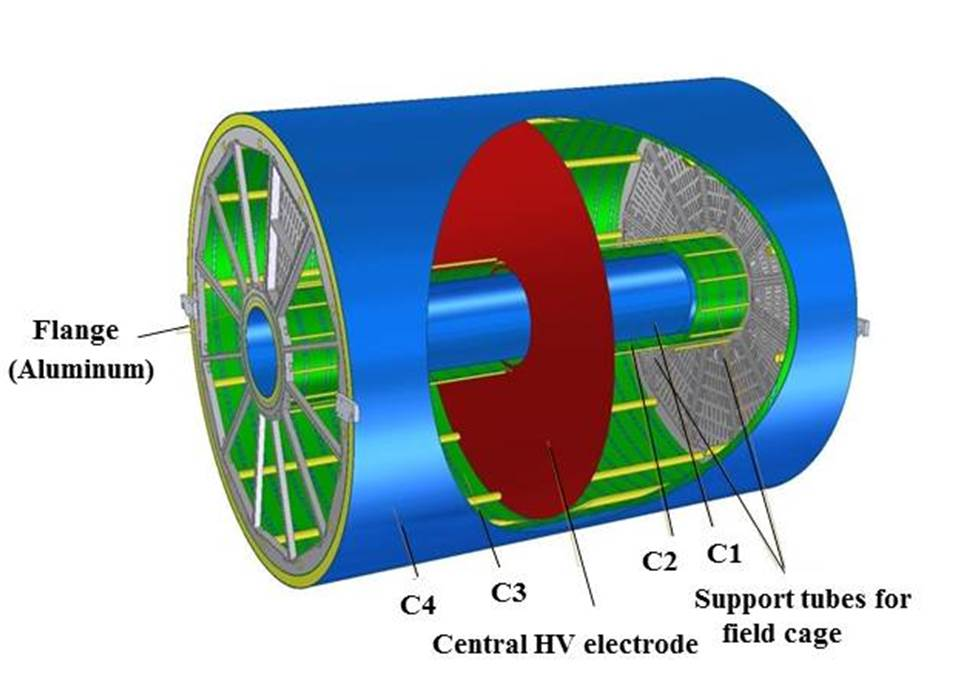}
  \includegraphics[width=0.9\columnwidth]{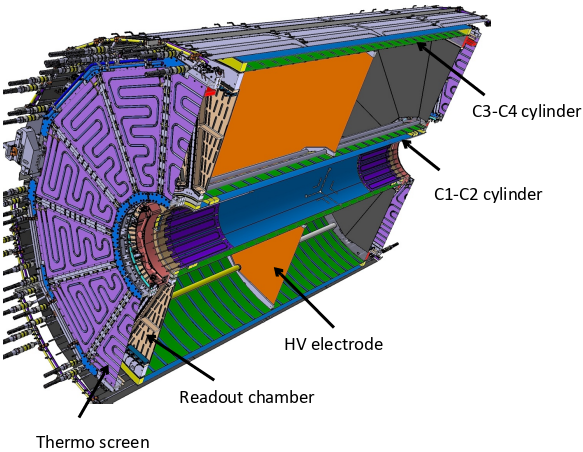}
  \caption{Schematic view of the main components of the MPD \acrshort{tpc}.}
  \label{fig:figtpc}
\end{figure}

The \acrshort{tpc} is the main tracking detector of the MPD central barrel. It is designed to perform three-dimensional precise tracking of charged particles and momentum measurements for transverse momentum $p_{\rm T}>50$~MeV/$c$. The track reconstruction is based on the drift time and R-$\phi$ cylindrical coordinate measurement of the primary ionization clusters created by a charged particle crossing the \acrshort{tpc}. Another important task for the \acrshort{tpc} is to identify the charged particles by measuring their specific ionization energy losses ($\langle {\rm d}E/{\rm d} x\rangle$) in the \acrshort{tpc} gas. 

\begin{table}[tbh]
\caption{\label{tab:tpcpar}
Key design parameters of the Time Projection Chamber of MPD.}
\centering
\begin{tabular}{|l|c|} \hline
Length & 340 cm \\
Vessel outer radius & 140 cm \\
Vessel inner radius & 27 cm \\
Drift vol. outer radius & 133 cm \\
Drift vol. inner radius & 34 cm \\
Drift vol. length & 163 cm (of each half) \\
HV electrode type & Central membrane \\
Electric field strength & $\sim$ 140 V/cm \\
Default magnetic field & 0.5 T \\
Drift gas mixture & 90\% Ar+10\% CH$_4$ \\
Pressure & Atm. pressure +2mbar \\
Gas amplification factor & $\sim$ 10$^4$ \\
Drift velocity & 5.45 cm/$\mu$s \\
Drift time & $< 30$ $\mu$s \\
Temperature stability & $<$ 0.5 $^{\circ}$C \\
Readout chambers & 24 (12 per end-plate) \\
Segmentation in $\phi$ & 30$^{\circ}$ \\
Inner pad size & 5x12 mm$^2$ \\
Outer pad size & 5x18 mm$^2$ \\
Total number of pads & 95232 \\
Pad row count & 53 \\
Maximum event rate & 7 kHz ($L = 10^{27}$ cm$^{-2}$s$^{-1}$) \\
Electronics shaping time & $\sim$ 180-190 ns \\
Signal-to-noise ratio & 30:1 \\
Signal dynamical range & 10 bits \\
Sampling rate & 10 MHz \\
Sampling depth & 310 time buckets \\
Two-track resolution & $\sim$ 1 cm \\ \hline
%
%
\end{tabular}
\end{table}

The main construction and operational parameters of the \acrshort{tpc} design are listed in Table~\ref{tab:tpcpar}. The overall schematic of the \acrshort{tpc} is shown in Fig.~\ref{fig:figtpc}. \acrshort{tpc} is a barrel with inner and outer radii of 27~cm and 140~cm, respectively and 340~cm length. The beam pipe traverses the \acrshort{tpc} along the barrel axis and the nominal interaction point is located at a geometrical center of the \acrshort{tpc}. The inner volume is divided in two halves by the central electrode, generating a uniform electrical field $\vec{E}$ of 140~V/cm along the axis. Read-out chambers (ROC) consisting of MultiWire Proportional Chambers are located at the end-caps, with 12 \acrshort{roc}s per end-cap. Each \acrshort{roc} has a trapezoidal pad plane with 53 pad rows (perpendicular to the radial direction). Pads have a width of 5~mm along the row. The row height is 12~mm for inner rows and 18~mm for outer rows. The MPD Magnet is designed to provide a highly uniform magnetic field $\vec{B}$ of up to 0.57~T of either polarity (the default setting being 0.5~T) along the $z$-axis inside the \acrshort{tpc} volume. Charged particles traversing this volume ionize the gas mixture of 90\%Ar+10\%CH$_4$ along helix-shaped trajectories. The ionization charge drifts to the end-caps and is collected by the \acrshort{roc}s, with time sampling (310 time buckets). This allows for three-dimensional trajectory reconstruction, as well as measurement of the specific ionization energy loss $\langle {\rm d}E/{\rm d} x\rangle$ on a track-by-track basis. The maximum design event rate for the \acrshort{tpc} is 7~kHz.

In order to minimize the uncertainty in the absolute track point position measurement by the \acrshort{tpc}, it is necessary to take into account both static and time-dependent distortions in the drift path of the ionization cloud. The static distortions are the result of non-uniformities in the magnetic $\vec{B}$ and electric $\vec{E}$ fields. The time-dependent distortions can result from the
residual space charge, ion feedback, or from spontaneous failures.
 
A calibration system that can reproduce fiducial tracks is needed to monitor the \acrshort{tpc} performance. The system has to provide on-line monitoring of the value of the drift velocity, which depends on the drift gas pressure changes (caused by changes of atmospheric pressure), the temperature, $\vec{E}\times \vec{B}$ non-collinearity and space charge effects. A UV laser system will be used to monitor the \acrshort{tpc} working regime parameters. The laser is part of the test and calibration system designed to produce a set of laser beam tracks at well-defined angles and positions.

\begin{figure}[t]
  \centering
  \includegraphics[width=0.95\columnwidth]{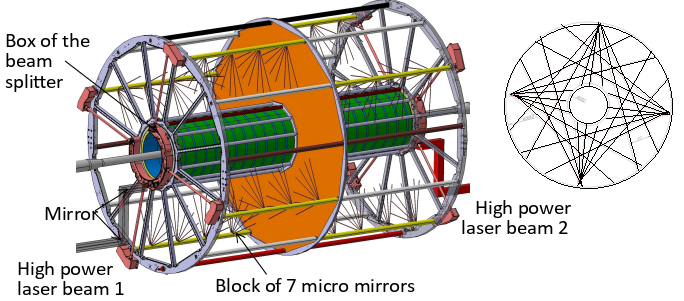}
  \caption{(Left) Scheme of the high power laser beam splitting into four equal power beams. (Right) \acrshort{tpc} quasi plane showing the positions of four mirror blocks emitting 28 beams within one plane.}
  \label{fig:figtpclaser}
\end{figure}

The system consists of two NL303HT-10-FH lasers with 110~mJ power pulsed with a 10~Hz repetition rate. A beam expander built inside the laser provides a beam diameter of the order of 18~mm at the entrance of the system. The 18~mm wide beams from each laser are split into four beams (see Fig.~\ref{fig:figtpclaser}) and then, through 4 tubes placed inside the drift volume of the \acrshort{tpc},  micro mirrors (with an active reflecting surface of 1.3 mm diameter) are illuminated to form 112 narrow calibration rays at each side of the \acrshort{tpc} HV membrane. These 112 rays are distributed into 4 equidistant quasi planes of 28 rays in each (see Fig.~\ref{fig:figtpclaser}), emitted from the 4 tubes within half of the active volume of \acrshort{tpc} (224 rays in the whole \acrshort{tpc}). This number of laser beams is sufficient to calibrate the \acrshort{tpc} at different azimuthal and polar angles and depth.

The \acrshort{tpc} vessel construction and the production of \acrshort{roc}s are well advanced. The auxiliary systems (gas system, laser calibration system, front-end electronics, HV, LV and cooling systems) are also at advanced production or commissioning stages. The whole \acrshort{tpc} is expected to be ready for commissioning with cosmic rays planned for 2022 and 2023.

\begin{figure}[hbt]
 \centering
  \begin{minipage}[h]{0.87\linewidth}
    \includegraphics[width=72mm,angle=0]{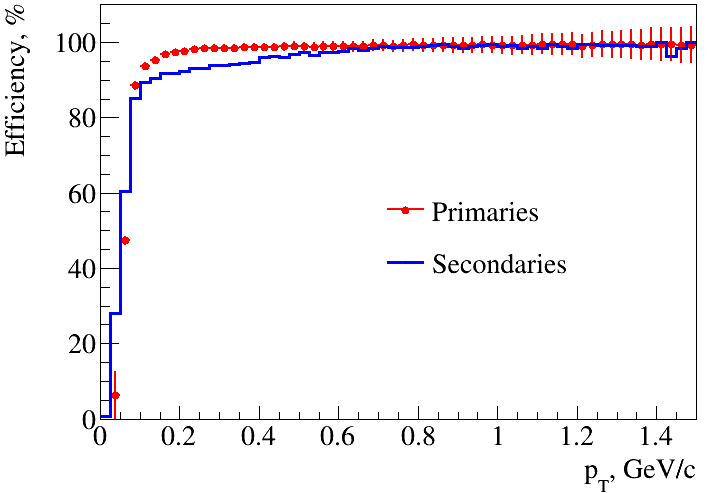}\\ \hspace*{4cm}
  \end{minipage}
  
  \begin{minipage}[h]{0.87\linewidth}
   \includegraphics[width=72mm,angle=0]{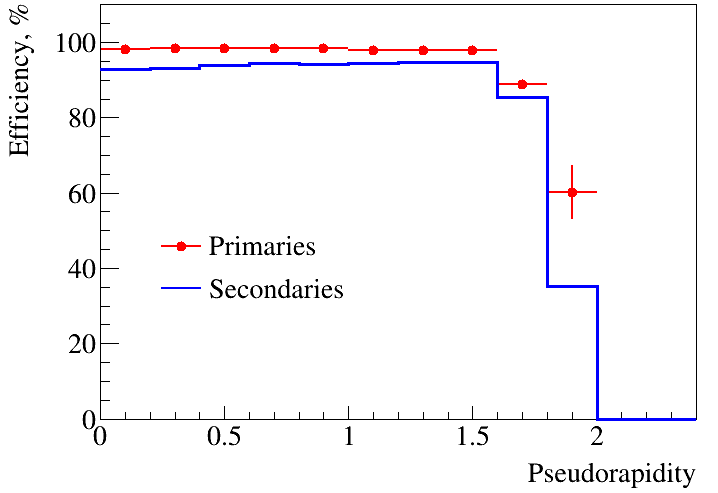}\\ \hspace*{4cm}
  \end{minipage}
  \caption{Track reconstruction efficiency for particles with the number of measured points in the \acrshort{tpc} (hits) greater than 14: (top) as a function of $p_{\rm T}$ for pseudorapidity $|\eta|<$ 1.3; (bottom) as a function of pseudorapidity $|\eta|$ for $p_{\rm T}>$ 0.1~GeV/$c$. Symbols and lines present primary and secondary particles, respectively. Secondary particles here were defined to be those produced within 50~cm from the interaction point. }
 \label{effic}
\end{figure}

\begin{figure}[hbt]
 \centering
 \begin{minipage}[h]{0.87\linewidth}
   \includegraphics[width=72mm,angle=0]{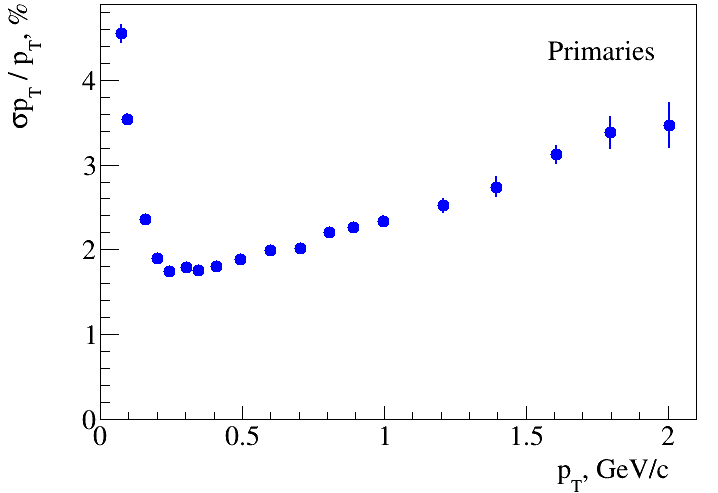}\\ \hspace*{4cm}
 \end{minipage}
 
 \begin{minipage}[h]{0.87\linewidth}
   \includegraphics[width=72mm,angle=0]{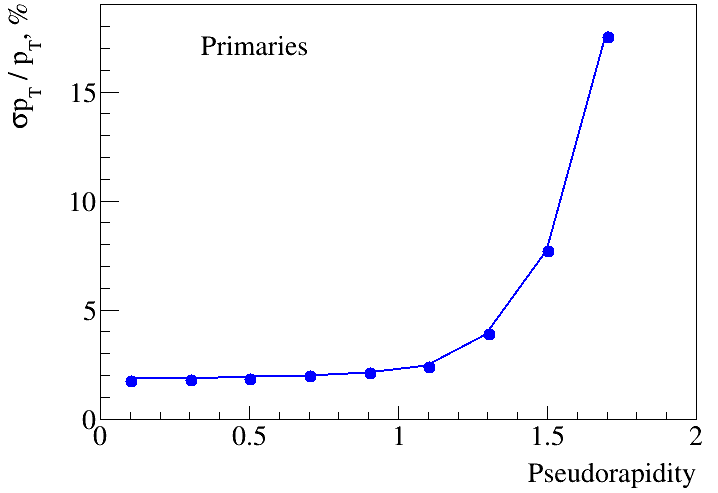}\\ \hspace*{4cm}
 \end{minipage}
 \caption{Relative transverse momentum resolution as a function of: (top) $p_{\rm T}$; (bottom) pseudorapidity $|\eta|$.}
 \label{dptpt}
\end{figure}

The expected performance of the \acrshort{tpc} was estimated by running full-event simulations of Au-Au colisions generated by \acrshort{urqmd}. The track and vertex reconstruction methods are based on the Kalman filtering technique 
(see, e.g. \cite{kalman,kalman-vertex}). The implementation details can be found in Ref.~\cite{reco}. The efficiency of track reconstruction and the relative transverse momentum resolution are plotted as function of the transverse momentum and pseudorapidity in Figs.~\ref{effic} and \ref{dptpt}, respectively, 
for charged particles with at least 15 measured points in the \acrshort{tpc}.
The reconstruction quality remains high up to $\sim |\eta|$=1.5 (with efficiency near 100\% and relative momentum resolution of $\sim$2\% at $p_{\rm T}$ of 1 GeV/$c$) for both the primary and secondary particles. The secondary track sample on the plots corresponds to particles produced within 50~cm of the interaction point.

\begin{figure}[hbt]
 \centering
 \begin{minipage}[h]{0.87\linewidth}
   \includegraphics[width=75mm,angle=0]{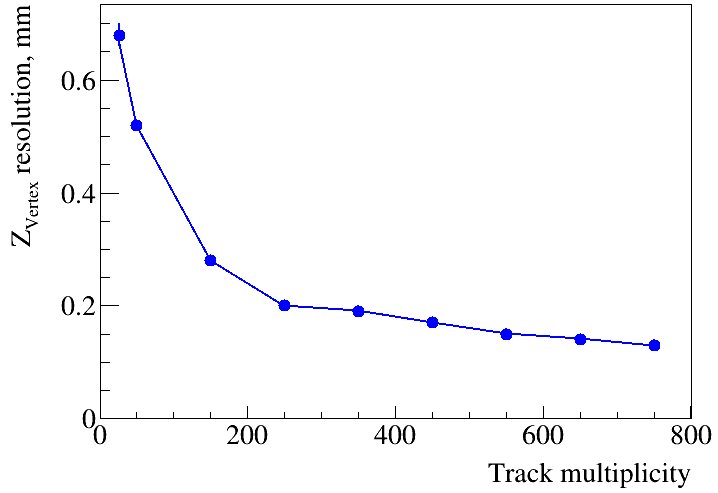}\\ \hspace*{4cm} 
 \end{minipage}
 
 \begin{minipage}[h]{0.87\linewidth}
   \includegraphics[width=75mm,angle=0]{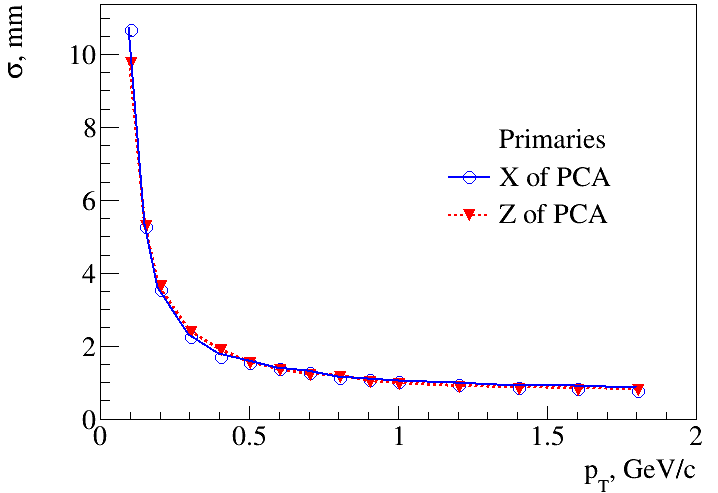}\\ \hspace*{4cm}
 \end{minipage}
 \caption{(Top) Uncertainty of the longitudinal position of the reconstructed primary vertex as a function of track multiplicity; (Bottom) transverse and longitudinal position uncertainties at the point of closest approach (\acrshort{pca}) to the interaction point for \acrshort{tpc} reconstructed primary tracks with $|\eta|<1.3$ versus particle transverse momentum. }
  \label{vertex}
\end{figure}

An important characteristic of the detector is its ability to reconstruct vertices. This depends on the achievable accuracy of the track direction pointing to
the production point. The top panel of Fig.~\ref{vertex} demonstrates the precision of the reconstructed interaction point (along the beam direction) as a function of the charged track multiplicity in the event. The uncertainty of the reconstruction of the position of the primary vertex  varies from about 150 to 700 $\mu$m in central and peripheral collisions, respectively. The bottom panel of Fig.~\ref{vertex} shows the transverse and longitudinal position uncertainty at the Point of the Closest Approach (\acrshort{pca}) to the interaction vertex as function of the track transverse momentum. Because of the relatively large distance between the interaction point and the first measured point inside the \acrshort{tpc} ($\sim$~40~cm) the extrapolation of low momenta tracks to the primary vertex has limited accuracy. To improve this, a silicon vertex detector is planned to be added to the MPD setup (see Sec.~\ref{sec:its}).  The main purpose of this detector is to increase the accuracy of reconstruction of the primary interaction vertex and secondary decay vertices of unstable particles, as well as to lower the detection threshold of charged particles with small transverse momenta~\cite{Zherebchevsky:2021fbi}.

\begin{figure}[t]
\centering
\begin{minipage}[h]{0.87\linewidth}
  \includegraphics[width=75mm,angle=0]{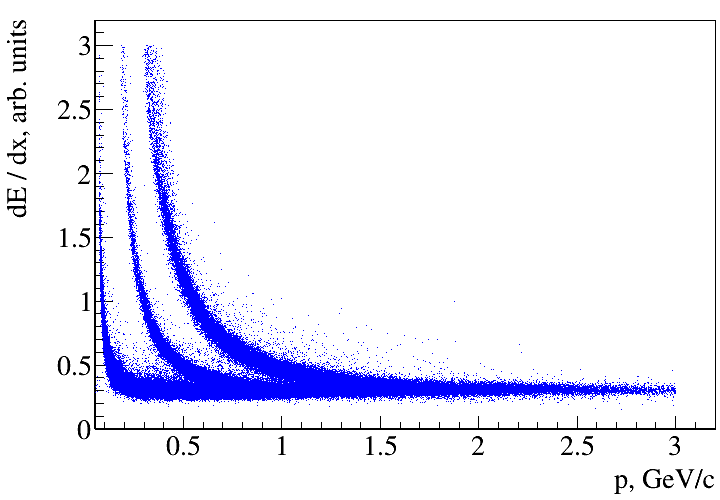}\\ \hspace*{4cm}
 \end{minipage}

\begin{minipage}[t]{0.87\linewidth}
 \includegraphics[width=75mm,angle=0]{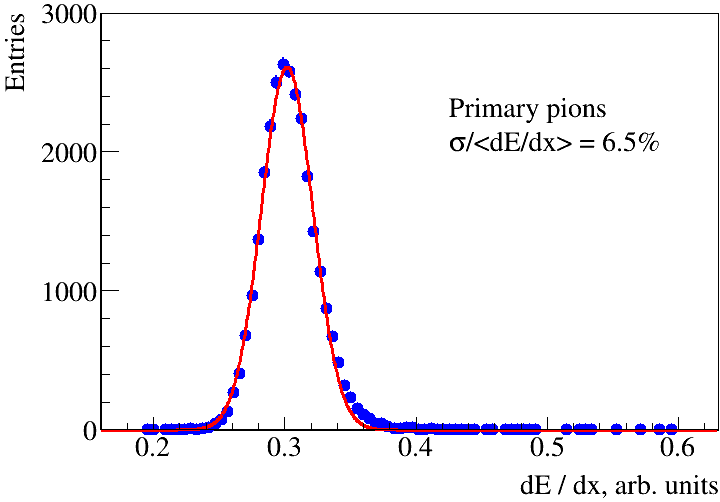}\\ \hspace*{4cm}
\end{minipage}
\caption{The specific energy loss $\langle {\rm d}E/{\rm d}x \rangle$ as a function of momentum (top panel); (bottom panel) - the
  energy loss distribution for pions with $p$ = 0.28-0.32~GeV/$c$ fitted to a Gaussian. }
 \label{dedx}
\end{figure}

Particle identification (\acrshort{pid}) in the \acrshort{tpc} will be achieved by using the information on the specific ionization energy loss ($\langle {\rm d}E/{\rm d}x\rangle$) in the \acrshort{tpc} gas, determined using a truncation value of 30\%, {\it i.e.} rejecting 30\% of the clusters associated to the track with the largest energy deposit before calculating the mean value. As demonstrated in Ref.~\cite{tpcsim}, the achieved accuracy of the energy loss $\langle {\rm d}E/{\rm d}x\rangle$  is 6-7\% (presented in Fig.~\ref{dedx}), allowing the discrimination of charged pions from kaons up to momenta of $\sim 0.7$~GeV/$c$ and kaons from protons up to $\sim 1.1$~GeV/$c$.

\subsection{\label{sec:tof}Time of Flight}

\begin{figure}[th]
  \centering
  \includegraphics[scale=1.0]{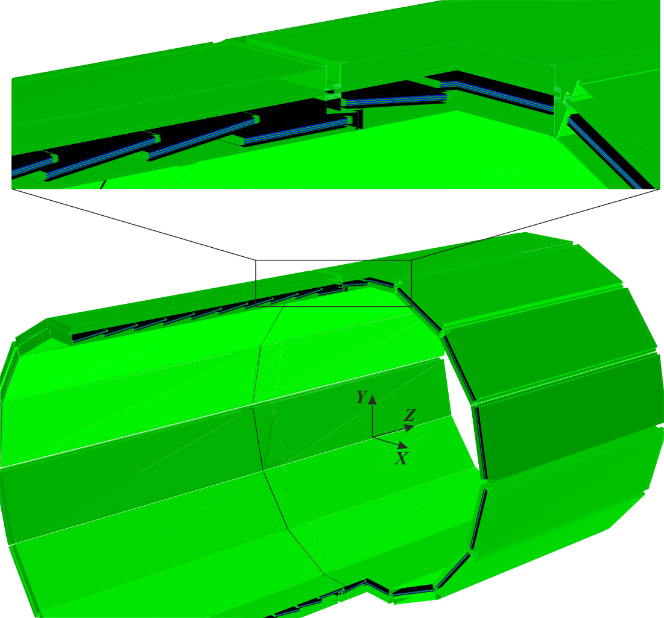}
  \caption{Schematic view of the spatial arrangement of the \acrshort{tof} sectors and modules and a cross-section of selected \acrshort{tof} modules. The separation between two half sectors is visible in the upper inset, and the location of the \acrshort{mrpc}s is visible in the cross-section.}
  \label{fig:figtof}
\end{figure}
The \acrshort{tof} system of the MPD, developed for the identification of charged hadrons in the intermediate momentum range, is based on the technology of Multigap Resistive Plate Chambers (\acrshort{mrpc}).  The detector is designed to provide both time and coordinate 
measurements with an accuracy of $\sim$80~ps and $\sim$0.5~cm, respectively~\cite{Babkin:2016xlr}.  Triple-stack \acrshort{mrpc}s with 5 gaps of 200~$\mu$m each are used. A time resolution of 50~ps for a single module has been achieved. The gas mixture is composed of 90\% of $C_2 H_2 F_4$, 5\% of $SF_6$ and 5\% of $i$-$C_4H_{10}$. In the basic configuration, the \acrshort{tof} is a barrel consisting of 14 plate sectors (Fig.~\ref{fig:figtof}). All sectors are formed by two modules of different types for convenience of installation. Every module contains 10 \acrshort{mrpc}s with 24 readout strips each. Signals are read from both sides of the strip. Thus, the total number of \acrshort{mrpc} detectors of the barrel \acrshort{tof} is 280 and the number of readout electronics channels is 13440.

\begin{figure}[t] \centering
 \begin{minipage}[t]{0.87\linewidth}
   \includegraphics[width=75mm,angle=0]{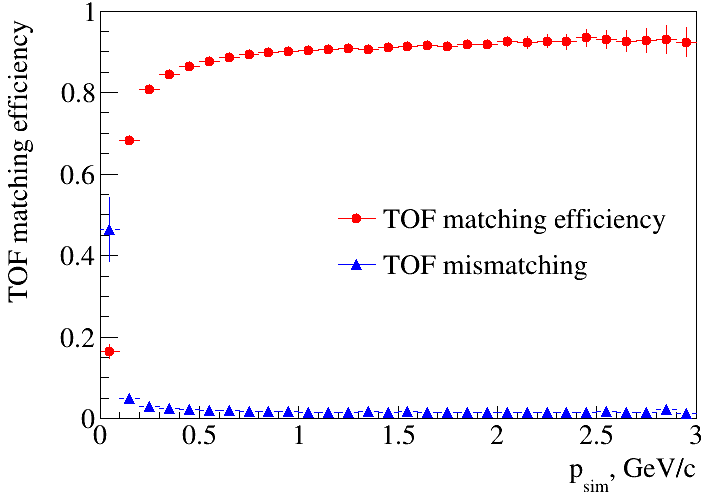} \\ \hspace*{4cm} (a)
 \end{minipage}
 
 \begin{minipage}[t]{0.87\linewidth}
   \includegraphics[width=75mm,angle=0]{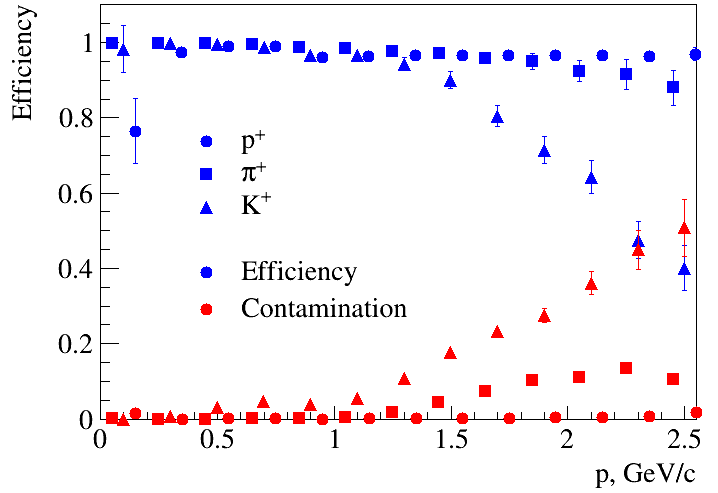} \\ \hspace*{4cm} (b)
 \end{minipage}
 \caption{(a) \acrshort{tof} matching efficiency; (b) Particle identification efficiency for positively charged hadrons (blue symbols)
and a fraction of wrongly identified species (red symbols) in Au+Au collisions at $\sqrt{s_{\rm NN}}$=9~GeV.}
 \label{pid}
\end{figure}

The commissioning is ongoing for all service subsystems, such as a gas system, a high-, and low-voltage power supply system, cooling and a slow control system. The most challenging task is the organization of gas supply to the \acrshort{tof} system. It should ensure a high purity and stability of the gas mixture with minimal leakage. A fully functional prototype of such a system has been developed for testing \acrshort{tof} modules with cosmic rays.

An accurate calibration and correction procedure is required to achieve the expected \acrshort{tof} system performance. As far as possible, calibrations should be performed before the start of the physics programme. There is a good opportunity to use particles from extended air showers initiated by cosmic rays for calibrations. The calibration strategy is as follows. First, we need to calibrate the electronics of data acquisition to assure its maximum time resolution. To determine the non-linearities of electronics, the main technique is the so-called statistical code density test based on filling one time cycle of TDC by random events~\cite{ALICE:tofman}. After calibrating the electronics, the position of all the \acrshort{tof} detectors will be determined. This is needed to make sure that the ideal geometry embedded in the software is updated with realistic detector positions. This will be done by determining the spatial coordinate of the passage of particles through the detectors with high accuracy. Tracks of cosmic air shower particles that have passed through the \acrshort{tpc} and the \acrshort{tof} will be used for this purpose before the start of data collection from ion collisions.

For the read-out of strip signals, front-end electronics with a digital LVDS output signal are used. The leading and trailing edges of this digital signal correspond to the moment of passage of the corresponding edges of the analog pulse through the discriminator threshold. The slope of the leading edge of the analog pulse depends on the amplitude of the signal. Therefore, the position of the leading edge of the digital rectangular pulse depends on its width. This dependence needs to be measured as accurately as possible for future use as a correction. Such a \lq\lq time-over-threshold'' correction gives the best time resolution of the \acrshort{tof} system. Finally, using cosmic radiation, we can test and optimize the algorithm of matching \acrshort{tpc} tracks with \acrshort{tof} hits. 

The \acrshort{tof} performance for identification of charged particles was estimated from the matching procedure of tracks reconstructed in the \acrshort{tpc} with hits in the \acrshort{tof} detector. The matching consists of  extrapolating the \acrshort{tpc} track to the \acrshort{tof} surface and finding a \acrshort{tof} hit nearest to the extrapolated point within a pre-set window (\lq\lq matching window''). The matching window size is taken as a compromise between the \acrshort{tof} intrinsic performance numbers (time and coordinate resolutions) and the overall \acrshort{tof} occupancy in heavy-ion collisions. In Fig.~\ref{pid} (a) the \acrshort{tof} (mis)matching efficiency is plotted as function of the total momentum. The results are obtained for central Au+Au collisions at $\sqrt{s_{\rm NN}}$=9~GeV and data points are averaged over the entire \acrshort{tof} acceptance $|\eta|<1.4$. The efficiency is defined as the fraction of tracks having produced a Monte Carlo point in the \acrshort{tof} and matched with any \acrshort{tof} hit according to the described procedure. If such a match includes a wrong hit, it is also considered as a mismatch. The overall efficiency is about 90\% and it drops below 80\% for track momenta below 250~MeV/$c$ because of the multiple scattering which makes the difference between the expected positions of the extrapolated tracks and the actual ones larger than the size of the matching window in some cases. The errors in the extrapolation for low momentum tracks also cause an increase of the number of wrongly matched \acrshort{tpc} track extrapolations and \acrshort{tof} hits (see triangles in Fig.~\ref{pid} (a) for mismatches). Nevertheless, for momenta above 200~MeV/$c$ (a typical low-momentum cutoff in the analysis) the fraction of the \acrshort{tpc}-\acrshort{tof} mismatches is below 3\%.

The best \acrshort{pid} performance for charged particles is achieved when the capabilities of the \acrshort{tpc} and \acrshort{tof} detectors are combined. For optimal performance, the \acrshort{pid} procedure should rely on a good knowledge of the detector characteristics such as the momentum dependence of the average energy loss as well as the variation of the $\langle{\rm d}E/{\rm d}x\rangle$ and mass-squared resolutions for each particle species.
Based on this information a vector of probabilities to be a particle of a particular sort is assigned to each track and the highest probability
defines the particle species. The MPD performance for discrimination of hadrons in minimum bias Au+Au collisions at $\sqrt{s_{\rm NN}}$=9~GeV is demonstrated in Fig.~\ref{pid} (b), where the fraction of correctly identified particles is shown as a function of momentum (blue symbols).
This fraction is above 90\% for protons and positively charged pions up to $p$=2.5~GeV/$c$, while the percentage of the cases with a wrong identification is below 10\%. With the chosen set of cuts,
charged kaons can be identified up to $p\sim 1.7$~GeV/$c$ with an approximately 80\% efficiency and 20\% contamination at the \acrshort{pid} limit. Making the selection criteria for kaons tighter, the achieved
contamination level can be decreased further resulting in a lower value for the \acrshort{pid} efficiency.

\begin{figure}[t]
  \centering
  \includegraphics[scale=0.4]{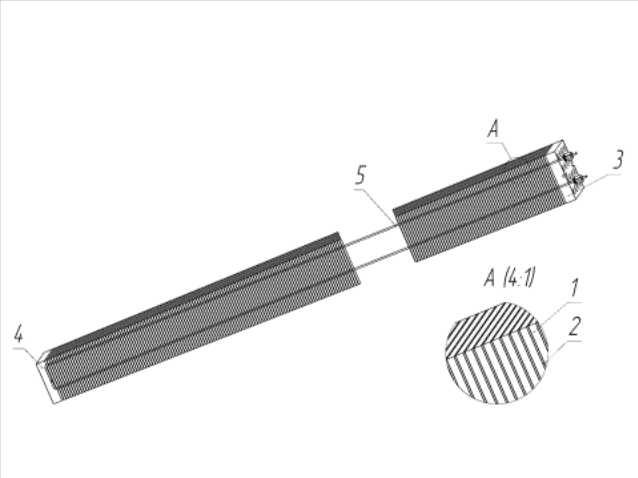}\\ \hspace*{4cm} (a)
  \includegraphics[scale=0.3]{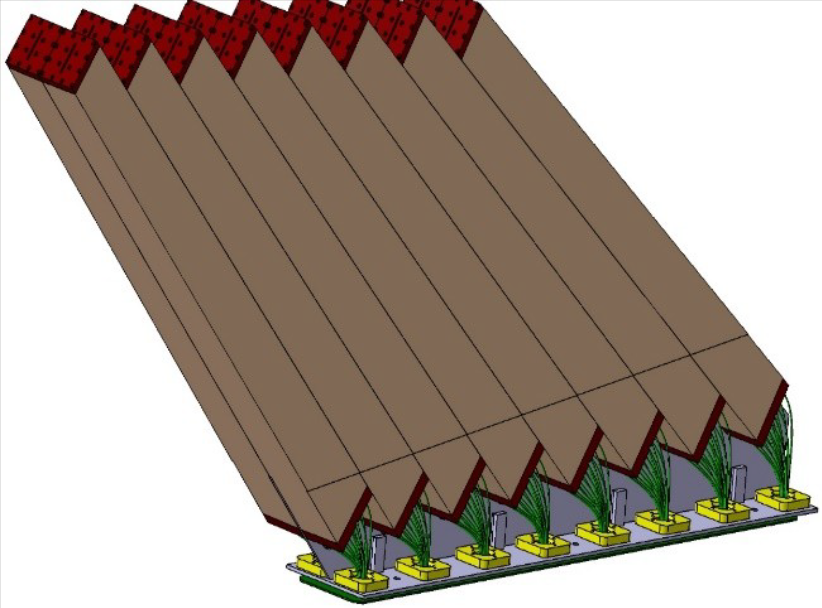}\\ \hspace*{4cm} (b)
  \caption{The schematic of a single \acrshort{ecal} tower (a) and one of the \acrshort{ecal} modules (b).}
  \label{fig:figecal1}
\end{figure}

\subsection{\label{sec:ecal}Electromagnetic Calorimeter}

The primary role of the electromagnetic calorimeter is to measure, with good resolution, the spatial position and total deposited energy of electromagnetic cascades induced by electrons and photons produced in heavy ion collisions. The \acrshort{ecal} will operate in the magnetic field of the MPD solenoid and will detect particles in the energy range from 10 MeV to a few GeV. The expected high multiplicity environment implies a high segmentation of the calorimeter. 

\begin{figure}[t]
  \centering
 \includegraphics[scale=0.50]{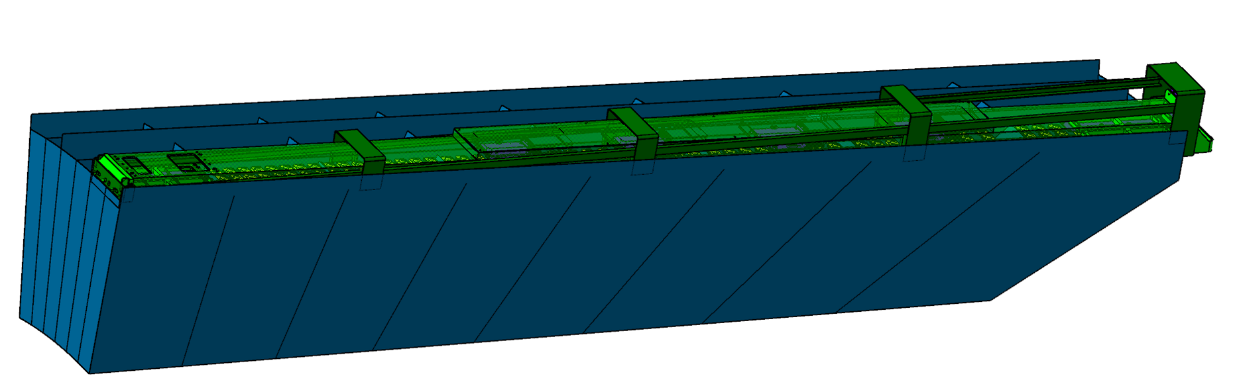}\\ \hspace*{4cm} (a)
 \includegraphics[scale=0.47]{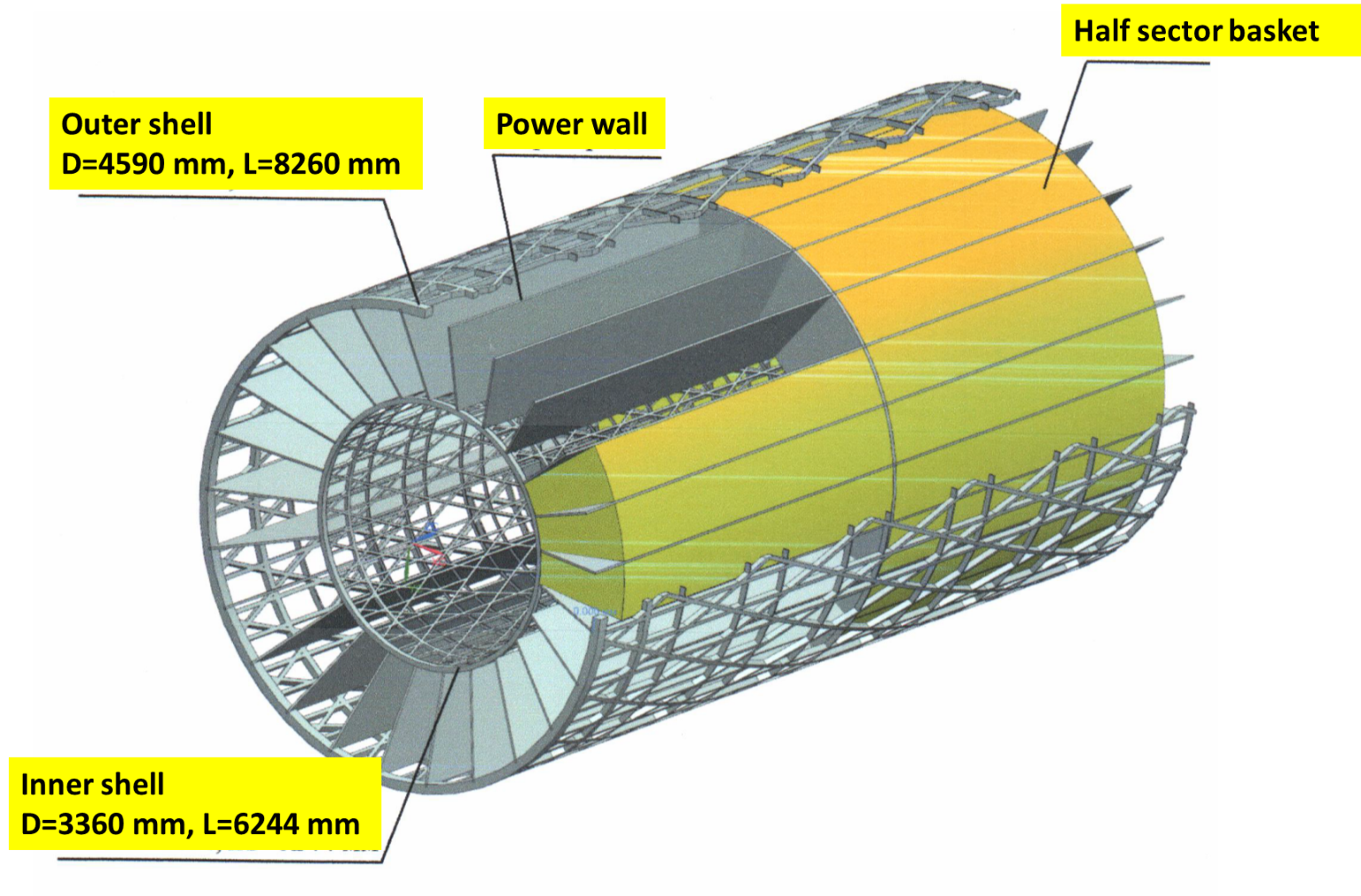}\\ \hspace*{4cm} (b)
  \caption{The schematic of a single \acrshort{ecal} sector (a) and structure of \acrshort{ecal} inside the support frame of MPD (b).}
  \label{fig:figecal2}
\end{figure}

\begin{figure*}
\includegraphics[width=1.0\columnwidth]{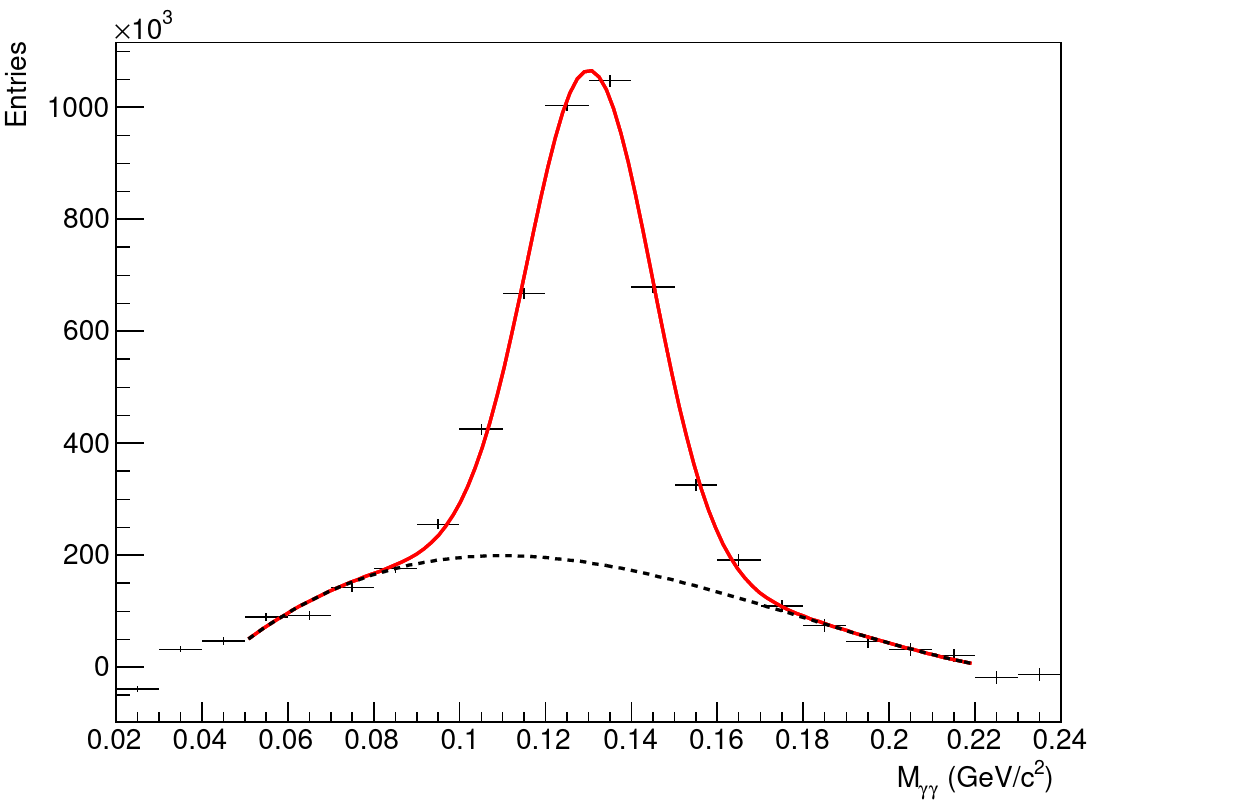}
\includegraphics[width=1.0\columnwidth]{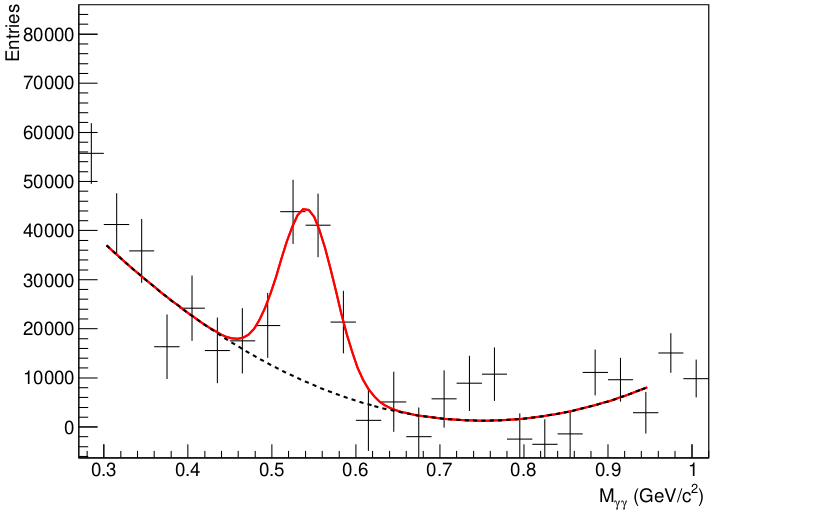}
\caption{\label{Pi0Eta_peaks}Reconstructed peaks from decays of $\pi^{0}$ and $\eta$ mesons in Au+Au collisions at $\sqrt{s_{\rm{NN}}}=11$~GeV after subtraction of the mixed-event background. The peaks are shown for $\pi^{0}$($\eta$) mesons in 0.3-0.4~GeV/$c$ (0.5-1~GeV/$c$) transverse momentum bins and are fitted to a combination of a Gaussian function for the signal and a polynomial for the remaining background.}
\end{figure*}

The MPD \acrshort{ecal} is a shashlik type calorimeter made of Pb-scintillator sandwiches. Geometrically it consists of 50 isolated half-sectors forming a 6-m-long (along the beam direction) cylindrical shell having inner and outer radii of about 1.710~m and 2.278~m, respectively. Each half-sector is a segment of that shell and has a length of about 3~m (half-length of the calorimeter). In the transverse plane, each half-sector covers an azimuthal angle range of about 360$^{\circ}$/25 =14.4$^{\circ}$. The primary support structure of the half-sector is a basket, which consists of a carbon composite material. A half-sector contains 48 calorimeter modules (8 modules of different types in the longitudinal direction) $\times$ (6 modules in the transverse direction) that are locked (glued) into the basket, so that the modules contribute to the rigidity of the whole half-sector. A schematic view of an \acrshort{ecal} half-sector is shown in Fig.~\ref{fig:figecal2} (top). Half-sectors will be installed in the MPD by inserting them inside a power frame made of carbon composite material, shown in Fig.~\ref{fig:figecal2} (bottom).

Each \acrshort{ecal} module consists of 16 towers that are glued together. The design of each module depends on the module’s z-coordinate location with respect to the beams interaction point to form a projective geometry. There are 8 types of modules. Schematic views of one shashlik tower and one module are shown in Fig.~\ref{fig:figecal1}. Each tower has a $40\times40$~mm$^2$ transverse cross-section and is a lead-scintillator sandwich that contains 210 tiles of Pb (0.3~mm thick each). The lead tiles are interleaved with 210 tiles of plastic scintillator (1.5~mm thick each). The total thickness of the tower is approximately 41~cm (about 11 radiation lengths).

The calibration for the \acrshort{ecal} usually consists of two steps: 1) equalize the gain of all calorimeter cells using an electron beam or cosmic ray measurements, and 2) find one (common for the whole calorimeter) \acrshort{adc}-to-energy coefficient from electron beam tests or reconstruction of narrow invariant mass distributions. This method does not take into account the geometrical difference of the calorimeter cells and dependency of the calorimeter properties on the energy of the primary particle and the distance from the particle hit. Attempts to correct for these effects result in additional labour-consuming measurements and \lq\lq phenomenological'' calibrations on impact particle energy, $p_{\rm T}$, hit-position, and other non-linear dependencies. The preliminary calibration approach consists of two steps. First, a connection will be established between the amplitude of the observed signals from \acrshort{adc} and the energy deposition in the active volume of the calorimeter (scintillators) for each calorimeter cell; this can be done via comparison of the calorimeter signals from cosmic air shower muons with predictions from corresponding detailed simulations. The second step contains computer simulations of a data base of the physical properties of the calorimeter (viz., sampling fractions and energy leaks) as function of primary particle energy, hit position, location of the cell in the calorimeter etc., to correctly convert the energy deposition in the calorimeter active volume to the impact particle energy, and to do it in a phenomenologically-free manner. With a detailed description of the \acrshort{ecal} structure in the simulation (viz., the exact structure, shape and location of the towers as well as the accurately measured light absorption in the \acrshort{wls} fibers), the preliminary calibration of the calorimeter will be performed with an accuracy of approximately 3\% or better. This is the accuracy expected for the first MPD run. The main task in this run is the study of calorimeter operation in real conditions (thresholds, noise, etc.). The next step will be a check of calorimeter calibration by means of $\pi^{0}$ reconstruction and reconstruction of the energy of electrons.

Feasibility studies for Au+Au collisions at $\sqrt{s_{\rm{NN}}}=11$~GeV simulated by \acrshort{urqmd} show the possibility to measure $\pi^{0}$($\eta$) meson differential yields in the momentum range  $0.05-0.1~$GeV/$c < p_{\rm T} < 3~$GeV/$c$  using a data sample of $10^{8}$ minimum bias events. Examples of the reconstructed invariant mass distributions for $\gamma\gamma$-pairs after subtraction of the mixed-event background are shown in Fig.~\ref{Pi0Eta_peaks}. Distinct peaks from decays of $\pi^{0}$ and $\eta$ mesons appear on top of the remaining correlated background.

\begin{figure}
\includegraphics[width=\columnwidth]{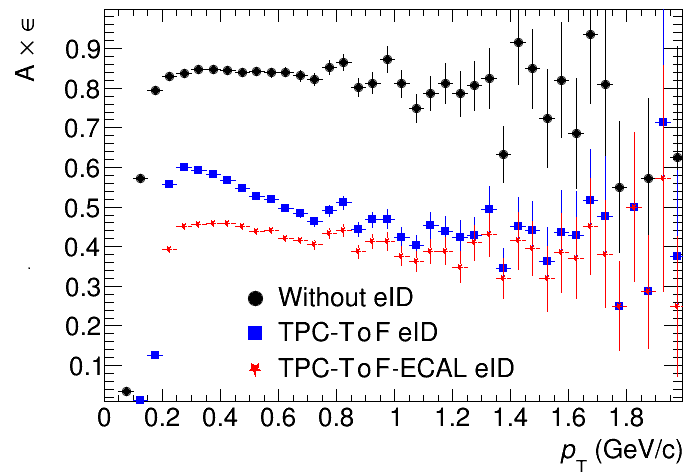}
\caption{\label{fig:ElEff} Reconstruction efficiency of electron tracks. Black markers correspond to tracks reconstructed in the \acrshort{tpc} and matched to the primary vertex. Blue markers correspond to tracks identified as electrons in the \acrshort{tpc} and \acrshort{tof}. Red markers correspond to tracks identified as electrons in the \acrshort{tpc}, \acrshort{tof} and \acrshort{ecal}.}
\end{figure}

\begin{figure}
\includegraphics[width=\columnwidth]{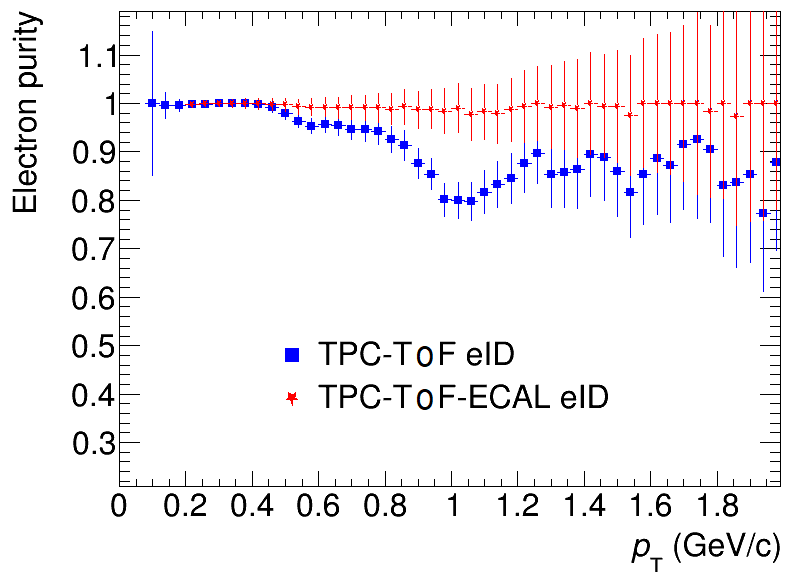}
\caption{\label{fig:ElPur} Electron purity as a function of particle transverse momentum with different
  electron selection options (see Fig.~\ref{fig:ElEff} for details).
}
\end{figure}

The \acrshort{ecal} also plays an important role in the identification of electrons. The MPD can effectively identify electrons by $\langle{\rm d}E/{\rm d}x\rangle$ measurements in the \acrshort{tpc}, by time-of-flight measurements in the \acrshort{tof} and by time-of-flight and $E/p$ measurements in the \acrshort{ecal}. For electrons, the $E/p$ ratio, where $E$ is the energy measured in the \acrshort{ecal} and $p$ is the total momentum measured in the \acrshort{tpc}, is expected to be $\sim 1$. The time resolution of the \acrshort{ecal} is inferior to that of the \acrshort{tof}, but the time measurements still help to reject kaons and protons in the range of measurements. The \acrshort{ecal} becomes effective for electron identification at $p_{\rm T}>200$~MeV/$c$ since lower momentum tracks just bend in the magnetic field and do not reach the \acrshort{ecal}. Figs.~\ref{fig:ElEff} and \ref{fig:ElPur} show the reconstruction efficiency for electron tracks and the electron purity evaluated for minimum bias Au+Au collisions at $\sqrt{s_{\rm{NN}}}=11$~GeV simulated by \acrshort{urqmd}. The electron identification in the \acrshort{tpc} and \acrshort{tof} helps to reject hadron contamination at low-to-intermediate transverse momentum. The \acrshort{ecal} helps to clean up the electron sample at higher momenta.

\subsection{\label{sec:HCAL}Forward Hadron Calorimeter}

\begin{figure}
\begin{center}
\includegraphics[width=0.8\columnwidth]{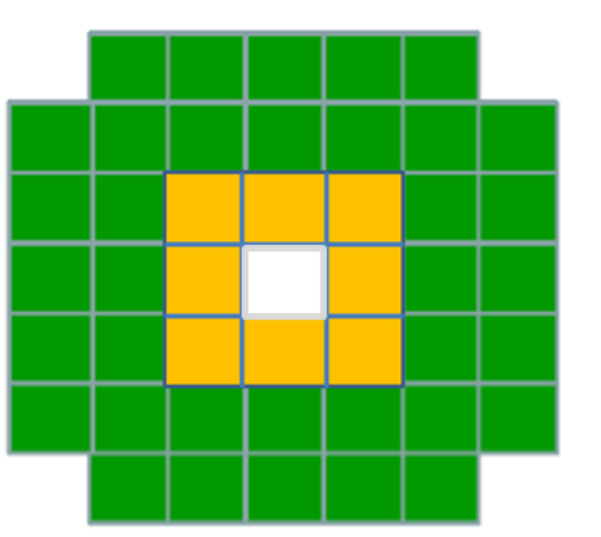}
\caption{\label{fig:fhcal}The modular structure of one \acrshort{fhcal}, in the plane perpendicular to the beam axis.}
\end{center}
\end{figure}

The \acrshort{fhcal} is intended for the measurements of global properties of heavy-ion collisions. It consists of two identical detectors, each consisting of 44 modules, placed approximately 3.2~m upstream and downstream from the center of the detector. The modular structure of one \acrshort{fhcal} in the plane perpendicular to the beam axis is presented in Fig.~\ref{fig:fhcal}.

The module transverse size of $15\times15$~cm$^2$ was chosen to match the size of the hadron showers. Each module includes 42 lead-scintillator sandwiches with a sampling ratio 4:1 (the thickness of lead plates and scintillator tiles is 16 mm and 4 mm, respectively). According to Geant-4 simulations, the sampling fluctuations provide the relative calorimeter energy resolution $\sigma_{E}/E\approx55\%/\sqrt{E}$ (GeV). Beam tests of the calorimeter with the same sampling confirm the results of the simulation. The 42 lead/scintillator layers are loaded into a box made of 0.5~mm stainless steel sheet and tied together in one block with a length of about 90~cm (4 nuclear interaction lengths) by a 0.5~mm stainless steel tape. After assembly, the module is covered by another similar stainless steel box. The two upper and lower boxes are spot-welded providing a mechanically stable construction. The weight of each module is about 200~kg.

Light readout is provided by \acrshort{wls}-fibers embedded inside grooves in the scintillator tiles. This ensures high efficiency and uniformity of the light collection over the scintillator tiles. \acrshort{wls} fibers Y-11(200) with double cladding and 1~mm diameter produced by the Kuraray Co. are used. To reduce the loss of light, the bending radius of the \acrshort{wls} fiber must be larger than 5 cm. Spiral grooves in the scintillator tiles provide slightly better parameters than the circular ones. They were selected for this reason to design the \acrshort{fhcal} modules. The end of the \acrshort{wls}-fiber inside the scintillator grove is mirrored by silver paint, thus improving the light collection by about 30\%.

Each scintillator tile is covered with a white reflector (TYVEK paper) to improve light collection. \acrshort{wls}-fibers from 6 consecutive scintillator tiles are collected together in the optical connector at the end of the module and polished to improve the optical contact with the photodetector. The longitudinal segmentation in 7 sections requires the same number of optical connectors and compensates for the non-uniformity of the light collection along the module caused by the different lengths of the \acrshort{wls}-fibers. In addition, seven compact photodetectors are coupled to the optical connectors at the rear side of the module. The use of \acrshort{sipm}s is an optimum choice due to their remarkable properties such as high internal gain, compactness, low cost and immunity to magnetic fields. \acrshort{sipm}s have no nuclear counter effect due to their pixel structure. Hamamatsu \acrshort{mppc} S12572-010C/P with a pixel size of $10\times10$~$\mu$m$^2$ were selected to ensure a high dynamic range of detected energies. The \acrshort{fee} used for the \acrshort{mppc} readout includes an amplifier and a shaper with differential output signals. Due to the shaper, the signal length is about $0.2$~$\mu$s which is a few times longer than the original signal width after the photodetectors. The necessity of a longer signal is related to the relatively low sampling frequency of the pipe-line \acrshort{adc} that digitizes the signal waveform. A 64-channel $62.5$~MS/s ADC64s2 board manufactured by the Dubna company AFI Electronics is used. All \acrshort{fhcal} modules are assembled and ready for installation at MPD. 

After assembly of the \acrshort{fhcal} modules, the light yield of all longitudinal sections was measured by using cosmic air shower muons crossing the longitudinal sections in a module. On average, cosmic muons deposit about 50 photoelectrons/section, which is enough to calibrate the energy scale of \acrshort{fhcal} modules during the calorimeter operation in the MPD setup. 


\subsection{\label{sec:ffd}Fast Forward Detector}

\begin{figure}[t]
  \centering
  \includegraphics[width=0.95\columnwidth]{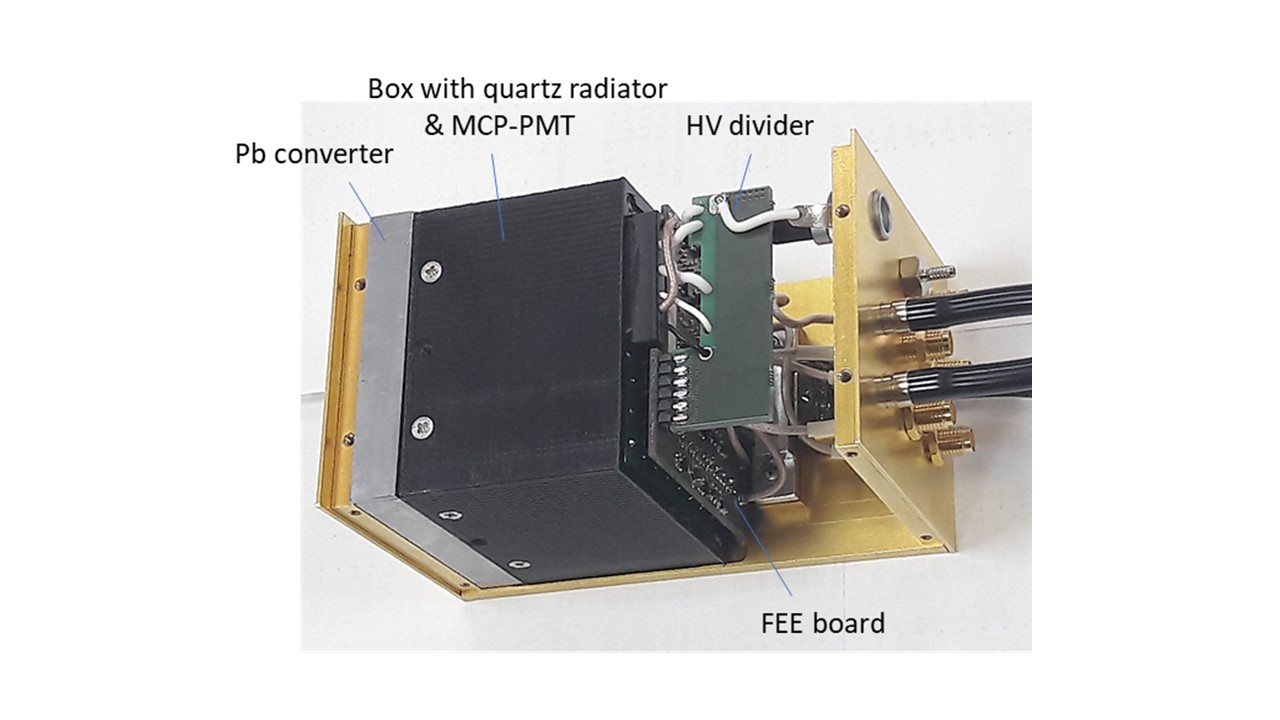} \\ 
  \includegraphics[width=0.95\columnwidth]{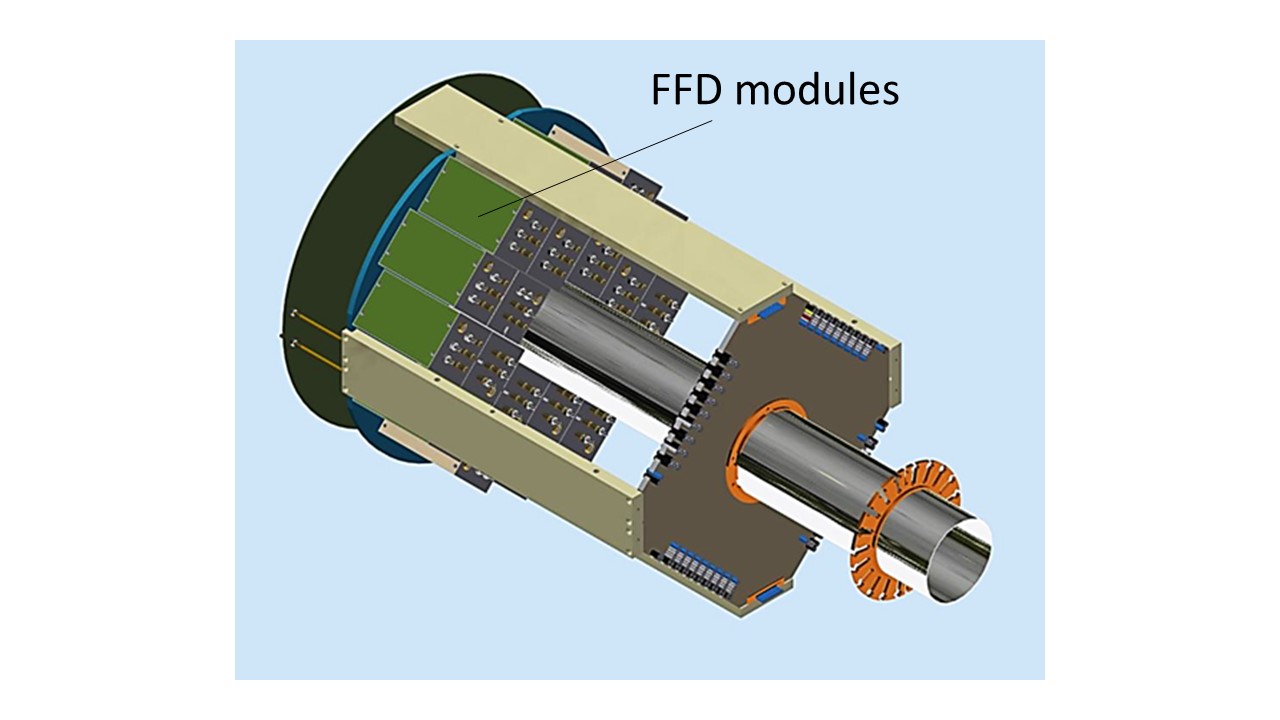} 
  \caption{An assembled \acrshort{ffd} module (top) and a design of one full \acrshort{ffd} (bottom).}
  \label{fig:figffd}
\end{figure}

The \acrshort{ffd} fulfills two important tasks for the MPD: it provides fast triggering of A+A collisions and generates the start-time (T0) pulse generation for the \acrshort{tof} detector with a time resolution better than 50~ps. Also, the \acrshort{ffd} is a useful instrument for the adjustment of beam-beam collisions in the center of the MPD setup with operative control of the collision rate and interaction point position during a run. 

The \acrshort{ffd} consists of two identical Cherenkov modular arrays FFDE and FFDW with large active area and picosecond time resolution which is achieved by the registration of relativistic charged particles and high-energy photons
produced in the collisions. The acceptance in pseudorapidity of the detector is $2.7<|\eta|<4.1$, which corresponds to a the polar angle range of $1.9^{\circ}<|\theta|<7.3^{\circ}$. 

Each \acrshort{ffd} consists of 20 Cherenkov modules shown in Fig.~\ref{fig:figffd}. One module consists of a 10~mm lead converter, a 15~mm quartz radiator, \acrshort{mcppmt}s XP85012/A1 (Photonis) and a \acrshort{fee} board.  The detector has 80 channel granularity, 400~mm outer diameter and a 96~mm diameter hole for the beam pipe. The FFDs are located 140~cm away from the MPD center point. Photons are detected by their conversion into electrons in a 10~mm lead converter. 

The modules are tested with laser and cosmic rays in a special stand developed for this purpose. 
The sub-detector electronics concept includes two VME crates with custom made backplane and electronics modules based on \acrshort{fpga} technology. 
The LV power supply of \acrshort{fee} boards are also developed as VME modules. The HV power supply is based on a WIENER crate with three 16-channel modules. The local readout electronics used for the control of the \acrshort{ffd} operation consists of 
8 modules of 5 GS/s digitizes mod. N6742 (CAEN) with optical readout. 
The global readout electronics, as part of the MPD readout system, consists of four modules of TDC72VHL produced at JINR.  
The detector control system (DCS) allows to perform and to control all functions required for operation of the \acrshort{ffd} using a custom developed GUI. 

\begin{figure}[t]
  \centering
  \includegraphics[width=0.8\columnwidth]{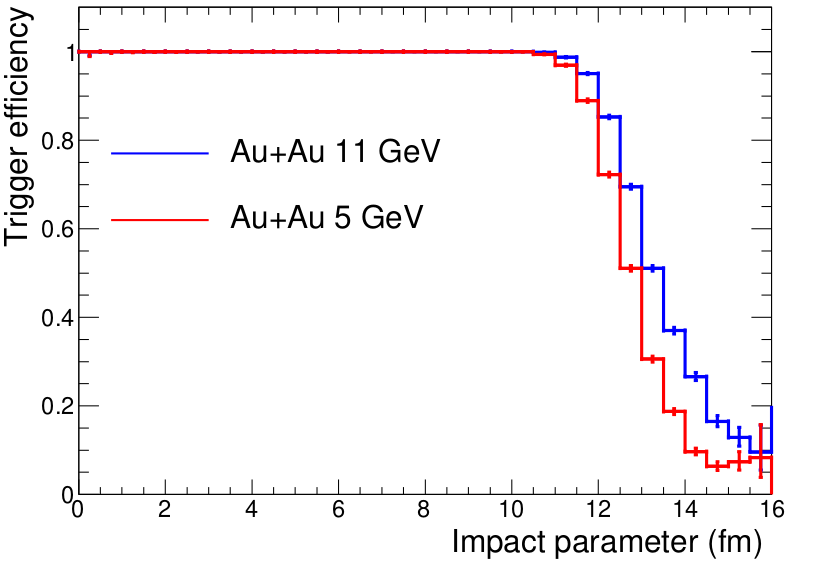}
  \caption{The L0 trigger efficiency as a function of the impact parameter of Au + Au collisions for two energies $\sqrt{s_{\rm NN}}=5$ (red) and $11$~GeV (blue).}
  \label{fig:trigeff}
\end{figure}

A laser system is used for \acrshort{ffd} calibration, test, and operation control. It consists of a PiLas laser, a special optical box with 120-fiber output, a system of optical cables, and a reference detector with \acrshort{mcppmt}.  

\subsection{Plans for additional detectors}
\label{sec:detectorfuture}

Several additional detector subsystems and upgrades are being considered for installation in the MPD apparatus. The Inner Tracking System is already in an advanced development stage. Part of it is planned to be installed after the initial commissioning of Stage I subdetectors, with the complete setup expected to be ready at a later date. The MPD Cosmic Ray Detector is in an advanced stage of preparation. A limited set of modules will be used in pre-installation tests of other major detector components. They can later be installed on the outside of the MPD Magnet Yoke. The Mini Beam-Beam counter is under consideration for the enhancement of triggering capabilities and determination of the start time. These systems are described in this section.

\subsubsection{The Inner Tracking System}
\label{sec:its}

\begin{figure}[b]
  \centering
  \includegraphics[width=0.8\columnwidth]{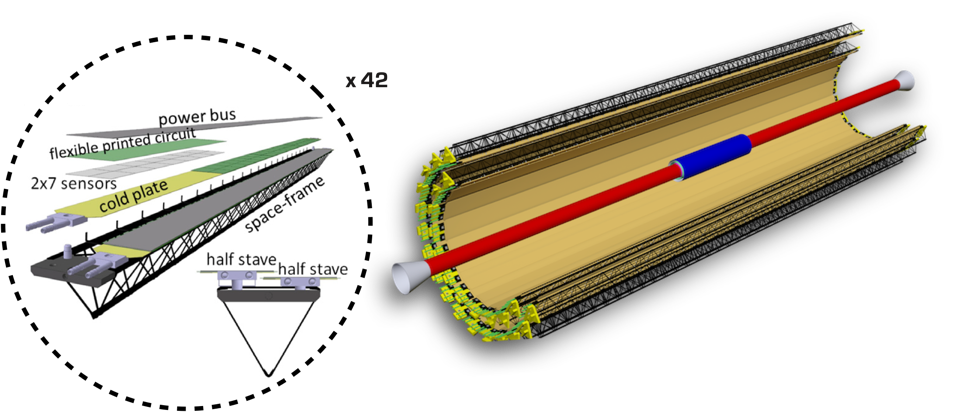}
  \caption{(Left) Breakdown of one Outer Barrel Stave.  (Right) Cut of the full MPD-ITS geometry.}
  \label{fig:its1}
\end{figure}

The Inner Tracking System (\acrshort{its}) of the MPD will be a vertex detector meant to complement the \acrshort{tpc} for the precise tracking, momentum determination and vertex reconstruction for hyperons ($\Lambda$, $\Xi$, $\Omega$) and D-mesons. It will be placed inside the bore of the \acrshort{tpc} and it will be composed of 5 layers of silicon Monolithic Active Pixel Sensors (MAPS) grouped in two barrels, with layers 1-3 on the inner barrel and layers 4-5 on the outer barrel (see Fig.~\ref{fig:its1} and Table~\ref{tab:itslay}), with a spatial resolution of less that 5 $\mu$m in a sensor plane and a material budget of less than 0.8\%$X_0$ for the entire \acrshort{its}. The project foresees the construction of the outer barrel in a first stage (2022/2023) based on the MAPS technology used for the outer barrel of the ALICE-ITS2~\cite{Mager:2016yvj}.
These are 15 mm $\times$ 30 mm $\times$ 100 $\mu$m silicon sensors (from TowerJazz 180 nm CMOS technology) with 1024 $\times$ 512 pixels. The addition of the inner barrel is planned for a later stage (2025/2026) with the intention of building it, based on 280 mm-long and 30 $\mu$m-thick bent sensors currently under development by the ALICE-ITS3 project at CERN~\cite{ALICE:alits}. 
Figure~\ref{fig:its1} shows a cut of the MPD-\acrshort{its} geometry with a breakdown of one of the 42 Stave structures that compose the outer barrel. Each one of these Staves is segmented into two identical structures (Half Staves) where 2 rows of 7 MAPS are attached to a Flexible Printed Circuit to create a structure called Hybrid Integrated Circuit (HIC). Seven of these HICs are glued to a multilayer composite graphite plate with embedded cooling pipes (Cold Plate). 
In central Au+Au collisions at $\sqrt{s_{\rm NN}}=9$~GeV simulated in MpdRoot~\cite{Zinchenko:2020bsf,Zinchenko:2021ffj} with \acrshort{its} with 2 outer barrel layers only, the \acrshort{tpc} and a beam pipe diameter of 64 mm, the signal extraction of reconstructed hyperons would be performed with an efficiency of 0.2\%. This is enough for assessing the identification ability of the system at debugging stage. On the other hand, only with the 5 layer setup of the \acrshort{its}, the \acrshort{tpc} and a beam pipe diameter of 40 mm, it will be possible to achieve a reliable detection efficiency of about 1\% for both multistrange and charmed particles.

\begin{table}[tbh]
\caption{\label{tab:itslay}
Geometrical parameters of the MPD-ITS layers.}
\begin{tabular}{c|c|c} \hline
Layer No. & Radius (cm) &
Length (cm) \\ \hline
1 & 2.2 & 75.0 \\
2 & 4.1 & 75.0 \\
3 & 6.0 & 75.0 \\
4 & 14.5 & 152.6 \\
5 & 19.4 & 152.6 \\ \hline
\end{tabular}
\end{table}

\begin{figure}[b]
  \centering
  \includegraphics[width=0.8\columnwidth]{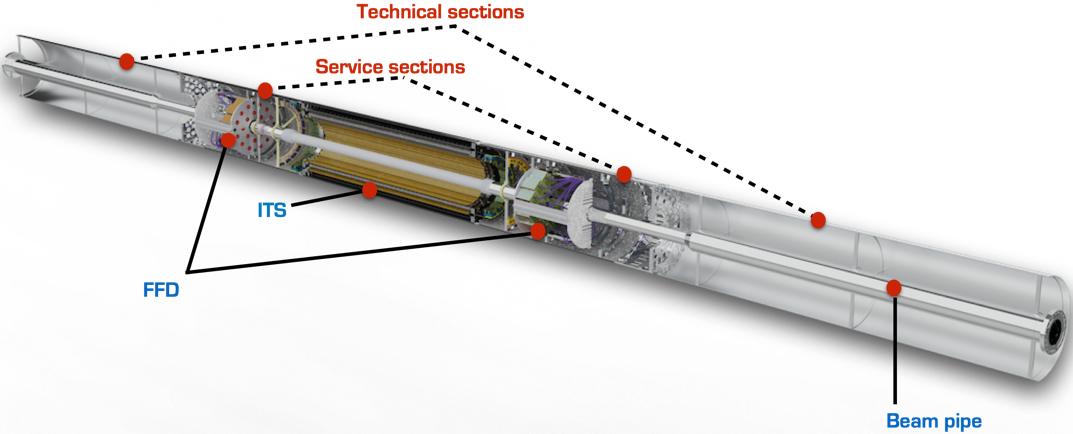}
  \caption{Insertion container for the integration of the \acrshort{its}, the \acrshort{ffd} and the Beam pipe into the \acrshort{tpc}.}
  \label{fig:its2}
\end{figure}

The very small clearance ($\sim$ 8 mm) for the integration of the \acrshort{its} and other components of the MPD inside the bore of the \acrshort{tpc}, prevents the implementation of the common solution of placing rails for the sequential insertion of each component. Instead, the integration scenario includes the use of a custom-designed insertion container (Fig.~\ref{fig:its2}) to be slided in as a whole and which includes the \acrshort{its} section, the services section hosting the \acrshort{ffd} and the cabling and cooling structures for both detectors, the technical sections that will be removed after the insertion into the \acrshort{tpc} bore as well as part of the beam pipe comprising the Beryllium-made portion.

\subsubsection{The miniBeBe Detector}
\label{sec:mbb}

The proposal for the construction of the mini Beam-Beam detector (miniBeBe)~\cite{Kado_2021} is under consideration by the MPD Collaboration.
Its main goal is to contribute with an additional wake-up trigger signal for the \acrshort{tof}, particularly for low multiplicity events. 
Since the shortest track length
within the \acrshort{tpc} is 1.5 m then, in order for pions, kaons and protons to be reliably separated over the entire particle momentum range, the \acrshort{tof} is expected to have an overall time resolution better than 100 ps. Thus, the wake-up trigger signal should be optimized to keep a time resolution below 100 ps. The expected miniBeBe time resolution is 30 ps at most.

\begin{figure*}[tb]
\centering
\includegraphics[scale=0.3]{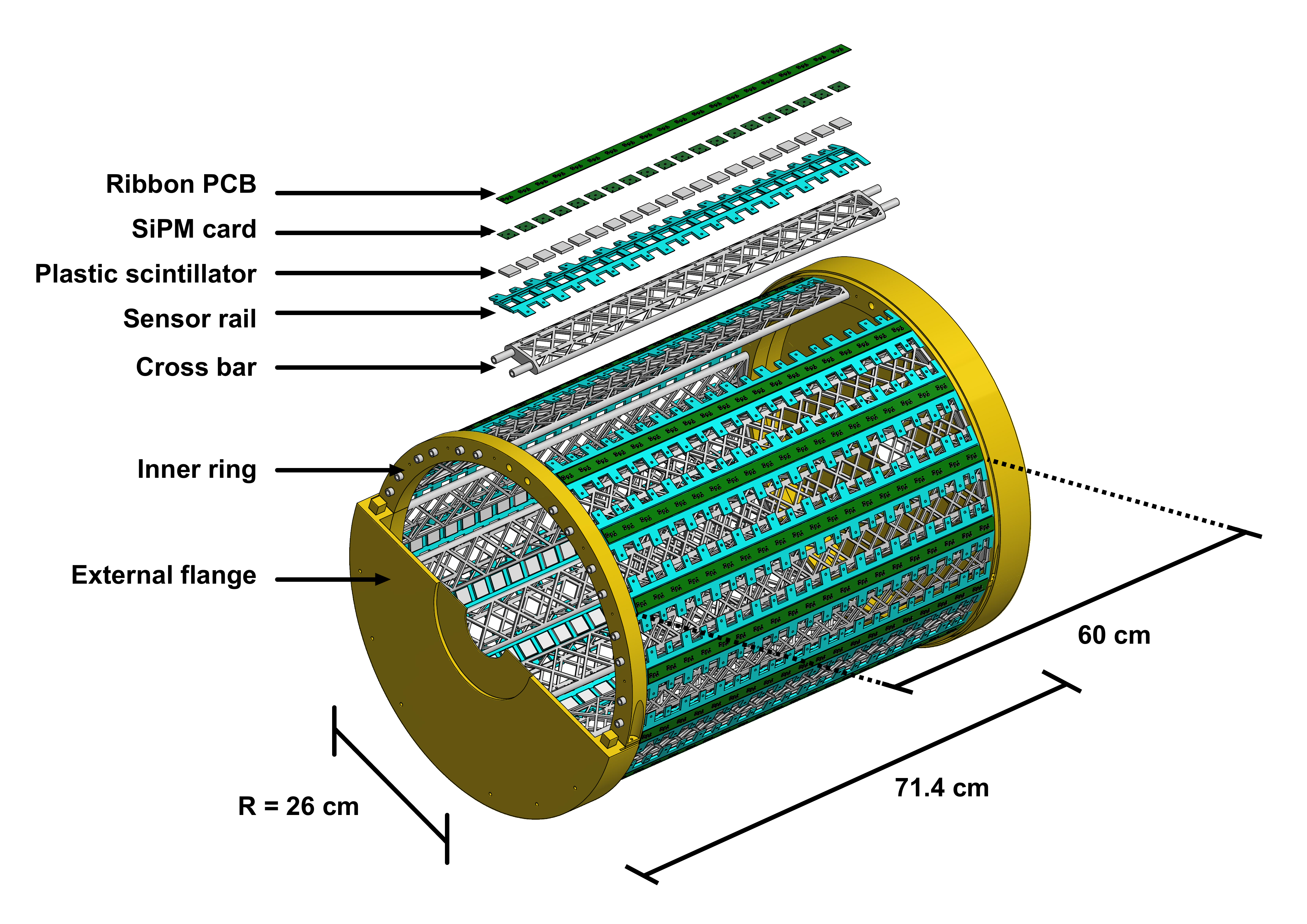}
\caption{Illustration of the miniBeBe detector. The structure holds sixteen 600 mm long strips mounted on a cylinder, with inner and outer radii of 220 and 260 mm, respectively, placed around the beam pipe. 
Each strip consists of 20 squared plastic scintillators with dimensions $20 \times 20 \times 3\  \mathrm{mm}^3$, with four \acrshort{sipm}s coupled to each cell. Taken from~\cite{Kado_2021}.}
\label{fig:minibebe}
\end{figure*}

The miniBeBe is a scintillator detector with a cylindrical structure covering the pseudorapidity range of $|\eta| < 1.44$. The baseline design of the miniBeBe consists of 16 strips, each one 600~mm long. Each strip consists of an array of 20 square-shaped plastic scintillator cells with dimensions $20 \times 20 \times 3 \ \mbox{mm}^3$. Each scintillator cell has four \acrshort{sipm}s coupled to the surface. In total, the miniBeBe consists of 320  plastic scintillator cells and 1,280 \acrshort{sipm}s covering an effective sensitive area of 128,000 $\mathrm{mm}^2$. Current work on the mechanical support for the detector and its integration in the MPD, has produced a design consisting of a hollow cylindrical structure, 260~mm in radius and 714~mm in length.
Fig.~\ref{fig:minibebe} shows the design plans for both the mechanical structure and the detector strips. Simulations show that the requested 30 ps resolution can be attained with an optimal design without smearing of the interaction point, where the same material budget is concentrated in a similar cylinder but with a 150 mm radius.
A prototype consisting of a two-cell detector made out of BC-404 plastic scintillator was tested. For the light sensors a Hamamatsu PMT R6249 and a SensL (C-60035-4P1521 EVB) \acrshort{sipm} were chosen. The test was carried out in the pion beam of the T10 facility at CERN using the well-tested and calibrated trigger and Data Acquisition (\acrshort{daq}) systems of the ALICE experiment~\cite{Alvarado:2018gbb}.

The best achieved time resolution was $68~\pm~5$~ps for the BC-404 hexagon coupled to the Hamamatsu PMT R6249 and $45~\pm~2$~ps for the BC-404 hexagon coupled to the SensL light sensor. These encouraging first results suggest that the desired time resolution for the MPD/NICA experiment are within reach~\cite{Zepeda-Fernandez:2020wnt}.

A system that can provide a digital signal from each scintillator cell, using an ultra-fast analog comparator HMC674 model from Analog Devices, is being currently designed. The comparator output is differential, thus it is less susceptible to noise and interference. With this in mind, a PCB card that allows for the signal transport from the scintillator cells to the interconnection zone using a processing card, has been designed. 

\subsubsection{\label{sec:mcord}The Cosmic Ray Detector}

The MPD cosmic ray detector (MCORD) aims at providing a trigger for cosmic showers (mostly muons). The estimated flux of cosmic muon events is on the order of 150 s$^{-1}$m$^{-2}$.
This detector could also be used for off-beam calibration of the MPD subsystems and to provide data for measurements of high-energy muon showers. 

The measurement of cosmic muons gives an opportunity to collect unique astrophysical observations. The ALEPH, DELPHI, and ALICE cosmic ray data~\cite{ALICE:2015wfa,Avati:2000mn,Abdallah:2007fk} contain information on muon production in extensive air showers (EAS) only for vertical showers (those with zenith angles not far from zero degree). The proposed MCORD along with the MPD \acrshort{tpc} have the unique capability of very precise measurement of atmospheric muon multiplicity distributions as a function of the zenith angle of Primary Cosmic Rays (\acrshort{prc}), up to nearly horizontal showers~\cite{Yashin:2003ew}. Such measurements have never been performed. Special attention is paid to muon groups of large multiplicity from \acrshort{eas}~\cite{Bogdanov:2018sfw,Neronov:2016iax}. Those data will be useful in the analysis of the possible discrepancies between current hadronic interaction models for extremely high energy, $\geq$10$^{15}$~eV, and to recognize the sky position of extra--galactic \acrshort{prc} sources with energy $\geq$10$^{19}$~eV~\cite{Kankiewicz:2016dha}.

\begin{figure}[t]
  \centering
 \includegraphics[width=0.8\columnwidth]{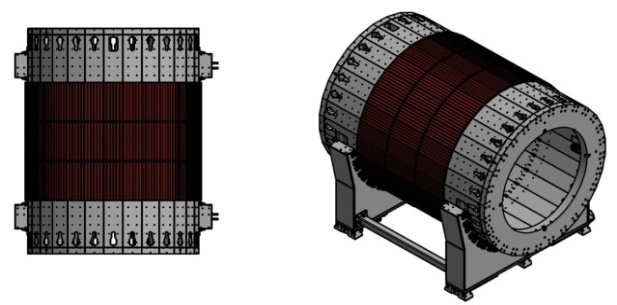}
 \includegraphics[width=0.8\columnwidth]{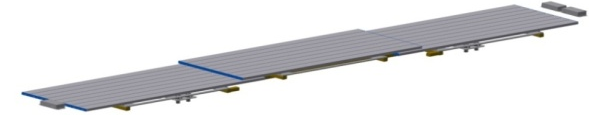}
  \caption{(Top) The MCORD is located on the surface of the MPD barrel. (Bottom) An MCORD module consists of three sections and a support frame, with the central section slightly elevated above the first one and third one. Each section consists of eight 1.65 m long scintillators. Passive hub boxes will be placed on both ends of the modules.}
  \label{fig:figmcord}
\end{figure}

MCORD is a cylindrical detector located around the MPD outer barrel~\cite{Bielewicz:2018pvt,Bielewicz:2019rnm}.
Plastic scintillators
with wavelength shifting (\acrshort{wls}) fibers are proposed to detect muons in MPD. The detector elements consist of Silicon Photo Multipliers (\acrshort{sipm}s) which will be placed on both sides of each plastic scintillator module~\cite{Grodzicka:2014yea,Grodzicka:2010jta}. The goal is to achieve a spatial resolution of about 7 cm and timing resolution below 1.5 ns (at FWHM), in the direction along the module. The design takes advantage of the small size, robustness and insensitivity to magnetic fields of the \acrshort{sipm}s. Detector modules will be arranged in a cylindrical shape around the central part of the MPD outer barrel (Fig.~\ref{fig:figmcord}). The MCORD will consists of 28 modules and it will comprise over 600 double-side readout detectors (see Fig.~\ref{fig:figmcord}).

Each \acrshort{sipm} will be directly connected to its own Analog Front-End module (\acrshort{afe}). The \acrshort{afe} consists of an amplifier, power supply, temperature compensation circuit and a unique electronic identification number. The signal will be sent by the control hub to the \acrshort{fpga} electronic analysis system based on \acrshort{mtca} crate. One \acrshort{mtca} crate receives signals from 384 \acrshort{sipm}s channels~\cite{Zabolotny:2017rkh,Zabolotny:2016cik}. 

MCORD will record signals induced by particles, mainly muons, crossing the entire MPD body, or coming from the collision vertex. The threshold for muons that may escape the MPD is approximately 1~GeV. Studying high energy muons coming from pion and kaon decays, muon-antimuon pairs coming from very rare meson decay processes, including the dark matter searches~\cite{Abegg:1994wx,Raggi:2015yfk,Arnaldi:2008er} are potential uses for the MCORD. The other potential fields of study are the decays of charmonium-like exotic states into $J/\Psi$ and then into muon pairs with kinetic energy of relative motion $\geq$~$1$~GeV~\cite{Barabanov:2016jjv,Barabanov:2019izj}. For cosmic muon measurement, a two stage coincidence signal will be used to generate a trigger signal for other MPD detection systems. 
In a first stage, two MCORD sections comprising 8 or 16 scintillators will be produced. These two sections will enable the testing of other subsystems (\acrshort{ecal}, \acrshort{tof}, \acrshort{tpc}) before their installation inside the MPD body. These sections could subsequently be installed around the MPD outer surface and be used for off-beam mode calibration of other detectors. The second stage of construction involves the delivery of 2-6 MCORD modules (each consists of 3 sections - see Fig.~\ref{fig:figmcord}). These modules could subsequently be installed around the MPD outer surface and be used for off-beam mode calibration of other detectors.

\subsection{\label{sec:construction}Infrastructure and support systems}

The MPD will be supported by several auxiliary systems, which are briefly described below.

\subsubsection{\label{sec:hall}MPD Hall}

The MPD experiment will be housed in the MPD Hall, which is an integral part of the NICA collider building. In the MPD Hall two main areas are foreseen: the south and the north side. The MPD Pit will be located on the south side. It will house the MPD after assembly. The NICA beam line crosses the MPD Pit at 1.5 m above ground level. The floor of the MPD Pit is therefore 3.19 m below ground level. The MPD, as well as the auxiliary support structure, the MPD Platform, will be placed on rails in order to allow the movement of the full structure. The MPD assembly will be performed in the SERVICE position, away from the beam line. After full assembly and initial commissioning, as well as after the initial beam tuning in NICA, the setup will be closed by endcaps and moved into the operational position around the accelerator beam line.

The north part of the MPD Hall, which will be separated by concrete blocks from the MPD Pit during beam operation, will provide additional assembly space as well as space for electronics racks, gas supply systems and other support systems for the MPD. The MPD Hall will be connected to the electrical grid by a pair of redundant transformers, which will provide electrical power to the full MPD setup. 
The MPD Hall is fully operational and the MPD assembly work is progressing.

\subsubsection{\label{sec:support}Mechanical integration and support structure}

\begin{figure}[t]
  \centering
  \includegraphics[width=0.95\columnwidth]{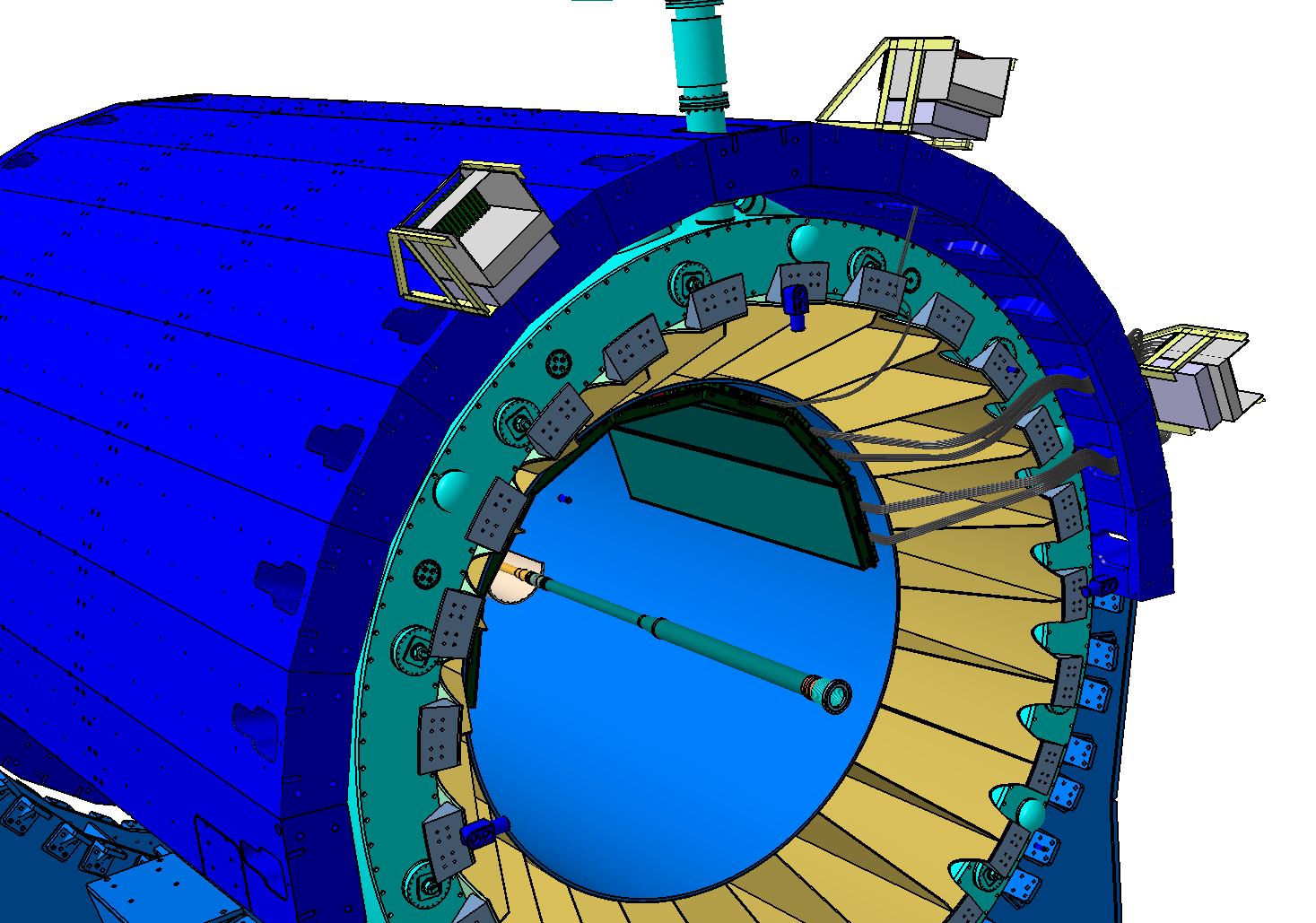}
  \caption{The MPD installation concept.}
  \label{fig:mpdstruct}
\end{figure}

The MPD Magnet provides the outer mechanical structure for the MPD. A carbon-fiber support structure, shown in Fig.~\ref{fig:mpdstruct} in yellow, is attached to the magnet yoke (blue) and provides support points for the major central barrel detector subsystems. Each Electromagnetic Calorimeter sector will be placed in a dedicated enclosure with carbon fiber walls between the sectors. On the inner wall of the structure, dedicated mounting points are provided for the 14 \acrshort{tof} sectors (28 modules). Two structural rails, placed at the level of the beam pipe on the inner wall of the structure, will support the \acrshort{tpc}. The beam pipe will be surrounded by a support structure, which may be used to attach additional detectors. The forward hadron calorimeter will have its own dedicated support structure, integrated with the magnet endcaps. The beam pipe in the initial operation is expected to have 64~mm in diameter in the central region of the MPD.

\subsubsection{\label{sec:platform}Support sytems}

The MPD requires many support systems, such as power supply (both high and low voltage), gas systems with strict requirements for purity, cooling liquids and temperature stabilisation. A Fire Protection System and an associated fast automatic fire extinguishing system are also important elements of the support systems. The MPD is protected by an Access Control System (ACS) as well as monitoring and management of cable connections, the monitoring of grounding, the level of interference with ionizing radiation, including neutrons and the intensity of uncontrolled electromagnetic fields.

The NICA-MPD adopted a service solution for moving the MPD in the MPD Hall from the RUN (Measuring) position to the SERVICE position about 11~m away. This solution required the design of a special mobile platform, the NICA-MPD-PLATFORM (wagon with a load capacity of about 150 tons), shown in Fig.~\ref{fig:mpdmagnet}. The NICA-MPD-PLATFORM will house all the MPD support systems forming a structure called the Experiment Control System, which can move between RUN and SERVICE positions.

The NICA-MPD-PLATFORM will have 32 RACK crates with dimensions of $600\times1200\times47$ U. To ensure compactness of the structure, it has been designed as a four-level construction. Each level is made of eight RACK units intended for MPD electronic support systems and five cooling units with dimensions of $300\times1200\times47$ U. The first (lowest) level is reserved for power distribution, including the main power line switch, as well as gas distribution and cooling medium distribution. The other floors are for power supply electronics control and data concentrators for the \acrshort{daq}.

The NICA-MPD-PLATFORM is powered by a two-line three-phase power network with a NEUTRAL and Protective Earth Cable (PEC). Each RACK can be loaded with $3\times25$ A current. The RACK is equipped with an autonomous fire extinguishing panel, connection management systems in the InteliPhy technology and its own Intelligent Power Distributor (\acrshort{ipd}). The task of the \acrshort{ipd} is to monitor up to 51 harmonics in the network, provide intelligent optimization of load balance (minimizing the current in the NEUTRAL cable), autonomous detection and switching off of damaged circuits within the available power allocation. 
The NICA-MPD-PLATFORM is managed under SCADA WinCC.

\subsection{\label{sec:electronics}Electronics}

\subsubsection{\label{sec:slowcontrol}Slow Control System}

All components of the MPD detection system will be under constant monitoring. Their monitoring includes, but is not limited to: high voltage levels and status, low voltage level and status, composition purity and flow of the gas mixtures for the \acrshort{tpc} and \acrshort{tof}, status of the \acrshort{daq} systems, status of the connections, operational parameters of the detectors, such as operational voltage, pressure, temperature, status of the detector cooling system. All those parameters will be stored in a dedicated database, in order to monitor the detector performance as well as to provide data for calibration purposes. Additional parameters, such as status of the superconducting magnet and the magnetic field, status of the beam conditions and the connection to the accelerator facility, fire-protection systems, radiation safety and others will be provided. The system will also use a sophisticated alarm framework to ensure safe, stable and reliable operation of the detector.

\subsubsection{\label{sec:daq}Data Acquisition}

\begin{figure}[t]
  \centering
  \includegraphics[width=0.9\columnwidth]{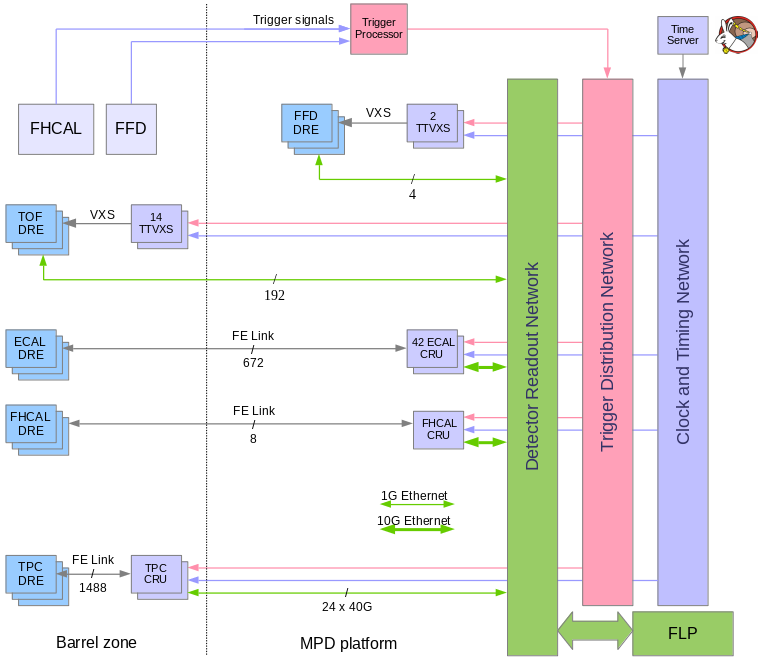}
  \caption{MPD \acrshort{daq} trigger, timing and data links.}
\label{fig:daq}
\end{figure}

The overall architecture of the \acrshort{daq}, shown in Fig.~\ref{fig:daq}, was designed as a data driven fully parallel push architecture. At each stage the content of the data stream drives autonomously its routing through the processing chain from source to destination. The throttling is implemented by back pressure from the destination to the source. For timing synchronization, White Rabbit is used. It provides sub-nanosecond accuracy and picosecond precision of synchronization for large distributed systems.

Elements of \acrshort{daq} will be installed both in the Barrel zone and the NICA-MPD-PLATFORM. All FLPs (first level processors) will be installed in \acrshort{mdc} (mini data center). \acrshort{mdc} is planned to be installed in the MPD hall, not far from NICA-MPD-PLATFORM. All FLPs, network infrastructure cluster and transient data storage will be located inside the \acrshort{mdc}.

The requirements to the \acrshort{daq} system follow from the physics tasks. Thus the average sustained event rate handled by the \acrshort{daq} system should be 7 kHz. The estimated data rate is about 6.5 GB/s.

Data from different sub-detectors will be combined together by the Event Builder system, whose components receive the data flow from FLPs, create events and store combined events in a transient data storage. 

\begin{figure}[b]
  \centering
  \includegraphics[width=0.95\columnwidth]{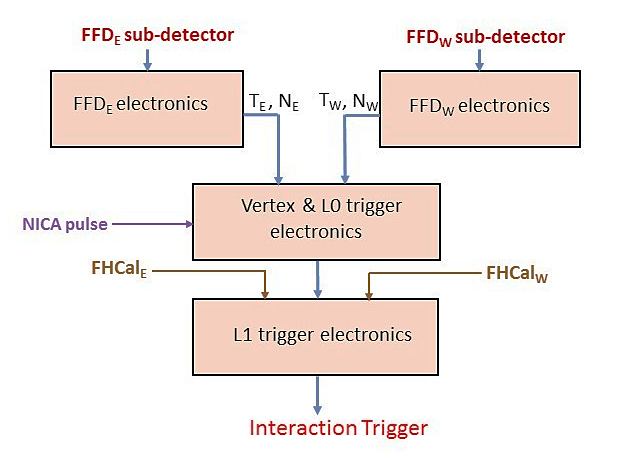}
  \caption{A scheme of the interaction trigger production.}
  \label{fig:trigsch}
\end{figure}

The fast vertex-trigger, provided by \acrshort{ffd}, is the level-0, L0, trigger of the MPD experiment. The fast determination of the $z$-position of the collision requires two pulses, TE and TW, produced by FFDE and FFDW, respectively. A “good vertex” signal in coincidence with NICA pulse generates the L0 trigger pulse. The L0 trigger efficiency for two energies of Au + Au collisions obtained in MC simulation (DCM-QGSM-SMM + GEANT4 code) is shown in Fig.~\ref{fig:trigeff}.

The L1 trigger uses additional information about energy deposition in FHCalE and FHCalW and it produces the interaction trigger for the MPD \acrshort{daq} and sub-systems. A scheme of the trigger generation is shown in Fig.~\ref{fig:trigsch}.

\section{\label{sec:itresources}Software development and computing resources for the MPD experiment}

\subsection{\label{sec:software}Software}

Development of software packages for simulation, reconstruction and physical analyses is an important part of the MPD experiment. The software framework for MPD, named MpdRoot~\cite{mpdroot}, has been developed and is routinely used by MPD collaborators. All the performance studies shown in this paper have been obtained using the MpdRoot framework. MpdRoot is in constant development by a dedicated software team as well as by individual collaboration members. This framework extends the FairRoot~\cite{FairROOT} classes via the inheritance mechanism and uses external programs from the FAIRsoft package. The latter includes tools for software development like ROOT, BOOST, GEANT4(3), DDS and ZeroMQ. All these packages are available under the LGPL license and therefore suitable for fundamental scientific research purposes. They are natively developed and primarily used on GNU/Linux-based operating systems, which are the de-facto standard in High Performance Computing for high energy physics.

The MpdRoot framework provides an interface to most Monte Carlo event generators, which model heavy ion collisions in the NICA energy range. 
Another important task is the transport of the produced particles through detector materials and the simulation of the resulting response of the detectors in the reconstruction stage. For these tasks the GEANT4~\cite{GEANT4} package is used. 

\begin{figure}[b]
  \centering
  \includegraphics[scale=0.2]{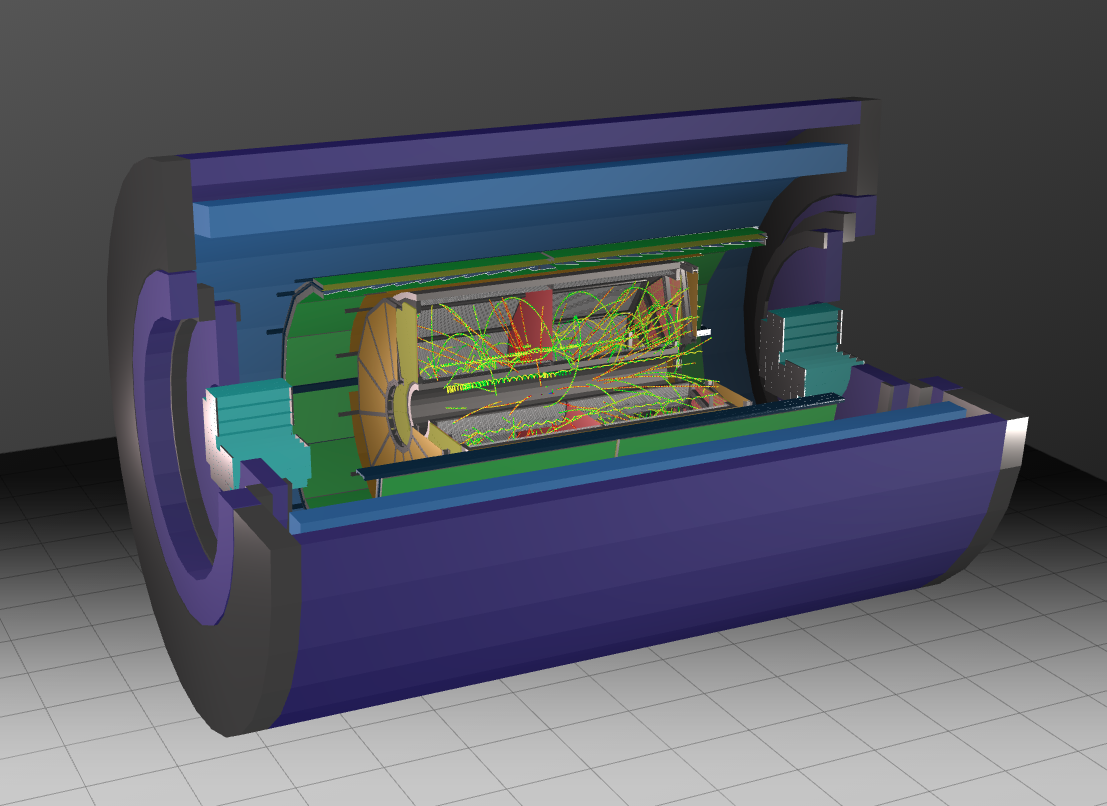}
  \caption{The event display for the MPD experiment with simulated tracks.}
  \label{soft:evdisp}
\end{figure}

The MpdRoot framework also provides methods for the reconstruction and analysis of data from simulated and real interactions. 
MpdRoot was used to carry out research and simulation of the technical design of all sub-detectors in order to optimize their designs. It was also used to assess the feasibility of specific physics studies.

A very useful and valuable tool to visualize properties of the detectors and the data is the event display, which shows the detectors structure with their responses and reconstructed data. Figure~\ref{soft:evdisp} presents a three-dimensional global event display showing the main detectors of the MPD and simulated tracks from a high multiplicity ion-ion collision.

\subsection{Computing}

\begin{figure}[t]
  \centering
  \includegraphics[scale=0.35]{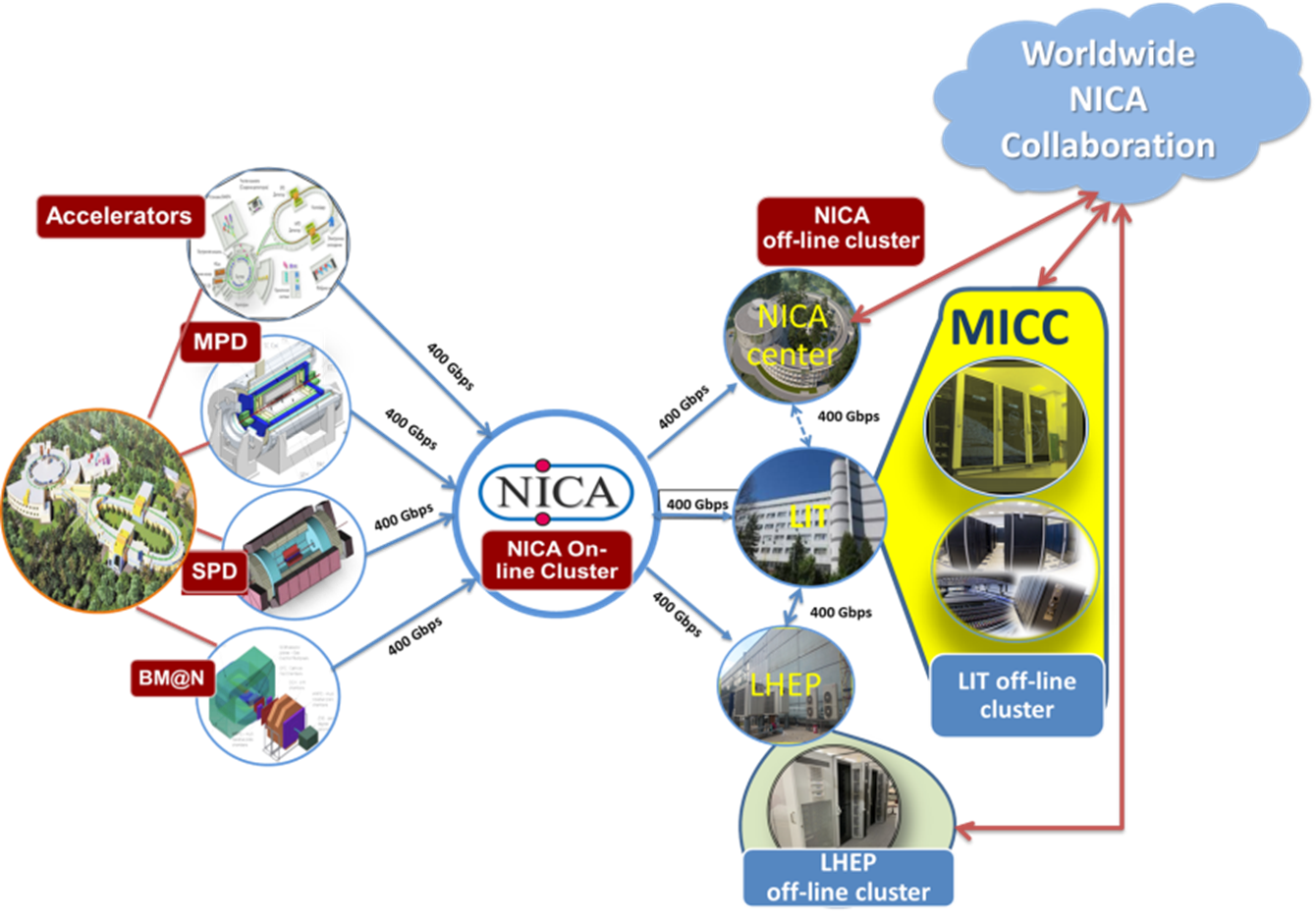}
  \caption{NICA computing resources.}
  \label{comp:res}
\end{figure}

The computing resources for the MPD experiment, shown in Fig.~\ref{comp:res}, are an integral part of the “NICA Complex” project. They are created, maintained and operated using modern technologies as a geographically distributed information and computing cluster, which satisfies the requirements for experimental data processing and analysis, as well as theoretical studies. 
This computer infrastructure aims at accumulating, transferring and storing physical data acquired from the main nodes of the “NICA Complex”, {\it i.e.} accelerators, BM\@{}N, MPD, and SPD detectors
as well as at processing data and analyzing them. All processes and systems of the infrastructure are constantly monitored.

The main technological elements of the basic configuration of the distributed information and computing cluster are located in four places, three of which are situated at the site of the \acrlong{lhep} (\acrshort{lhep}) and one in the \acrlong{lit} (\acrshort{lit}). The \acrshort{lit} NICA cluster is part of the JINR Multifunctional Information and Computing Complex (\acrshort{micc}). This structure presupposes its connection with computing complexes of other organizations which will be involved in the “NICA Complex” project in the future.

\begin{figure}[t]
  \centering
  \includegraphics[scale=0.22]{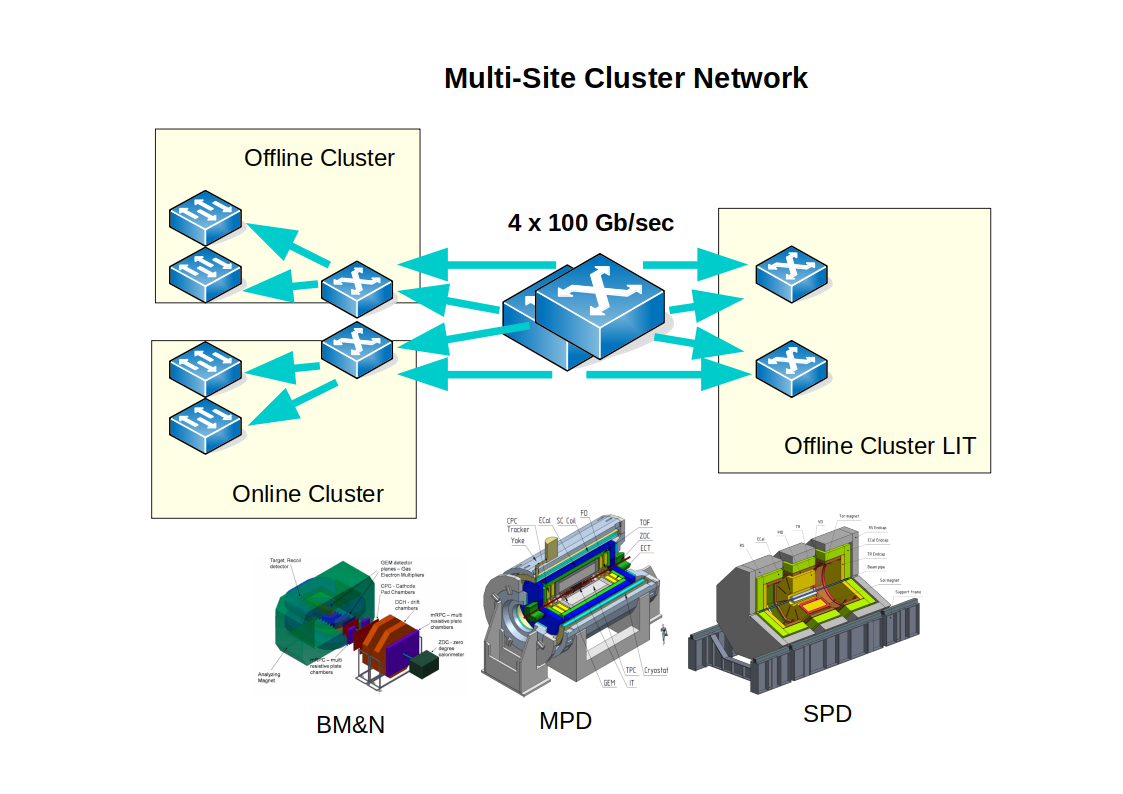}
  \caption{NICA network topology scheme, showing interconnections between physically separate computing centers in Laboratory of Information Technology (\acrshort{lit}) and LHEP.}
  \label{comp:net}
\end{figure}

The most important stage in the development of the \acrshort{lhep} network infrastructure is the organization of optical backbones between the \acrshort{lit} and \acrshort{lhep} sites with a bandwidth of $4 \times 100$ GB/s. Figure~\ref{comp:net} shows the scheme of the information and computing network implemented between the two JINR sites.

The mass production of simulated and experimental data is provided by the DIRAC Interware package. DIRAC aims at integrating heterogeneous computing and data storage resources into a unified platform. Resource integration is based on the use of standard data access protocols (xRootD, GridFTP, etc.) and pilot jobs. Thanks to this, a unified environment is created in which it is possible to run jobs, manage data, build processes and control their implementation. In the framework of DIRAC, batch processing systems, grid computing elements, clouds, supercomputers and even separate computing nodes, can act as computing resources. Storage resources are limited only to those that support file transfer protocols used in grid systems.

In addition to the computing resources of the “Govorun” supercomputer, the computing resources allocated for NICA on the Tier1 and Tier2 components of the \acrshort{micc}, are integrated using the DIRAC Interware. The procedure for recording and reading data using the mass storage facility (tape robot) was checked and tested.

\subsection{\label{sec:mcprod}Preparation for data taking}

A centralized large-scale production of events simulated with Monte Carlo generators, coupled to detector response simulations and event reconstructions in the MpdRoot, has been organized. All available MPD computing resources have been involved in this test. More than $10^{7}$ events were produced with the full simulation and reconstruction chain, and more than $5\times10^{8}$ generator-only events were also provided. Several physics event generators have been integrated within MpdRoot. In particular, the \acrshort{urqmd}, PHSD, vHLLE, LAQGSM, SMASH and PHQMD generators have been successfully used for large-scale Monte Carlo productions. 

The analysis codes for specific analysis tasks are being prepared within the MPD Physics Working Groups, and are developed to work with the official MPD data formats, such as MPD DST (Data Summary Tape) and mini-\acrshort{dst} (a slimmed-down version of the MPD \acrshort{dst} with focus on data for physics analyses). The codes are available to all collaborators in a shared repository and code integration and software release rules are applied. The physics analysis code will also be used for high-level monitoring of the data quality, via the preparation of general performance plots as well as reconstruction of specific physics observables, such as hyperon mass and width, specific ionization energy losses, uniformity of angular acceptance and others.

\section{Examples of physics feasibility studies}\label{sec:observables}

An important feature of the NICA complex is the capability to deliver high luminosity and to vary the collision energy and the collision system. The MPD is capable of detecting a large variety of probes such as charged hadrons, electrons and photons.  Furthermore, the MPD will be capable of providing three-dimensional tracking and accurate \acrshort{pid} given its large-volume gaseous tracking chamber, its \acrshort{tof}  and calorimetry components. The MPD will be able to produce event-by-event information on charged particle tracks for both primary and secondary vertices as well as on the collision centrality by measuring the number of participants. This will allow us to measure the yields of charged particles of different types,  their multiplicity, and their transverse momenta and rapidity distributions as a function of collision centrality.

In this section, we present several examples of physics feasibility studies performed by the MPD Collaboration in 2019-2020.
The examples are not intended to deliver a single consistent physics message, instead, they are
provided to illustrate the MPD capabilities to fulfill its physics program.

Several state-of-the-art transport models such as \acrshort{urqmd}~\cite{Bleicher:1999xi,Bass:1998ca},
AMPT~\cite{Lin:2004en}, PHSD~\cite{Cassing:2008sv,Cassing:2009vt}, PHQMD~\cite{Aichelin:2019tnk},
SMASH\cite{Weil:2016zrk} and  DCM-QGSM-SMM~\cite{Baznat:2019iom}
have been selected to be used as event
generators for the present studies. These models provide physically well-motivated scenario for A+A  collisions in the NICA energy range.  The generated  events then serve as the input
for the full chain of  realistic simulations of the MPD
subsystems, based on the GEANT3/GEANT4 platforms~\cite{GEANT4} and
reconstruction algorithms, built in the MpdRoot~\cite{mpdroot}.
Realistic procedures have been implemented for detector response simulation, centrality estimates, particle identification and event plane reconstruction.

\subsection{\label{sec:centrality}Centrality determination}
Centrality is a key parameter for defining the collision system size in relativistic heavy-ion collisions.
Usually, the nucleus-nucleus collisions are characterized by the particle multiplicity measured in a pre-defined pseudorapidity intervals or by the energy of spectator fragments detected  in the forward rapidity region. 
In the MPD experiment, two procedures are considered for the centrality determination. The first one is based on the charged particle multiplicity information provided by the \acrshort{tpc} and is planned to be used in the first measurements of heavy-ion collisions at NICA. The second one uses the energy of spectator fragments provided by the \acrshort{fhcal} information. A combined approach using both procedures is under study.
 
A sample of 1 million minimum-bias Au+Au and Bi+Bi collisions at
$\sqrt{s_{\rm NN}}$ = 7.7  GeV generated with the cascade model  \acrshort{urqmd} (version 3.4)~\cite{Bass:1998ca}
has been  used  as input for the analysis of the multiplicity of charged particles in the \acrshort{tpc}.
The MPD response was simulated using the GEANT4 transport code built 
in the MpdRoot framework along with the reconstruction software.

\begin{figure}[t]
\centering
 \begin{minipage}[h]{0.91\linewidth}
	\includegraphics[width=0.91\linewidth]{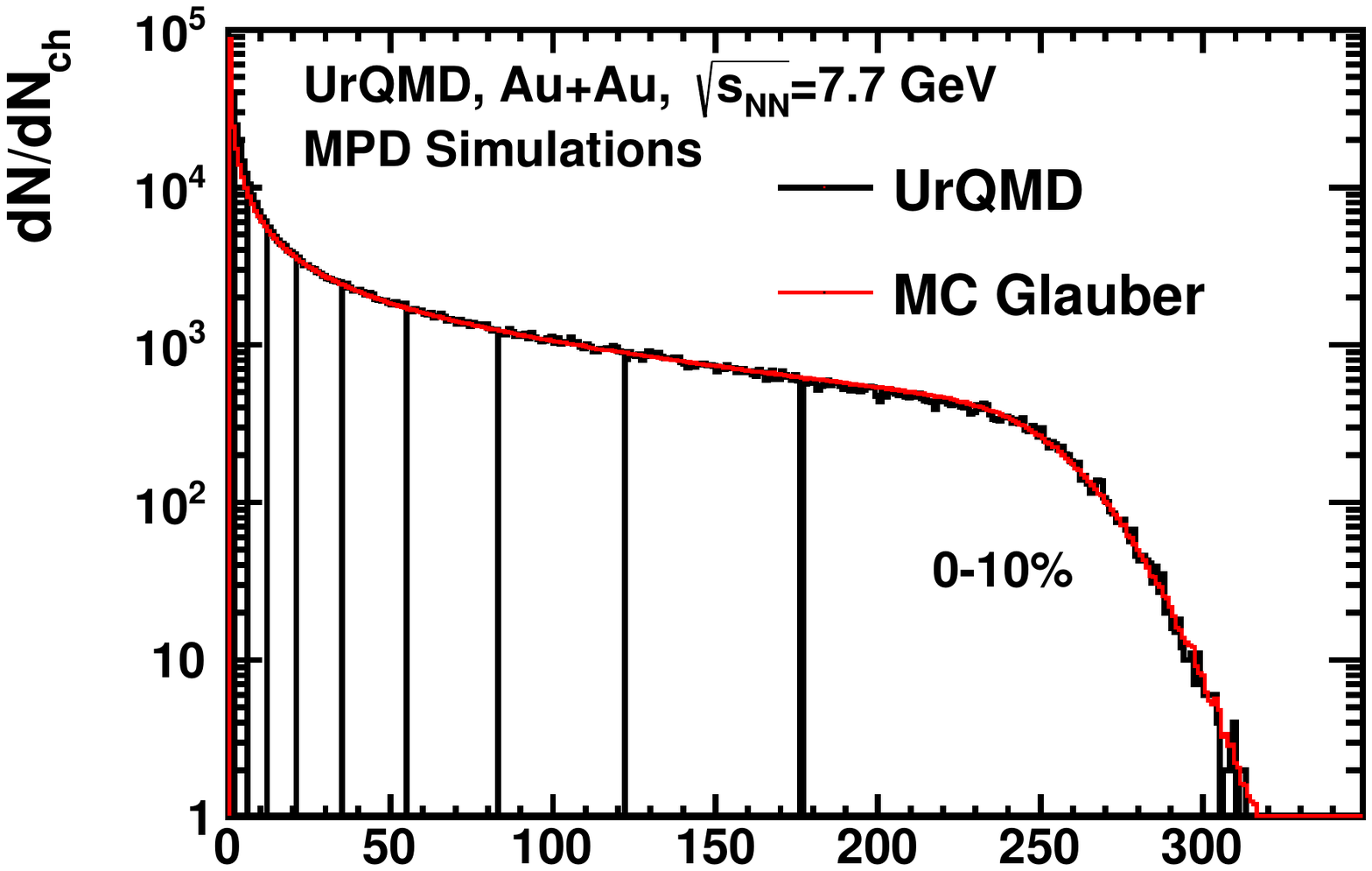} \hspace*{4cm} (a)
 \end{minipage}
 
 \begin{minipage}[h]{0.91\linewidth}
	\includegraphics[width=0.91\linewidth]{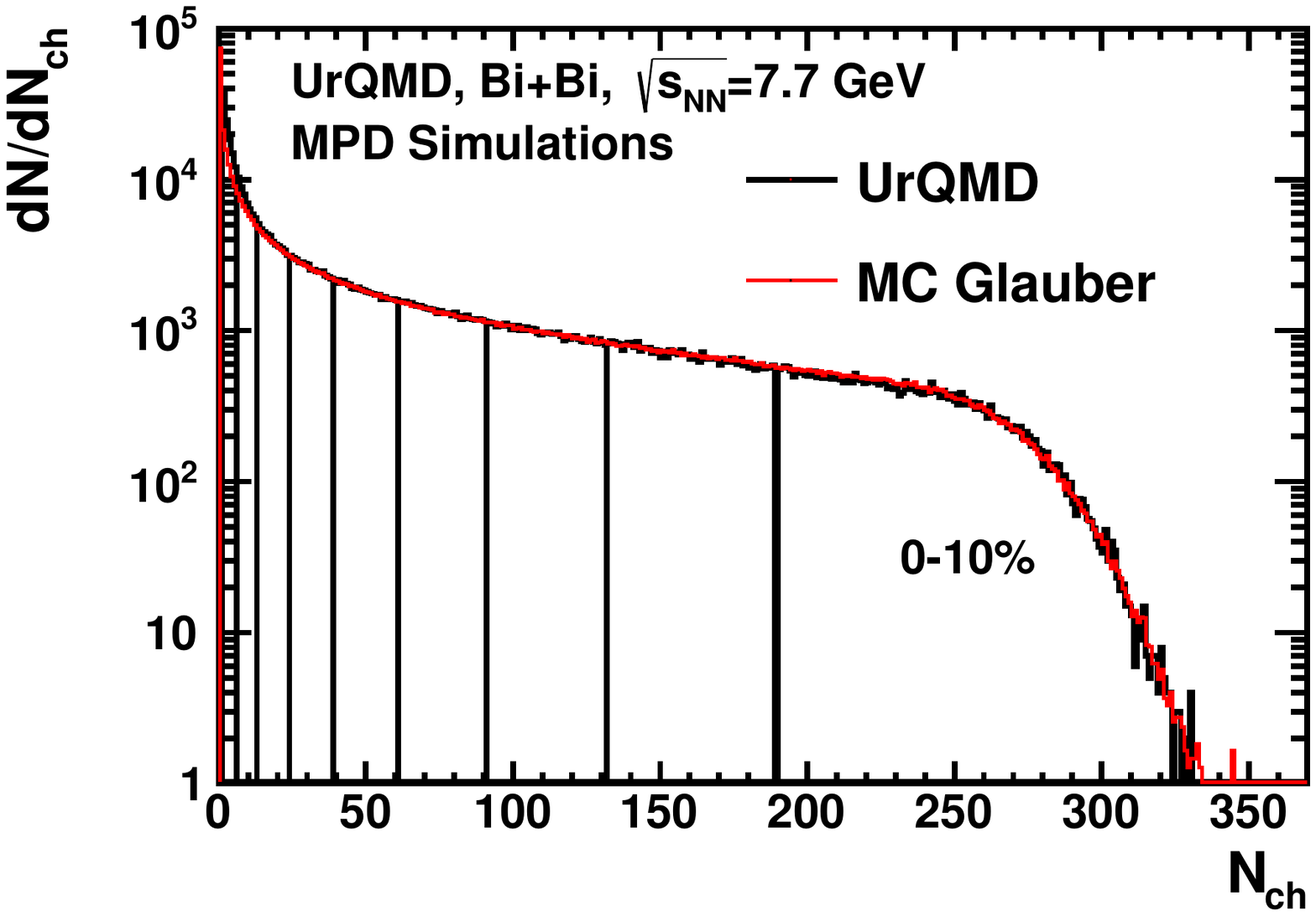} \hspace*{4cm} (b)
 \end{minipage}	\centering
 \caption{Track multiplicity distribution from fully reconstructed \acrshort{urqmd} events  for
         Au+Au  (a) and Bi+Bi (b) collisions 
         at $\sqrt{s_{\rm NN}}$ = 7.7 GeV compared to the fitted distributions using MC-Glauber approach (red line).
       10\% centrality classes defined with MC-Glauber normalization are indicated by black vertical lines.}
	\label{fig:CentralityFramework_refMultFit}
\end{figure}

\begin{figure}[t]
	\centering
 \begin{minipage}[h]{0.91\linewidth}
	\includegraphics[width=0.91\linewidth]{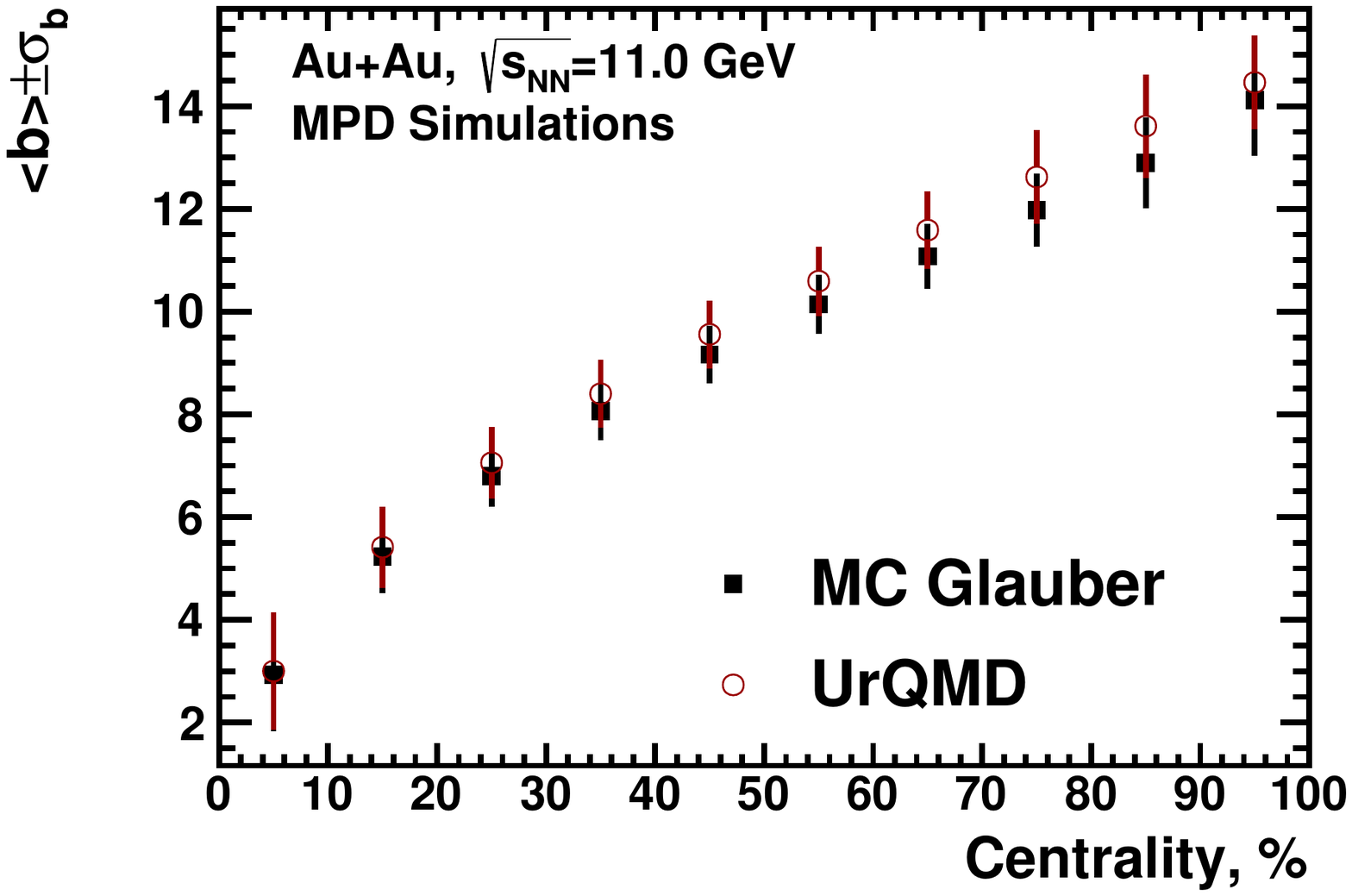}
 \end{minipage}
 
 \begin{minipage}[t]{0.91\linewidth}
	\includegraphics[width=0.91\linewidth]{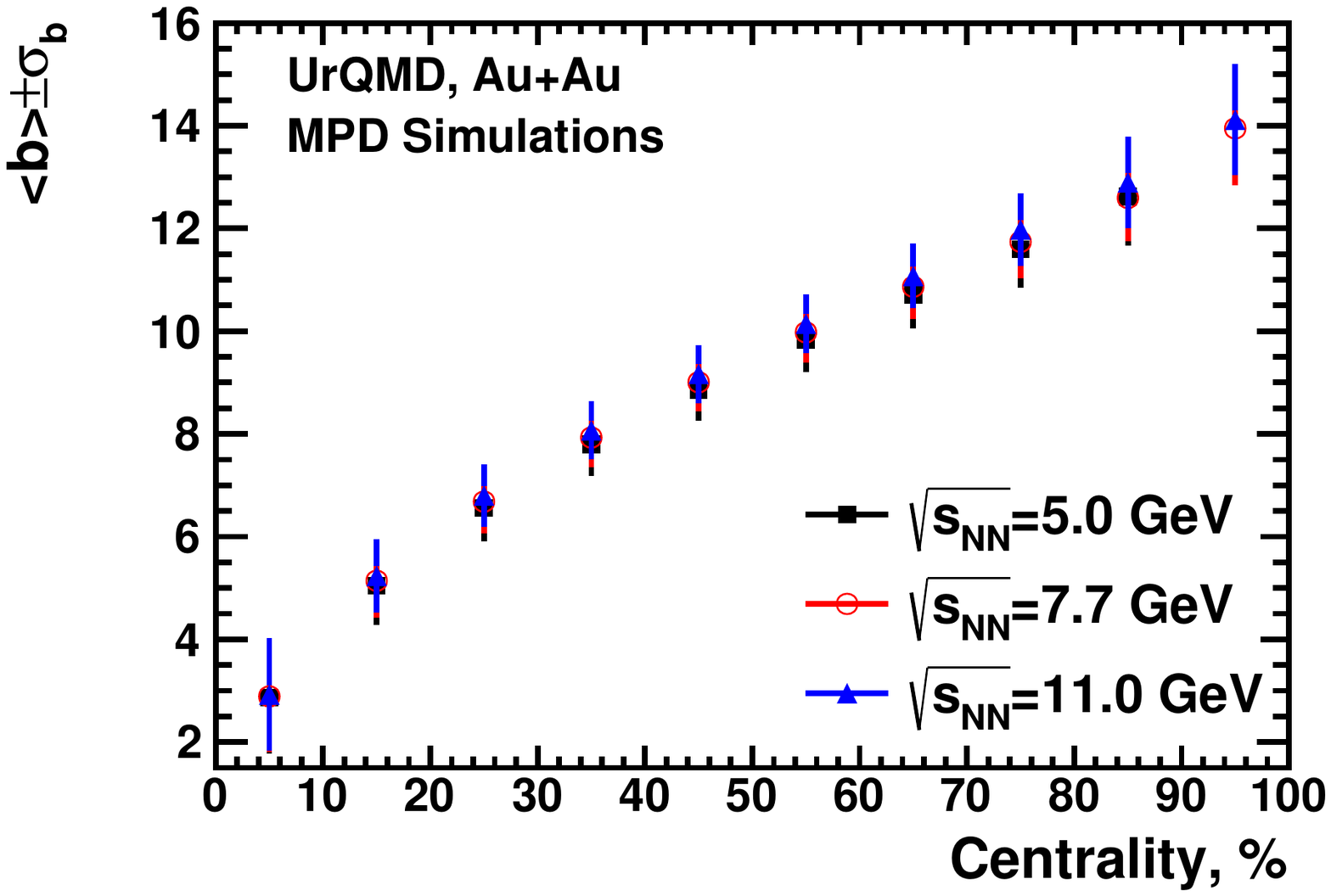}
 \end{minipage}	\centering
 \caption{Averaged impact parameter $\langle b \rangle$ for all centrality classes.
   (Top): Comparison of the results for $\langle b \rangle$ from MC-Glauber with the one from the \acrshort{urqmd} model for
   Au+Au collisions at $\sqrt{s_{\rm NN}} = 11$~GeV.
   (Bottom): Comparison of the results of MC-Glauber for different $\sqrt{s_{\rm NN}}$.
   RMS of the corresponding impact parameter distribution are denoted as error bars.}
	\label{fig:CentralityFramework_Results}
\end{figure}

As an example, Fig.~\ref{fig:CentralityFramework_refMultFit} shows the multiplicity distributions of charged
tracks ($N_{\rm ch}$) in the \acrshort{tpc} with $|\eta| < 0.5$ for Au+Au  (a) and Bi+Bi (b) collisions at $\sqrt{s_{\rm NN}}$ = 7.7 GeV.
Only tracks with at least 16 \acrshort{tpc} hits and transverse momentum $p_{\rm T}>0.15$ GeV/$c$ have been used in the analysis.
To reproduce the experimental multiplicity distribution,
a Monte Carlo Glauber Model (MC-Glauber or MG here in the text)~\cite{Loizides:2014vua} was  coupled with  a model for particle production, based on a
negative binomial distribution (NBD)~\cite{Kharzeev:2000ph,Adamczyk:2012ku}.
In this approach, the multiplicity $M_{MG}$ of particles in  a heavy-ion collision is modeled
as the sum of particles produced from a set of $N_a$ independent emitting sources (ancestors)~\cite{Klochkov:2017oxr,Klochkov:2017mpn}. Each
ancestor produces particles according to a NBD  $P_{\mu,k}$ with mean value
$\mu$ and width $k$: $ M_{MG}(N_{a},\mu,k)=P_{\mu,k}\times  N_{a}$.
The number of ancestors $N_{a}$ can be parameterized by :
\begin{equation}
N_{a} = [f N_{part} + (1-f)N_{coll}],
\end{equation}
where $N_{\rm part}$ is the number of participants and $N_{\rm coll}$ is the number of binary collisions simulated with the MC-Glauber Model.
The track multiplicity distribution  $M_{MG}$  of charged particles in the \acrshort{tpc} is simulated for
an ensemble of events and for various values of the NBD parameters $\mu$, $k$, and the $N_a$ parameter $f$.
A minimization procedure is applied to find the optimal set of parameters which result in the smallest value of $\chi^2/{\rm N_{\rm dof}}$.
The result of the procedure is shown
in Fig.~\ref{fig:CentralityFramework_refMultFit}  by the red line. There is good agreement between the 
initial $N_{\rm ch}$ and the  fitted $M_{MG}$ distributions.
After the fitting procedure, the MC-Glauber model is used to estimate the initial geometry of the centrality classes. Mean values and RMS of the MC-generated impact parameter distributions are shown in
Fig.~\ref{fig:CentralityFramework_Results} as a function of centrality.

The upper part of Fig.~\ref{fig:CentralityFramework_Results}  shows the very good agreement between
the average impact parameter $\langle b \rangle$ directly obtained from the \acrshort{urqmd} model and the one estimated from the MC-Glauber model.
The lower part of Fig.~\ref{fig:CentralityFramework_Results}  shows the MC-Glauber Model $\langle b \rangle$ for
Au+Au collisions at different energies, which exhibits a very small energy dependence of $\langle b \rangle$ as a function of centrality.

\begin{figure}[th]
	\centering
 \begin{minipage}[h]{0.91\linewidth}
	\includegraphics[width=0.91\linewidth]{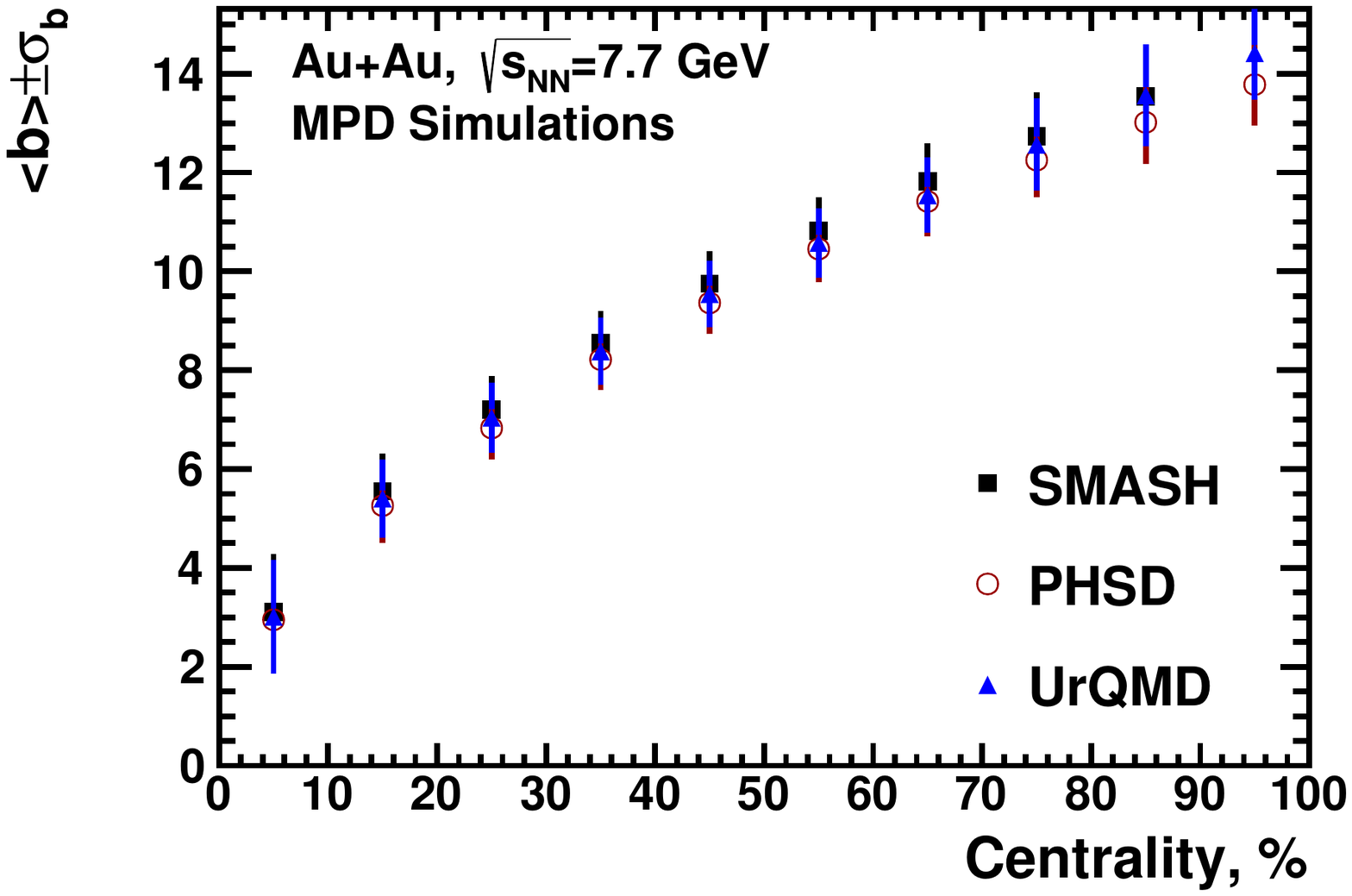}
 \end{minipage}
 
 \begin{minipage}[h]{0.91\linewidth}
	\includegraphics[width=0.91\linewidth]{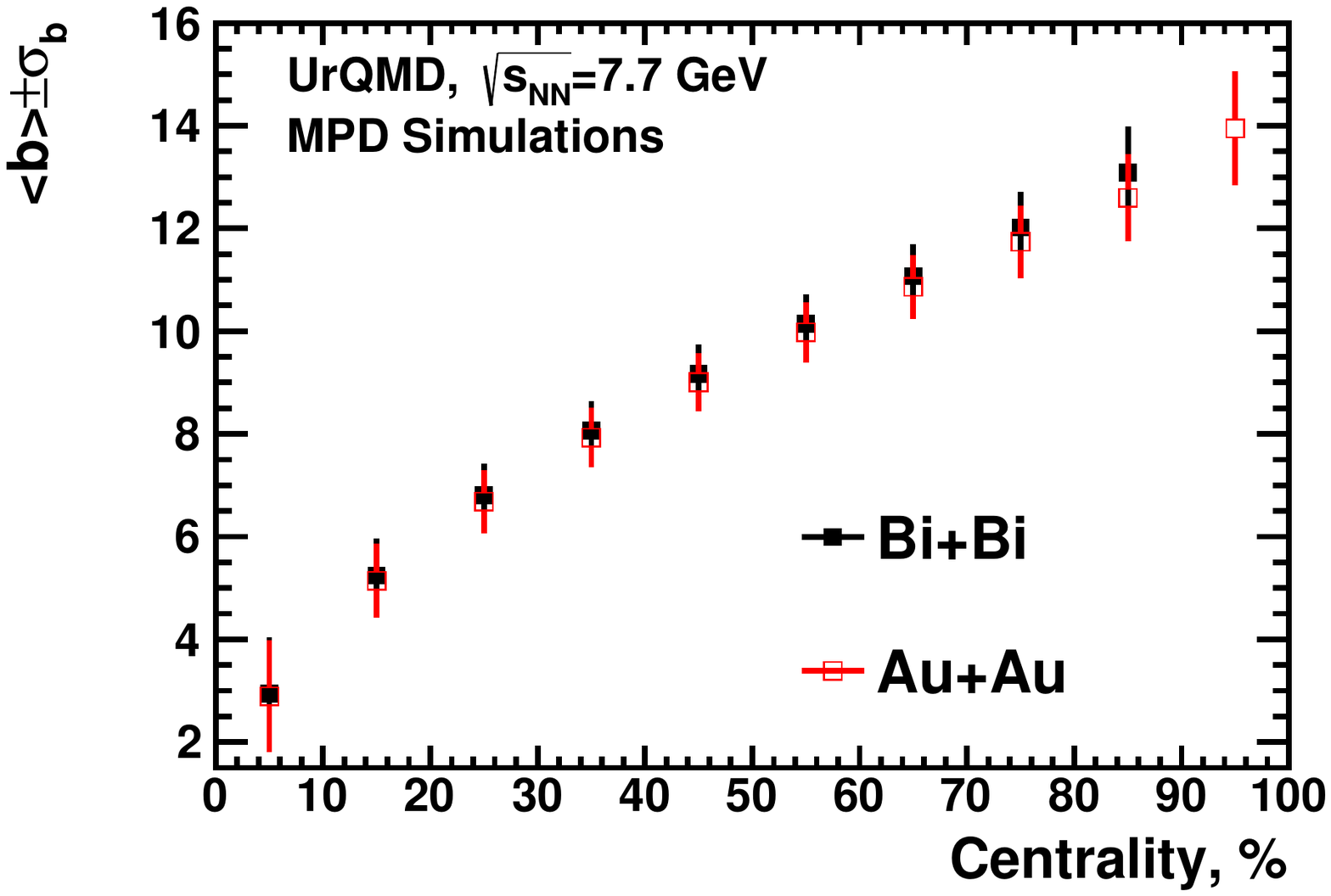}
 \end{minipage}	\centering
 \caption{ (Top) A comparison of the results from \acrshort{urqmd}, PHSD and SMASH models for the centrality dependence of the averaged impact parameter for Au+Au collisions at
   $\sqrt{s_{\rm NN}} = 7.7$~GeV. 
   (Bottom) The comparison of results for Au+Au and Bi+Bi collisions at $\sqrt{s_{\rm NN}} = 7.7$~GeV for \acrshort{urqmd} events. }
	\label{fig:CentralityFramework_ModelComparison}
\end{figure}

The centrality determination procedure was also done with the PHSD~\cite{Cassing:2008sv,Cassing:2009vt} and
SMASH~\cite{Weil:2016zrk} models.
The upper panel of Fig.~\ref{fig:CentralityFramework_ModelComparison} shows 
$\langle b \rangle$ as a function of centrality for the \acrshort{urqmd}, PHSD and SMASH models for Au+Au collisions at
$\sqrt{s_{\rm NN}} = 7.7$~GeV.  The centrality dependence of the average impact parameter $\langle b \rangle$ is very similar in all models shown.
The lower panel of Fig.~\ref{fig:CentralityFramework_ModelComparison} shows the comparison of the results for
Au+Au and Bi+Bi collisions at $\sqrt{s_{\rm NN}} = 7.7$~GeV obtained for \acrshort{urqmd} events. 

\begin{figure}[th]
  \centering
  \includegraphics[width=0.9\columnwidth]{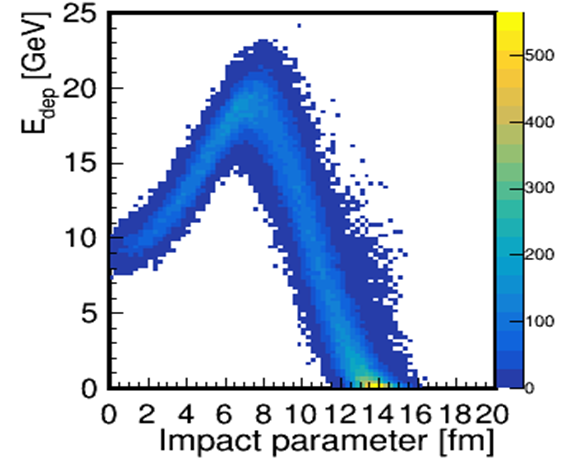}
  \includegraphics[width=0.9\columnwidth]{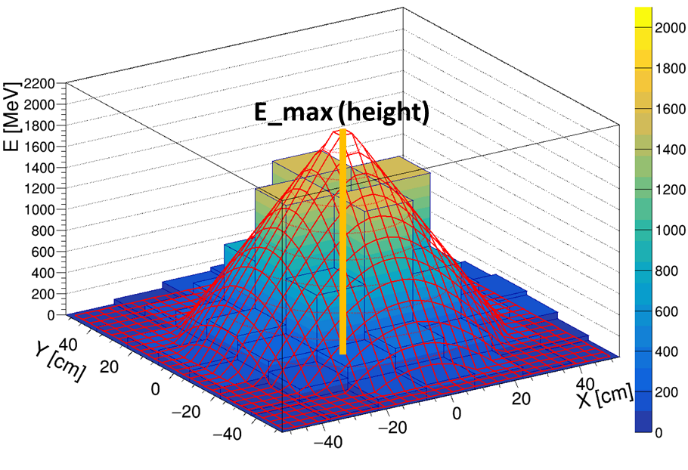}
  \caption{Top: dependence of the energy deposition in the \acrshort{fhcal} on the MC-generated impact parameter for events, simulated with the DCM-QGSM fragmentation model. Bottom: the two-dimensional linear fit (line mesh) of example single-event energy distribution in the \acrshort{fhcal} modules (histogram).}
  \label{fig:fhcal_cent_1}
\end{figure}

\acrshort{fhcal} provides a different way to determine the centrality classes of ion collisions by measuring the energy deposition and space distribution of spectators, i.e. non-interacting nuclear fragments~\cite{FHCal_TDR}. Unfortunately, most of the bound fragments escape through the beam hole in the \acrshort{fhcal} center. Therefore, the impact parameter dependence of energy deposition in the \acrshort{fhcal} has a non-monotonic behavior, and the detected spectator energy can be the same for central and peripheral events. This is illustrated in Fig.~\ref{fig:fhcal_cent_1}, top for events simulated with the DCM-QGSM fragmentation model.

To resolve this ambiguity in the energy spectrum, the three-dimensional   energy distribution in the \acrshort{fhcal} modules was fitted by a linear function (by a cone in the two-dimensional case, Fig.~\ref{fig:fhcal_cent_1}, bottom). This fit provides additional parameters, such as the height of the cone, its radius and volume. The height estimates the energy of free spectators in the \acrshort{fhcal} beam hole. The cone radius reflects the scattering angle of free spectators and the cone volume provides an estimate of the total spectator energy. The correlations of these fit parameters with the energy deposition in the \acrshort{fhcal} provides an opportunity to distinguish the central and peripheral events in spite of their similar energy depositions in the calorimeter.

\begin{figure}[th]
  \centering
  \includegraphics[width=0.9\columnwidth]{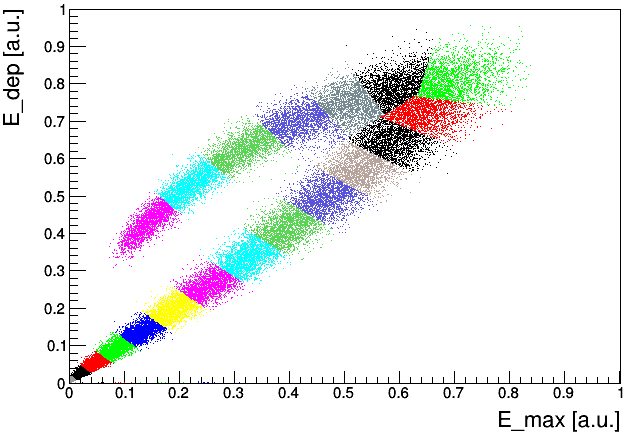}
  \includegraphics[width=0.9\columnwidth]{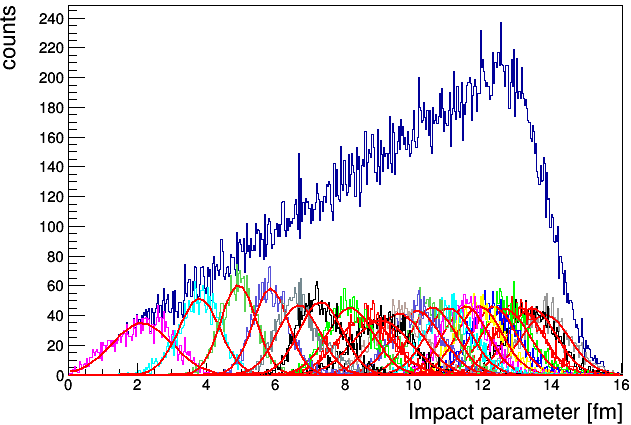}
  \caption{Top: correlation of the energy deposition in  the \acrshort{fhcal} and the height of the cone, obtained from the linear fit of the two two-dimensional energy distributions in the \acrshort{fhcal} modules.  The different colors indicate groups of events within 5\% centrality ranges.  Bottom: distributions of the MC-generated impact parameters for each 5\% group of events fitted to a Gaussian.}
  \label{fig:fhcal_cent_2}
\end{figure}

In Fig.~\ref{fig:fhcal_cent_2}, top, such correlation is shown for the height of the cone, $E_{max}$. This correlation was fitted by a polynomial and then the perpendiculars to the envelope were plotted to select 20 groups of events. Each group includes 5\% of all events and is indicated by a different color. The distribution of the MC-generated impact parameters for each 5\% group of events was fitted by a Gaussian as shown in Fig.~\ref{fig:fhcal_cent_2}, bottom. 

\begin{figure}[th]
  \centering
  \includegraphics[width=0.9\columnwidth]{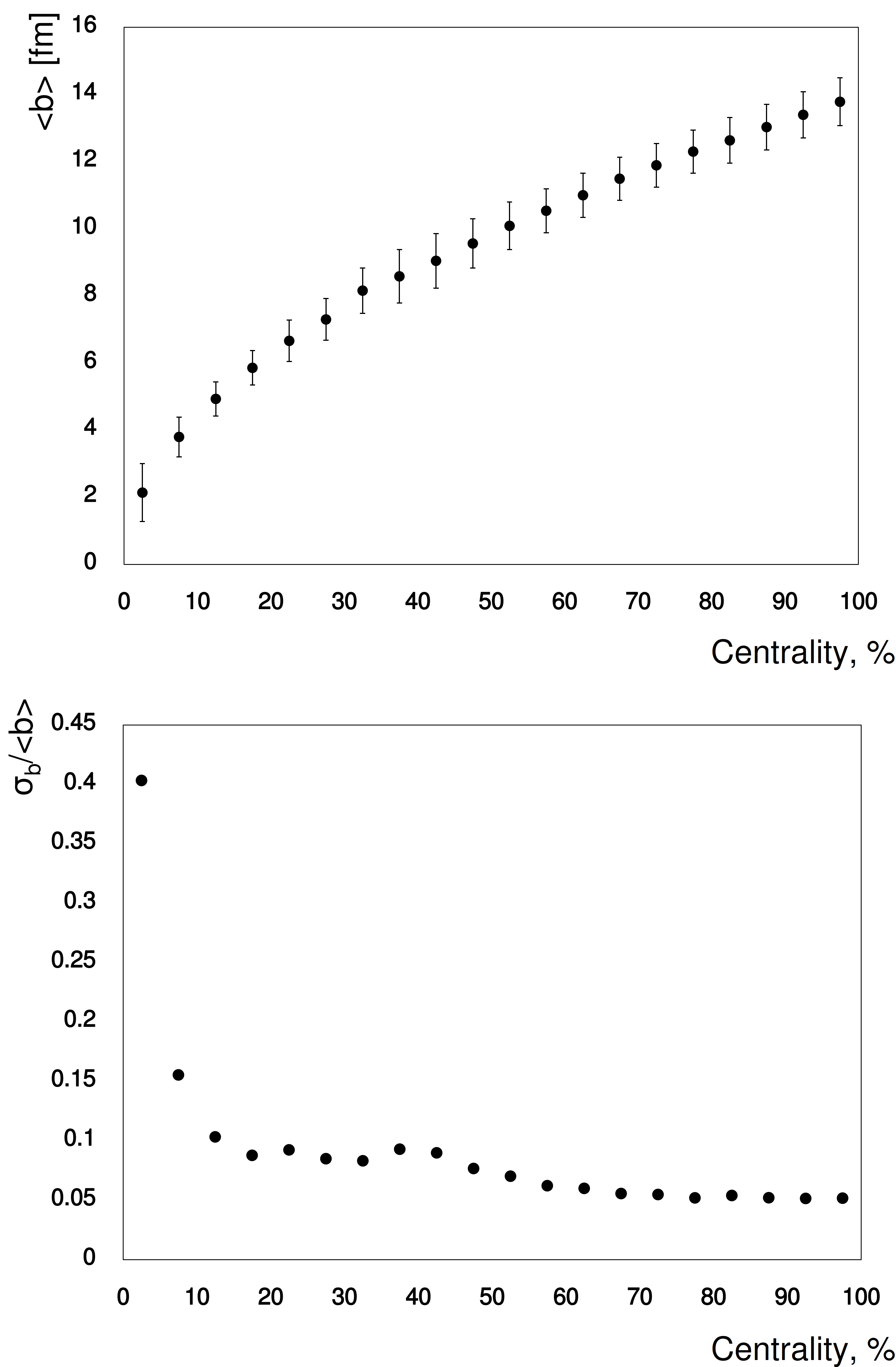}
  \caption{Dependence of the mean value of the energy deposition (top) and the width (bottom) of the Gaussian fit to the group of events with 5\% centrality.}
  \label{fig:fhcal_cent_3}
\end{figure}

The results of the fits are presented in Fig.~\ref{fig:fhcal_cent_3}. The top and bottom plots show the centrality dependence of the mean value and the width of the Gaussian fit, respectively. 
The width values, shown in Fig.~\ref{fig:fhcal_cent_3} (bottom), indicate that the current value of 5\% for the bin width is arbitrary.
In order to minimise the trivial fluctuations due to mixing of events with a broad impact parameter distribution, the individual width of each class will be optimized for central and peripheral events using the approach developed  earlier in Ref.~\cite{Drozhzhova:2016njd}. The accuracy of the mean impact parameter value and its resolution is currently widely discussed (see Refs.~\cite{Das:2017ned,Kurepin:2020goe}).
The resolution of the centrality determination combining the \acrshort{fhcal} information with the \acrshort{tpc} multiplicity measurement is under study~\cite{Golubeva:2013hla}.

\subsection{Bulk properties: hadron spectra, yields and ratios}
\label{sec:bulkspectra}

\begin{figure}[hbt]
\begin{center}
    \includegraphics[width=0.42\textwidth]{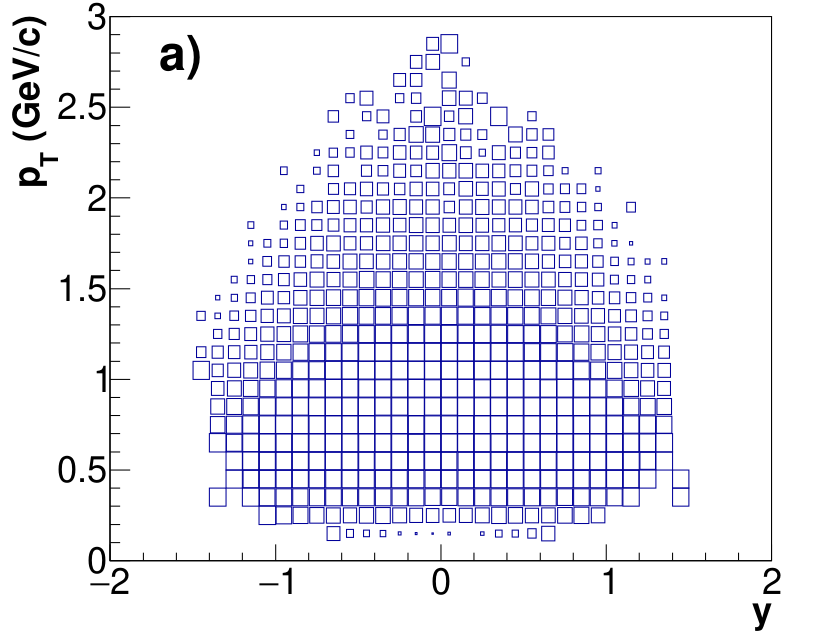}
 \hspace{5mm}
    \includegraphics[width=0.42\textwidth]{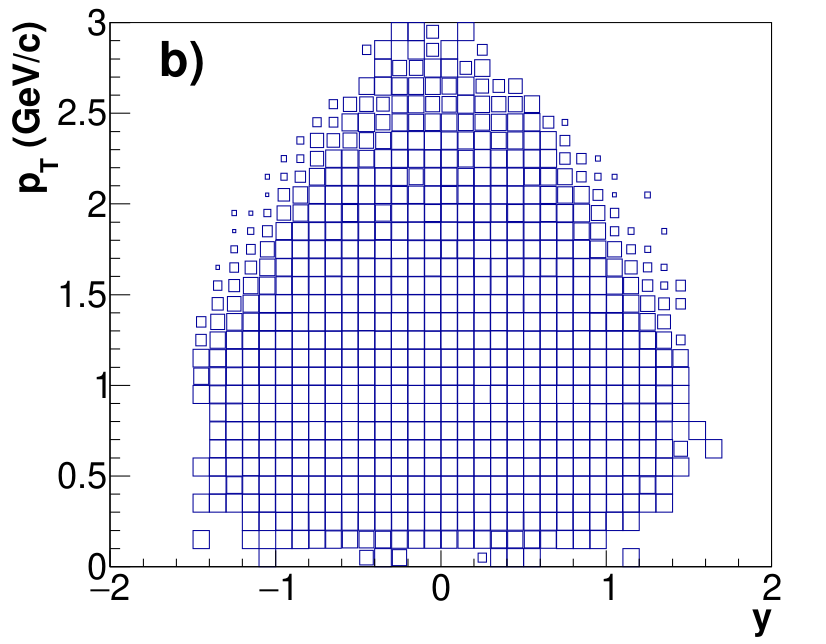}
\caption{The MPD phase space coverage in terms of $y$ and $p_{\rm T}$ for Au+Au collisions
 at $\sqrt{s_{\mathrm NN}}$\,=\,8.8~GeV
 for identified $K^+$ (a) and $\pi^+$ (b). Note that the CM beam rapidity at this energy is $y_{\rm beam}$\,=\,2.2.
 }
 \label{pika_phasespace}
 \end{center}
\end{figure}

The MPD detector has excellent particle identification capabilities with a large and uniform acceptance (see Fig.~\ref{pika_phasespace}). This will allow us to perform systematic studies of particle yields and ratios as a function of collision energy, centrality, and kinematic variables (rapidity, transverse momentum, and azimuthal angle). The results of this multi-parametric scan within the NICA energy range will be compared to model calculations and to worldwide available data.

\begin{figure}[hbt]
\begin{center}
   \includegraphics[width=0.45\textwidth]{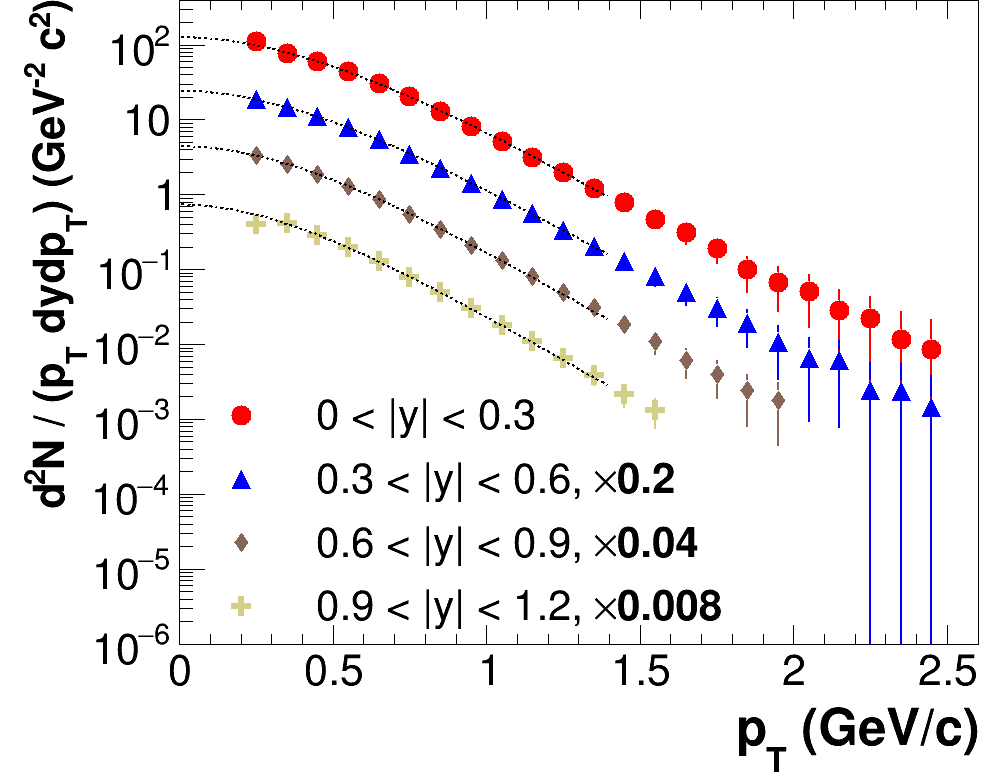} 
\caption{Invariant $p_{\rm T}$ spectra of $K^+$ from central Au+Au collisions at $\sqrt{s_{\rm NN}}$\,=\,8.8 GeV.
The midrapidity spectra
are drawn to scale. Other distributions are scaled down by successive powers of 5 for visibility.
The dashed lines indicate the fit function.}
 \label{ka_spec}
\end{center}
\end{figure}

One of the first planned measurements is the production of charged pions and kaons near the maximum in the strangeness-to-entropy ratio ($\sqrt{s_{\rm NN}}=$8-9 GeV).  
In order to estimate the sensitivity of the MPD setup to this measurement, a detailed Monte Carlo study has been performed with the PHSD event generator,
implementing both deconfinement and chiral symmetry
restoration (\acrshort{csr}) effects~\cite{Ehehalt:1996uq}. The details of the analysis and the results are briefly outlined here and described in detail in  Ref.~\cite{pika_ratio}.
Data sets of 0-5\% central Au+Au events at 5 collision energies
($\sqrt{s_{\mathrm{\rm NN}}}=4,6.2,7.6,8.8,12.3$~GeV) of $5\times10^4$ events each were used. At each energy, the PHSD events were generated with the \acrshort{csr} effects switched on and off, thus, 
the yields of $K^+$ and $\pi^+$ were analyzed in a total of 10 data sets. In addition, negatively charged pions and kaons as well as $\Lambda$
were studied at $\sqrt{s_{\rm NN}}$\,=\,7.6~GeV with the \acrshort{csr} effects switched on. 
As an example, invariant transverse momentum spectra of identified $K^+$
in several rapidity bins at  $\sqrt{s_{\rm NN}}=8.8$~GeV are shown in Fig.~\ref{ka_spec}.
With the MPD setup, in particular using the combination of \acrshort{pid} capabilities of \acrshort{tpc} (see Sec.~\ref{sec:tpc}) and \acrshort{tof}  (see Sec.~\ref{sec:tof}) kaon spectra are measured up
to $p_{\rm T}$\,=\,2.5~GeV/$c$, with part of the low-$p_{\rm T}$ spectra outside of acceptance.
In order to obtain the particle density, $dN/dy$, the reconstructed $p_{\rm T}$ spectra need to be extrapolated into the unmeasured transverse
momentum regions, exploiting information on the spectral shape. For this, the spectra were compared against a hydrodynamically inspired blast-wave (\acrshort{bw}) model~\cite{Schnedermann:1993ws}.
The results of a \acrshort{bw}-analysis of the kaon spectra are shown by dotted curves in Fig.~\ref{ka_spec}.

\begin{figure}[hbt]
\begin{center}
\includegraphics[width=0.45\textwidth]{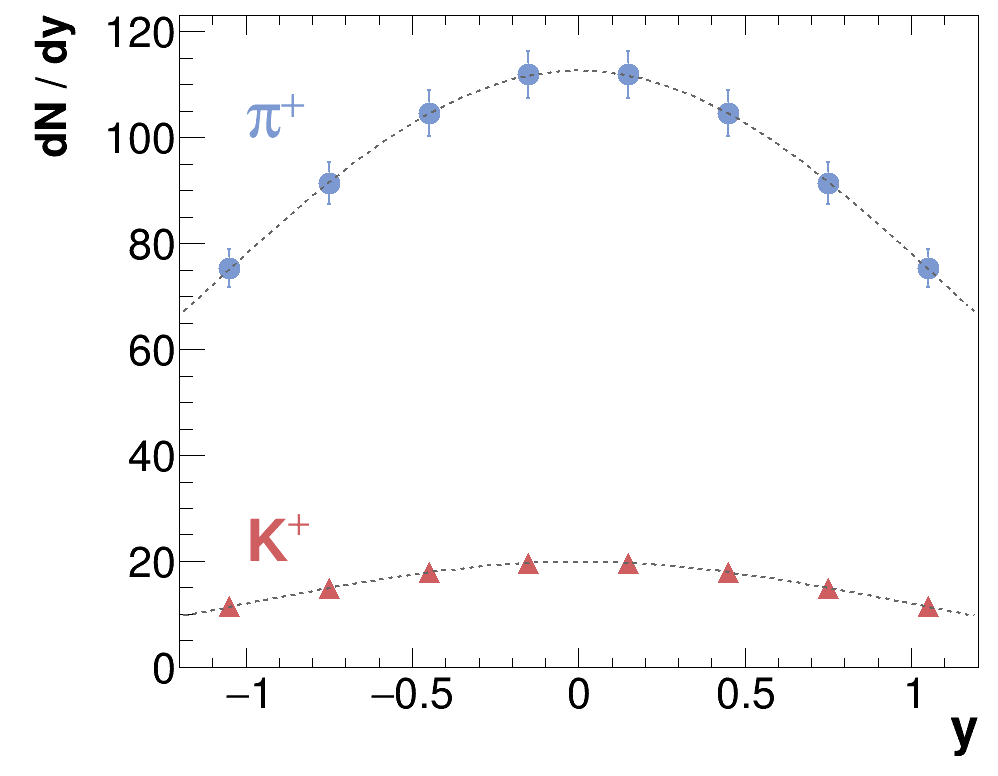}
\caption{Rapidity distributions of $\pi^+$ and $K^+$ from central Au+Au collisions
 at $\sqrt{s_{\rm NN}}$=8.8~GeV.
The dashed lines indicate the Gaussian fits used to extrapolate
to 4$\pi$ yields (see text for details).}
 \label{dndy_pika}
 \end{center}
\end{figure}

The $dN/dy$ results as a function of rapidity for positively charged pions and kaons from Au+Au collisions at $\sqrt{s_{\rm NN}} = 8.8$~GeV are plotted in Fig.~\ref{dndy_pika}. The indicated uncertainties are the quadratic sums
of the statistical and systematic uncertainties.
The results are well reproduced by a Gaussian function (see dashed lines) from which we derive that 70\% to 90\% of the 4$\pi$ yield
for the case of  $K^+$ (60\% to 80\% for the case of $\pi^+$) can be covered by the measurements at different collision energies.
Moreover, in the extrapolation to the full yield, an error in the mean multiplicity of less than 4\% for kaons and 3\% for pions is achieved. These errors
were estimated as the difference between the value from the full integral of the rapidity spectra
and the true multiplicity from the event generator.

\begin{figure}[t]
\begin{center}
  \includegraphics[width=0.42\textwidth]{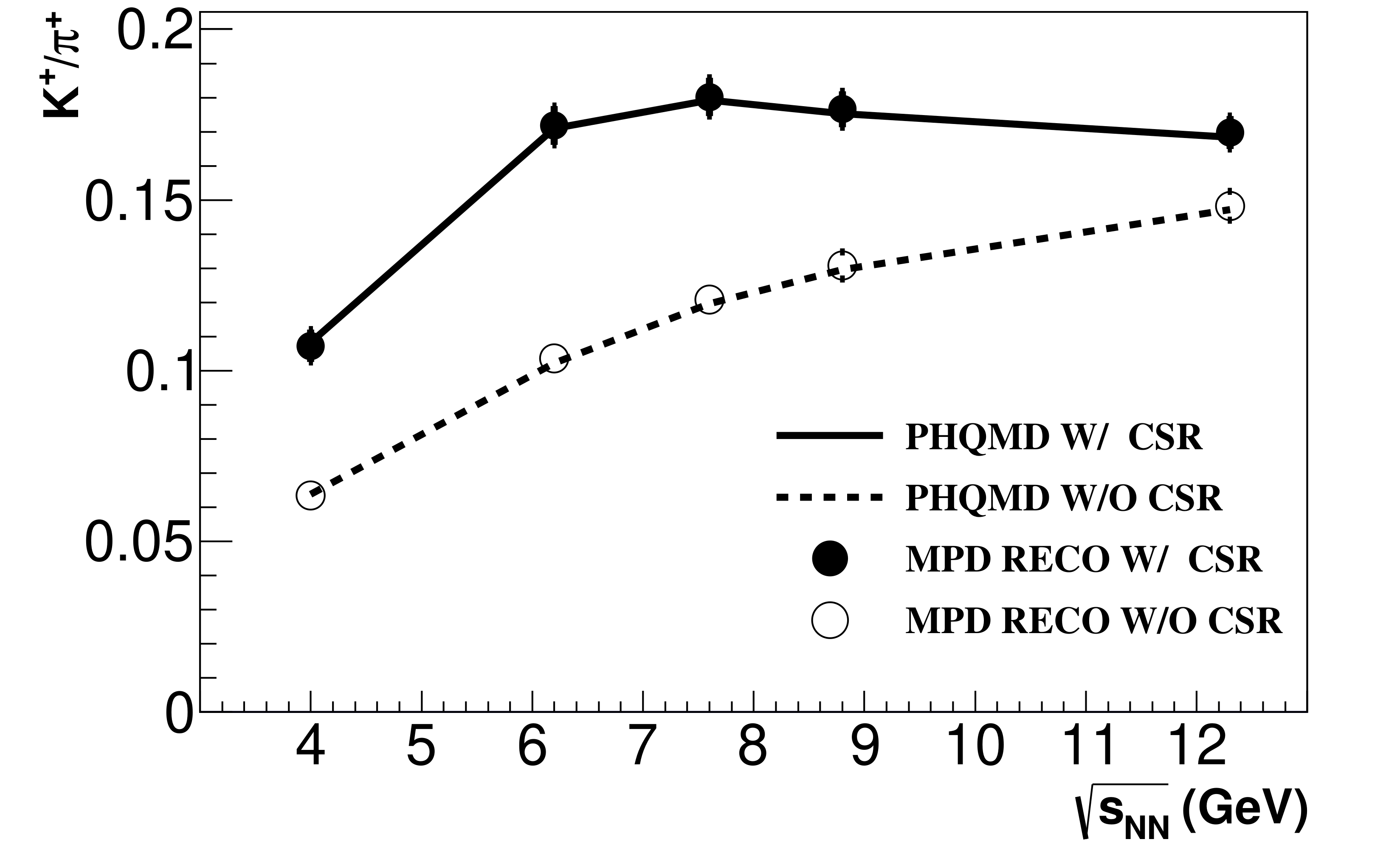}
\caption{The midrapidity $K^+$/$\pi^+$-ratio as a function of collision energy.
 PHSD predictions for chiral symmetry restoration switched on (solid line)
  and off (dashed line) are compared to the numbers, which are obtained in the hadron
  spectra analysis described in this study. The PHSD predictions at different energies are connected with straight lines.}
 \label{kapi_ratio}
 \end{center}
\end{figure}

The excitation function of the mid-rapidity $K^{+}/\pi^{+}$ ratio is shown in Fig.~\ref{kapi_ratio}. Predictions of the PHQMD model for chiral symmetry restoration
switched on and off are shown by the solid and dashed line, respectively.
The mid-rapidity yields are taken from the Gaussian fits shown in Fig.~\ref{dndy_pika}.
The corresponding errors for the midrapidity yields are also taken from the fits.
The results of the performance study indicate that the suggested experimental
setup can allow us to measure the hadron yield ratios with 
small enough uncertainty: the difference between the reconstructed
$K$/$\pi$-ratio and the true one (from Monte Carlo) is below 3\% at all energies.

\subsection{Hyperon reconstruction}
In this section we illustrate the MPD capabilities for the reconstruction of hyperons.

\subsubsection{\label{sec:ALXi-}$\Lambda$, $ \bar{\Lambda}$ and $\Xi^-$ reconstruction}

A PHSD data set of $2\times10^{6}$ Au+Au minimum bias events at 
$\sqrt{s_{\rm NN}}$ = 11~GeV was partitioned into four centrality bins in terms of impact parameter intervals.
\begin{table*}[t!]\centering
\begin{tabular}{|l|c|c|}\hline
       & \multicolumn{2}{c|}{Efficiency, \%}\\ \cline{2-3}
Factor & $\Lambda\rightarrow p+\pi^-$ & $\Xi^{-}\rightarrow\Lambda+\pi^-$ \\ \hline 
Branching ratio                                        & 63.4 &  63.3\\
$p$,$\pi^-$,$|\eta|<1.3$                        & 30.4 &  26.8\\
$p$,$\pi^-$,$|\eta|<1.3$,$p_{\rm T}>0.05$ GeV/$c$ & 28.7 &  24.3\\
$p$,$\pi^-$, $|\eta|<1.3$ $p_{\rm T}>0.1$ GeV/$c$  & 22.4 &  14.6\\
$p$,  $\pi^-$, $|\eta|<1.3$, $p_{\rm T}>0.2$~GeV/$c$  & 8.8  &   2.9\\
Reconstructed $p$,$\pi^-$, $|\eta|<1.3$          & 22.4 &  15.5\\ 
Reco+PID                                             & 21.3 &  14.5\\ 
Reco+PID+selection+significance        & 7.2  &   2.0\\ \hline
\end{tabular}
\caption{Factors affecting hyperon reconstruction efficiency.}
\label{lambtable}
\end{table*}

\begin{figure*}[t]
\begin{center}
\includegraphics[width=143mm]{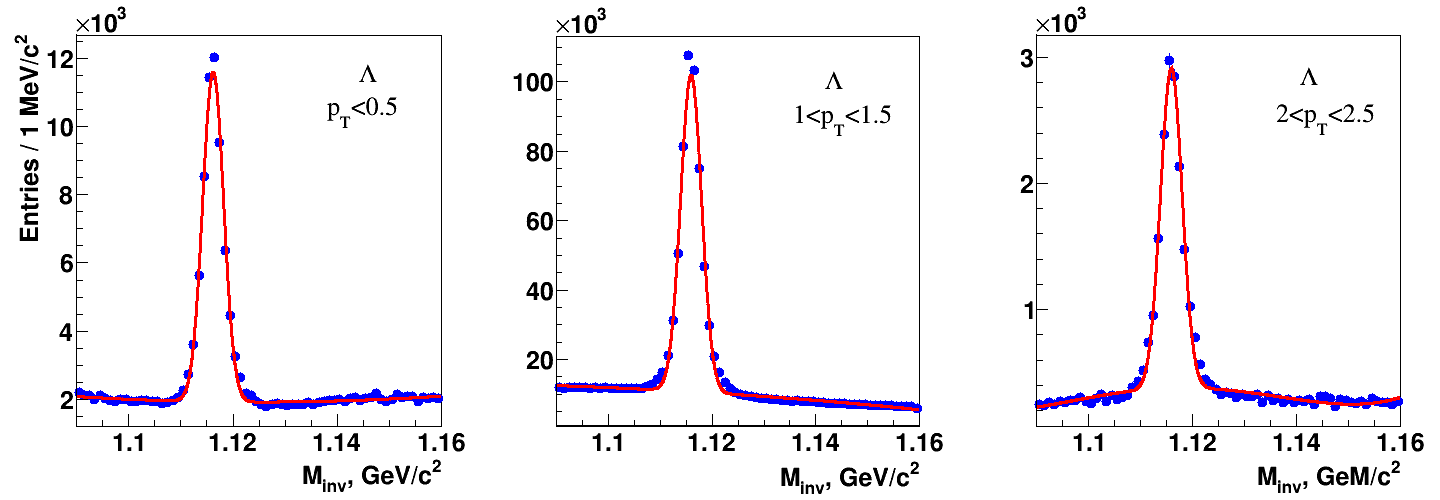}
\end{center}
\caption{\label{lam_raw_pt} 
Invariant mass spectra of charged pions and protons in the vicinity of $\Lambda$ mass, in selected $p_{\rm T}$ ranges. }
\end{figure*}

\begin{figure*}[t]
\begin{center}
\includegraphics[width=143mm]{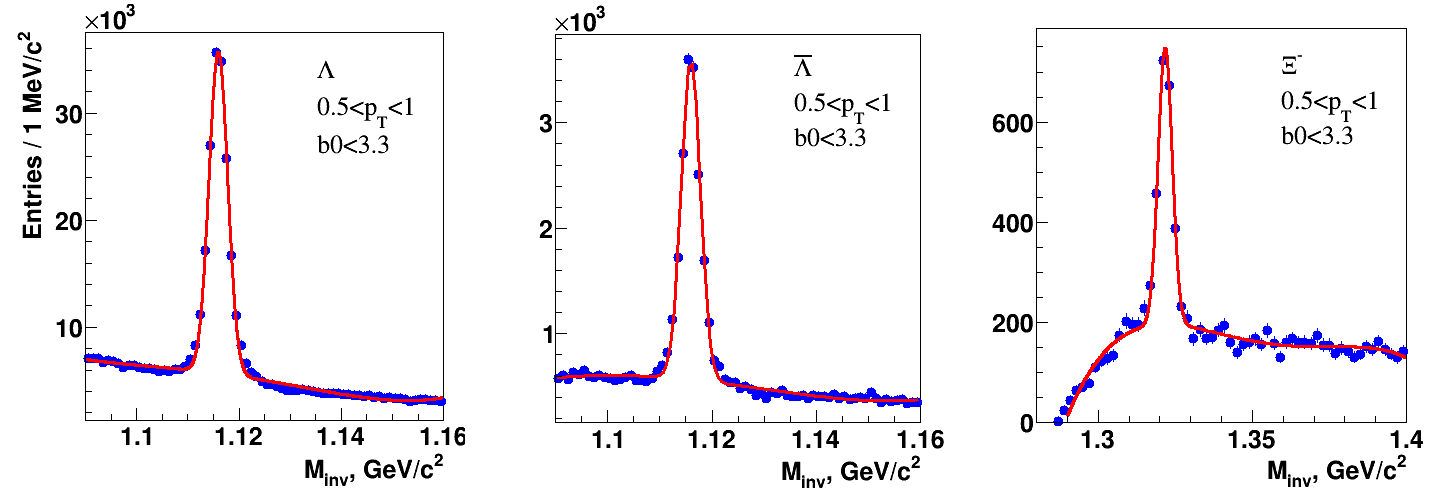}
\includegraphics[width=143mm]{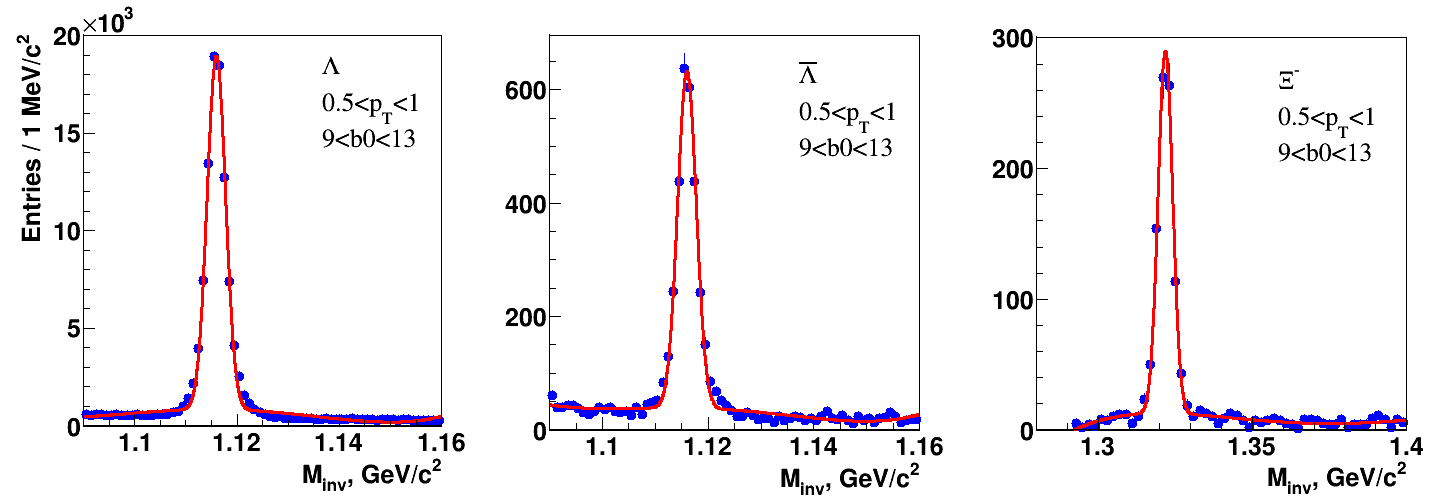}
\end{center}
\caption{\label{lam_raw_b} 
Invariant mass spectra of $\Lambda, \bar{\Lambda}$, and $\Xi^{-}$ candidates in the most central, $b_{0}<2.3$~fm (top row) and
the most peripheral  $9<b_{0}<13$~fm (bottom row) impact parameter bins. }
\end{figure*}

The hyperon reconstruction analysis was performed in transverse momentum intervals of 0.5 GeV/$c$ width. In Fig.~\ref{lam_raw_pt}, invariant mass spectra of charged pion and proton candidates in several $p_{\rm T}$ bins are shown.
In Fig.~\ref{lam_raw_b}, the invariant mass spectra of $\Lambda, \bar{\Lambda}$, and $\Xi^{-}$ candidates in the most central (top) and the most peripheral (bottom) bins are shown. 

\begin{figure*}[t]
\begin{center}
\includegraphics[width=143mm]{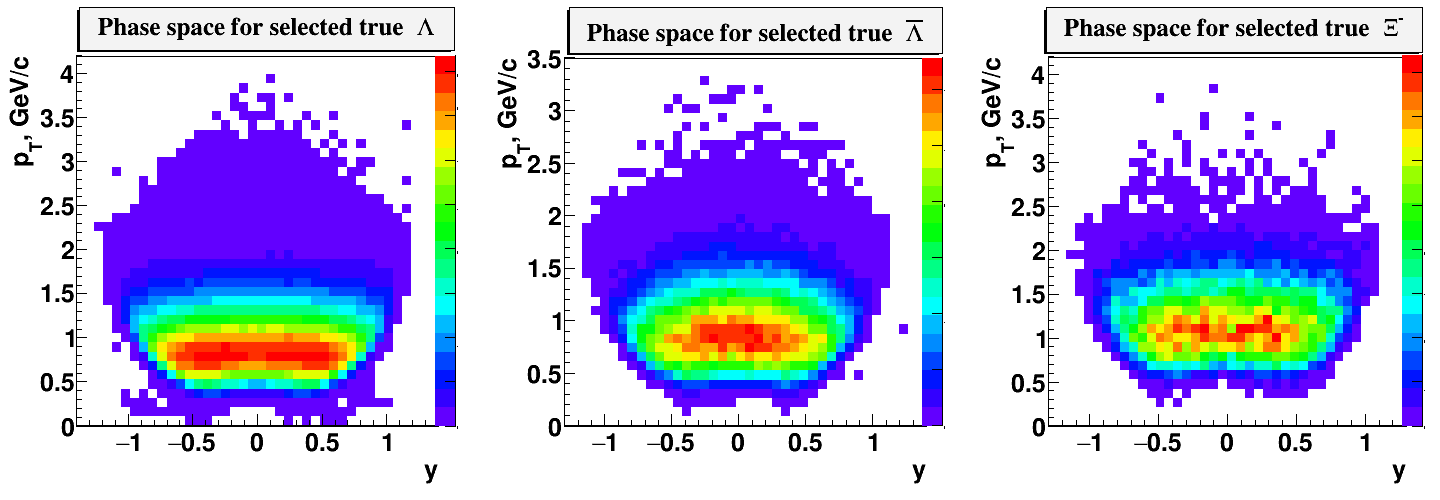}
\end{center}
\caption{\label{hyp_phasespace} 
MPD detector $y-p_{\rm T}$ phase-space for $\Lambda$ (left panel), $\bar{\Lambda}$ (center panel),
and $\Xi^-$ (right panel).}
\end{figure*}

\begin{figure}[b]
\begin{center}
\includegraphics[width=70mm]{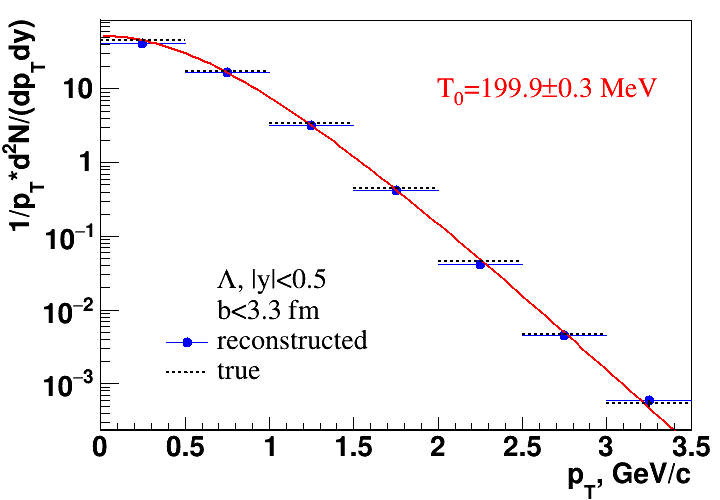}
\includegraphics[width=70mm]{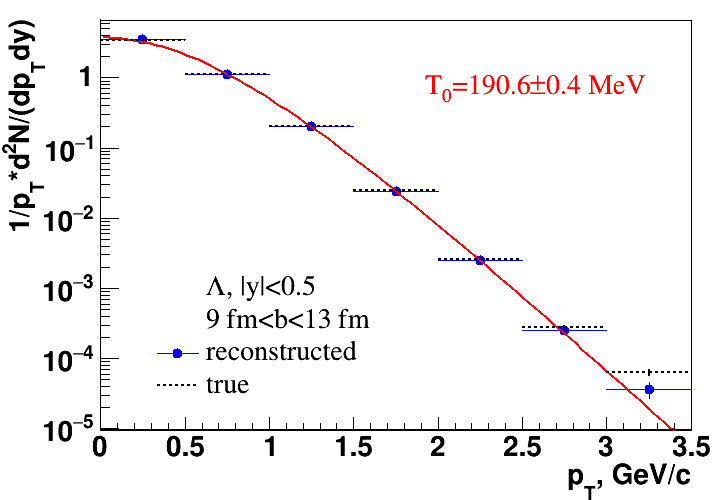}
\end{center}
\caption{\label{lam_pt} 
The transverse momentum spectra of $\Lambda$ within the rapidity range ($|y| < 0.5$)
for 0-5\% and 30-80\% central Au+Au collisions at $\sqrt{s_{\rm NN}}$=11~GeV.
The reconstructed data are indicated by symbols, while dashed-line histograms represent
the spectra from the model. }
\end{figure}

The factors affecting the hyperon reconstruction efficiency are shown in Table~\ref{lambtable}.
Fig.~\ref{hyp_phasespace} shows the MPD phase-space coverage for $\Lambda$, $\bar{\Lambda}$, and $\Xi^-$. The detector setup has
a sufficient coverage to study both the rapidity and the transverse momentum distributions of (anti)hyperons.
As an example, in Fig.~\ref{lam_pt} the invariant transverse momentum spectra are shown for $\Lambda$ in the most central and the most peripheral centrality bin. 
The distributions were obtained after correcting the raw particle yields in $p_{\rm T}$ bins for acceptance and efficiency effects. The figure shows the reconstructed yields plotted with symbols and the generated spectra drawn as histograms. 
The averaged point-by-point difference between the two does not exceed 2\%.

The extrapolation to low and high $p_{\rm T}$ contributes between 10\% and 25\% to the rapidity density
value, depending on the particle species and centrality. 

\subsubsection{\label{sec:Xi+Omp}$\Xi^{+}$ and $\Omega^\mp$ reconstruction}

In order to evaluate the expected statistics and phase space coverage of the experiment start-up
for rare hyperons, we used a PHSD minimum bias event sample of 8 million Au+Au collisions at $\sqrt{s_{\rm NN}}$=11~GeV.

\begin{figure}[hbt]
  \centering
  \begin{minipage}[h]{0.87\linewidth}
    \includegraphics[width=75mm,angle=0]{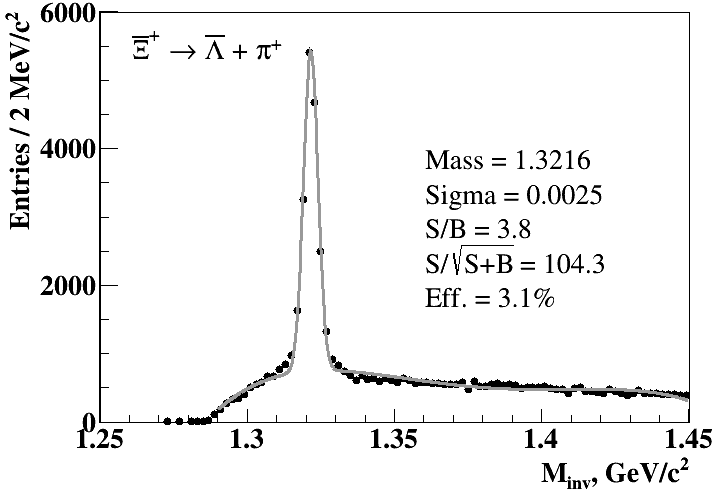} \\ \hspace*{4cm} (a)
  \end{minipage}
  
  \begin{minipage}[h]{0.87\linewidth}
    \includegraphics[width=75mm,angle=0]{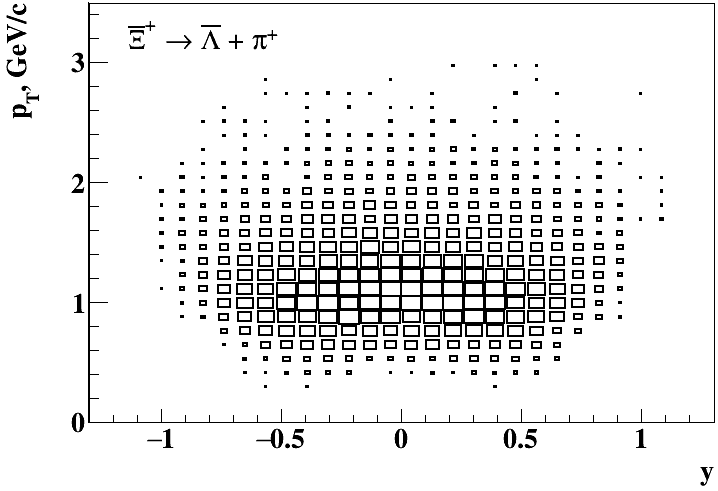} \\ \hspace*{4cm} (b)
  \end{minipage}
  \caption{Reconstructed invariant mass of $\bar{\Xi}^{+}$ (a) and $y-p_{\rm T}$ phase space coverage (b), PHSD model, Au+Au collisions at $\sqrt{s_{\rm NN}}=11$~GeV.}
 \label{aXi}
\end{figure}

\begin{figure}[hbt]
 \centering
 \begin{minipage}[h]{0.87\linewidth}
   \includegraphics[width=75mm,angle=0]{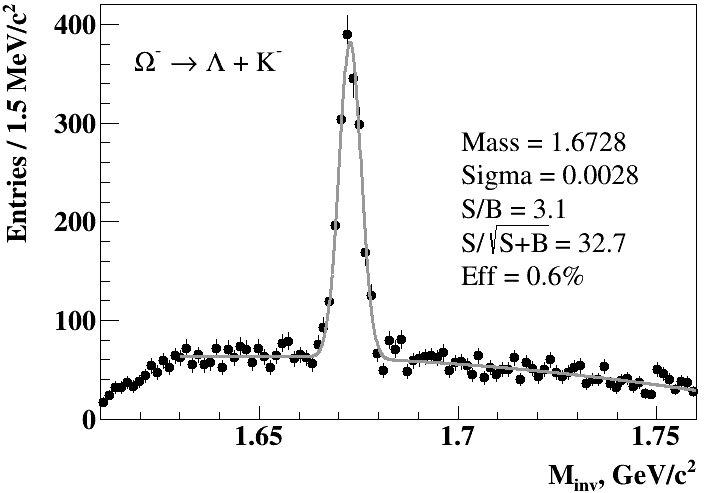} \\ \hspace*{4cm} (a)
 \end{minipage}
 
 \begin{minipage}[h]{0.87\linewidth}
   \includegraphics[width=75mm,angle=0]{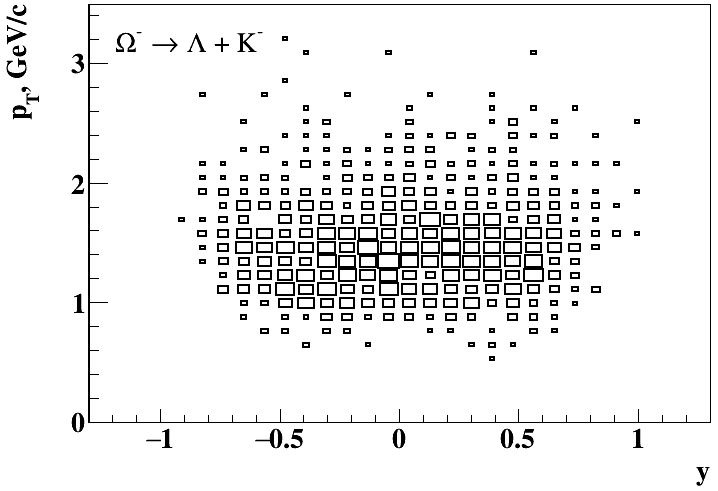} \\ \hspace*{4cm} (b)
 \end{minipage}
 \caption{Reconstructed invariant mass of $\Omega^{-}$ (a) and $y-p_{\rm T}$ phase space coverage (b), PHSD model, Au+Au collisions at $\sqrt{s_{\rm NN}}=11$~GeV.}
 \label{Om}
\end{figure}

The same analysis procedure, as for $\Lambda$ and $\Xi^-$ hyperons, was used to reach the maximum statistical significance of the signal determined as the square root of the signal to background ratio. The invariant mass spectra are shown in Figs.~\ref{aXi} and~\ref{Om},
along with the rapidity and transverse momentum phase space acceptance for correctly
reconstructed hyperons, after final selection. The invariant mass peaks are clearly visible. The detector provides good hyperon coverage at mid-rapidities. Table~\ref{efftable} shows, similar to Table~\ref{lambtable}, features that include quite
large signal losses when requiring maximum significance of the invariant mass peak. Efficient background suppression is a high priority task in this analysis.

Physics analyses will be aimed at multi-differential hyperon studies, so signal statistics 
must be high.
We can estimate the expected hyperon yields for a reasonable running 
scenario using the following considerations. Assuming that the start-up luminosity will be approximately twenty times lower than the design value due to the absence of the electron cooling system, the collision rate will be about 350 Hz. Using this assumption, the number of minimum bias events for 2 weeks will be $\sim$ 450 million. For comparison, the processed MC event sample (8 million events) corresponds to about six hours of running time. Table~\ref{yields} shows the expected reconstructed hyperon yields in 2 weeks of running time. The estimation is based on the assumption of 100\% beam time.

\begin{table*}[th]\centering
\begin{tabular}{|l|c|c|c|}\hline
       & \multicolumn{3}{c|}{Efficiency, \%}\\ \cline{2-4}
Factor &\rule{0pt}{13pt} $\bar{\Xi}^+ \rightarrow\bar{\Lambda}+\pi^+$  & $\Omega^-\rightarrow \Lambda+K^-$ & 
$\bar{\Omega}^+\rightarrow \bar{\Lambda}+K^+$  \\ \hline 
Branching ratio                                        & 62.4 & 41.6 & 42.8\\
$p$ and $\pi (K)$ at $|\eta|<1.3$                        & 27.6 & 14.4 & 18.8\\
$p$ and $\pi (K)$ at $|\eta|<1.3$ and $p_{\rm T}>0.05$ GeV/$c$ & 24.9 & 13.3 & 17.4\\
$p$ and $\pi (K)$ at $|\eta|<1.3$ and $p_{\rm T}>0.1$ GeV/$c$  & 15.3 & 9.3 & 12.6\\
$p$ and $\pi (K)$ at $|\eta|<1.3$ and $p_{\rm T}>0.2$ GeV/$c$  & 2.8 & 2.3 & 3.4\\
Reconstructed $p$ and $\pi (K)$ at $|\eta|<1.3$          & 14.0 & 6.7 & 8.9\\ 
Reco + \acrshort{pid}                                             & 13.2 & 6.3 & 8.3\\ 
Reco + \acrshort{pid} + sel. at maximum significance              & 3.1 & 0.6 & 1.0\\ \hline
\end{tabular}
\caption{Factors affecting multistrange hyperon reconstruction efficiency.}
\label{efftable}
\end{table*}

\begin{table*}[th]\centering
\begin{tabular}{|l|c|c|c|}\hline
\rule{0pt}{13pt} Particle & $\bar{\Xi}^+ \rightarrow\bar{\Lambda}+\pi^+$ & $\Omega^-\rightarrow \Lambda+K^-$ & 
$\bar{\Omega}^+\rightarrow \bar{\Lambda}+K^+$ \\ 
Expected yield & $7.2\times 10^5$ & $7.4\times 10^4$ & $2.3\times 10^4$ \\ \hline
\end{tabular}
\caption{Expected multistrange hyperon yields in minimum bias Au+Au collisions for 2 weeks of running time at initial NICA luminosity.  The estimation is based on the assumption of 100\% beam time.}
\label{yields}
\end{table*}


\subsection{\label{sec:resonances}Reconstruction of resonances}

As pointed out in Sec.~2.6, short-lived resonances are sensitive to the properties of the medium produced in heavy-ion collisions. The UrQMD event generator was used to simulate Au+Au collisions at different energies, $\sqrt{s_{\rm NN}} = 4, 7.7$ and 11 GeV. All simulated particles have been tracked through the detector materials using GEANT4.  
Realistic simulation of each detector subsystem response, track reconstruction and signal extraction were performed within the MpdRoot framework.
The following basic track selection cuts were used: tracks with a rapidity cut of $|\eta| < 1.0$ and at least 24 hits in the \acrshort{tpc} out of a maximum of 54 hits, primary particles matching to the primary vertex within $3\sigma$, transverse momentum $p_{\rm T} > 50$~MeV/$c$, final state charged particles identification based on $\langle{\rm d}E/{\rm d}x\rangle$ measurements in the \acrshort{tpc} and the time-of-flight measurements in the \acrshort{tof}. For the reconstruction of weakly decaying daughter particles such as $K_{S}^{0}$ and $\Lambda$, a set of topology cuts for the reconstruction of secondary vertices was used. In total, $10^{7}$ events were simulated at $\sqrt{s_{\rm NN}} = 11$~GeV and $5 \times 10^{6}$ events at $\sqrt{s_{\rm NN}} = 4$ and 7.7 GeV.

\begin{figure*}
\includegraphics[width=0.66\columnwidth]{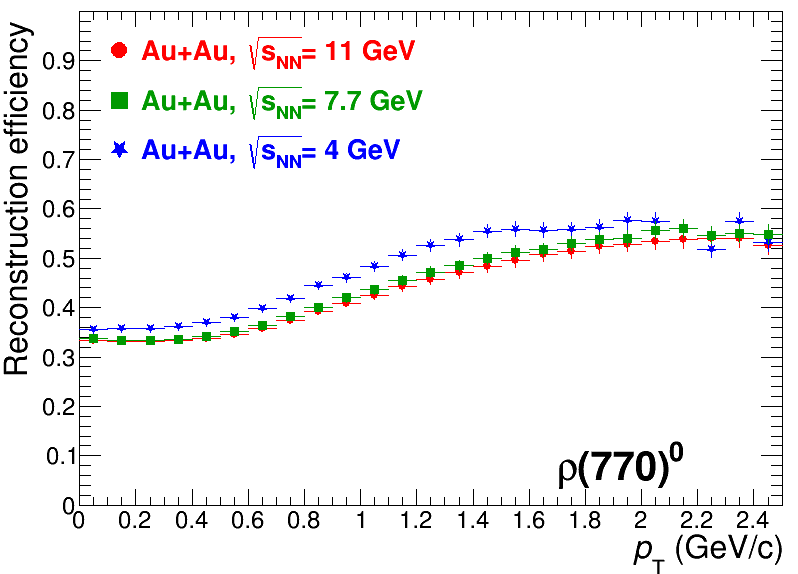}
\includegraphics[width=0.66\columnwidth]{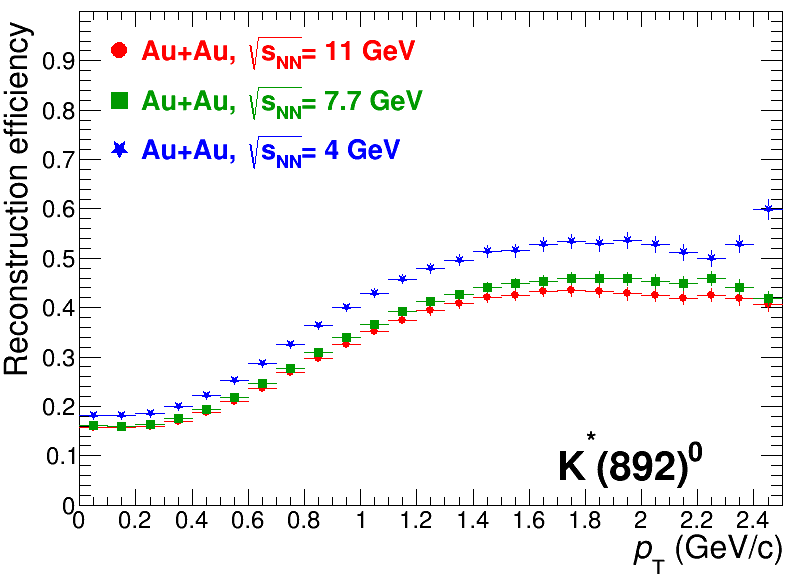}
\includegraphics[width=0.66\columnwidth]{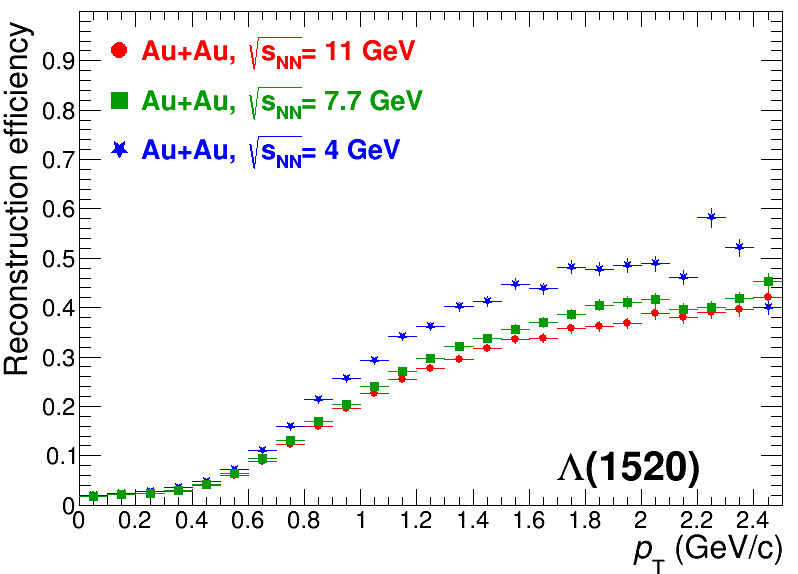}
\caption{\label{ResRecEff} $p_{\rm T}$-dependent reconstruction efficiencies evaluated for  $\rho(770)^{0}$, $K^{*}(892)^{0}$ and $\Lambda(1520)$ in Au+Au collisions at $\sqrt{s_{\rm NN}} = 4, 7.7$ and 11 GeV.}
\end{figure*}

Figure~\ref{ResRecEff} shows examples of $p_{\rm T}$-dependent reconstruction efficiencies evaluated for $\rho(770)^{0}$, $K^{*}(892)^{0}$ and $\Lambda(1520)$ in Au+Au collisions at $\sqrt{s_{\rm NN}} = 4, 7.7$ and 11 GeV.  For all particles, reasonable efficiencies are observed at mid-rapidity. A modest multiplicity dependence of the reconstruction efficiencies is observed, which is responsible for the variation of efficiencies with the collision energy.

The same simulated data samples were used to estimate the MPD mass resolution, which was found to be gradually increasing with transverse momentum: from 16 to 26 MeV/$c^{2}$ for $\rho(770)^{0}$, from 11 to 17 MeV/$c^{2}$ for $K^{*}(892)^{0}$, from 5 to 7.5 MeV/$c^{2}$ for  $K^{*}(892)^{\pm}$, from 2 to 3 MeV/$c^{2}$ for $\phi(1020)$, from 4 to 5 MeV/$c^{2}$ for $\Sigma(1385)^{\pm}$, from 7 to 10 MeV/$c^{2}$ for $\Lambda(1520)$ and from 2 to 4 MeV/$c^{2}$ for $\Xi(1530)^{0}$ with very modest dependence on the multiplicity. The excellent mass resolution offers the possibility to perform line shape studies even for narrow resonances like $\phi(1020)$ meson. A mass resolution of $\sim$ 2 MeV/$c^{2}$ does not completely smear the natural (Breit-Wigner) shape of the reconstructed peaks. 

\begin{figure*}
\includegraphics[width=0.99\columnwidth]{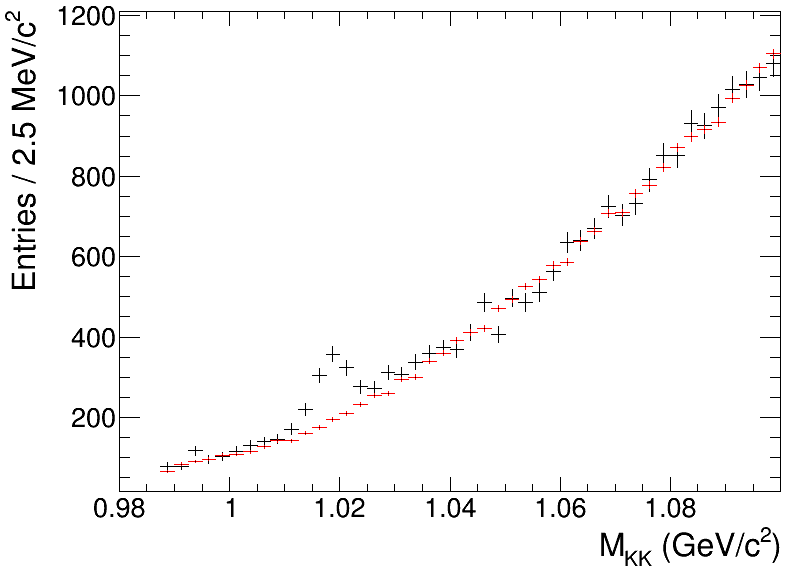}
\includegraphics[width=0.99\columnwidth]{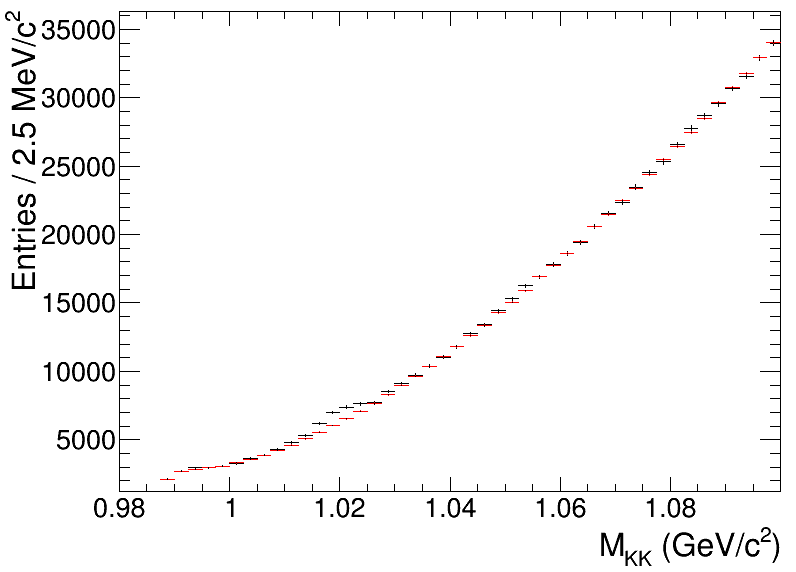}\\
\includegraphics[width=0.99\columnwidth]{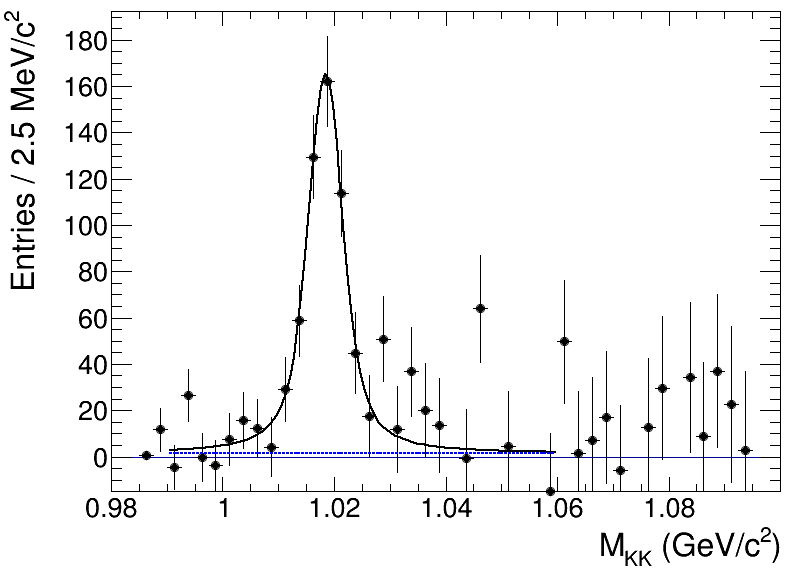}
\includegraphics[width=0.99\columnwidth]{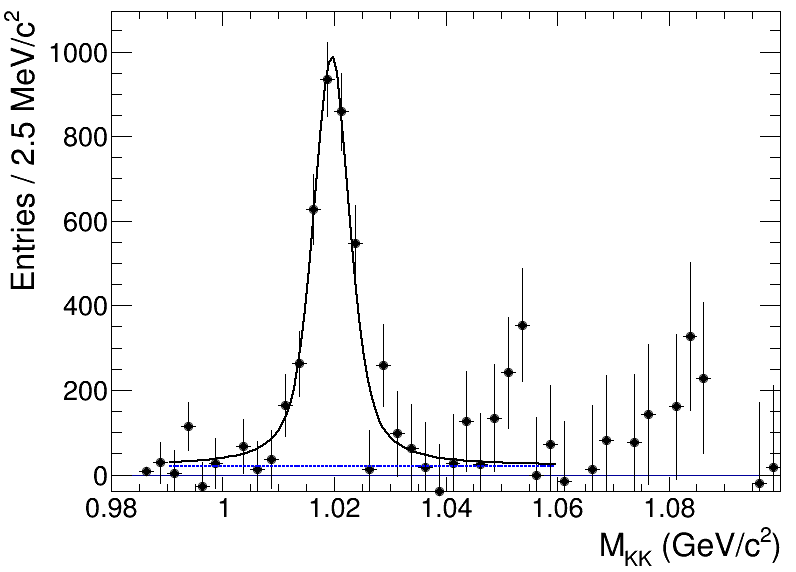}
\caption{\label{ResRaw} Examples of the invariant mass distributions for $K^{+}K^{-}$ pairs in Au+Au collisions at $\sqrt{s_{\rm NN}} = 4$ (left) and 11 GeV (right). Examples are shown before (top) and after (bottom) the mixed-event background subtraction. The distributions after background subtraction are fitted to a combination of a Voigtian function for the signal and a polynomial for the remaining background.}
\end{figure*}

Figure~\ref{ResRaw} shows invariant mass distributions reconstructed for $K^{+}K^{-}$ pairs in Au+Au collisions at $\sqrt{s_{\rm NN}} = 4$ and 11 GeV. Examples are shown for a low-$p_{\rm T}$ region, 0.2 -- 0.4~GeV/c. The upper panels show the foreground and the mixed-event background normalized to the foreground at high masses. The lower panels show what is left after subtraction of the mixed-event background.  At both collision energies, clear peaks from $\phi(1020) \rightarrow K^{+}K^{-}$ decays are observed on top of a relatively small residual correlated background. The signal-to-background ratio gets worse with increasing centrality and collision energy due to larger final state multiplicities and combinatorial background. To extract the resonance raw yields, the distributions are fit to a combination of a Voigtian function for the signal and a polynomial function of second degree for the remaining background. Examples of the fits are also shown in the plots.  

Similar invariant mass analyses were also performed for the following decays: 
$\rho(770)^{0} \rightarrow \pi^{+}\pi^{-}$, 
$K^{*}(892)^{0} \rightarrow \pi^{\pm}K^{\mp} $, 
$K^{*}(892)^{\pm} \rightarrow \pi^{\pm}K_{s}^{0} $, 
$\Sigma(1385)^{\pm} \rightarrow \pi^{\pm}\Lambda$, 
$\Lambda(1520) \rightarrow pK^{-}$ and 
$\Xi(1530)^{0} \rightarrow \pi^{+}\Xi^{-}$.
In all cases, clear peaks from the decays of the corresponding resonances were observed after subtraction of the mixed-event background. The particle raw yields were extracted by fitting the spectra to a combination of a Voigtian function and a polynomial function.

\begin{figure*}
\includegraphics[width=0.66\columnwidth]{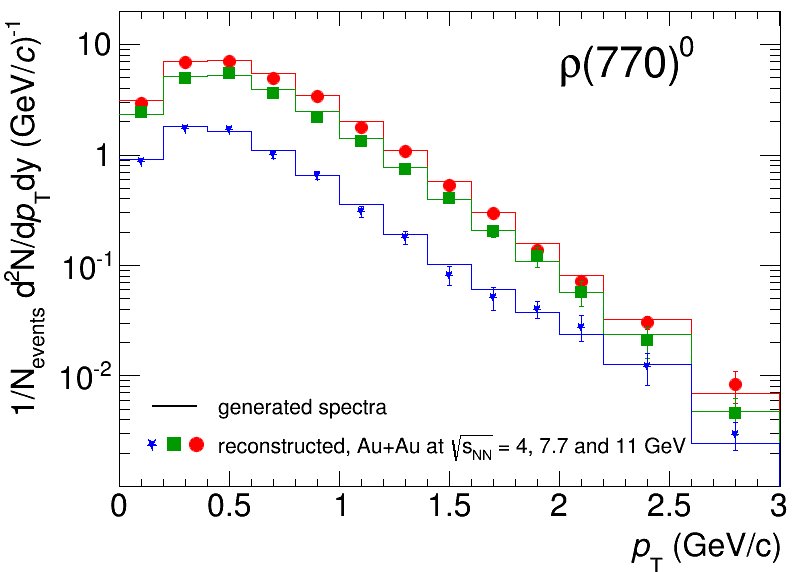}
\includegraphics[width=0.66\columnwidth]{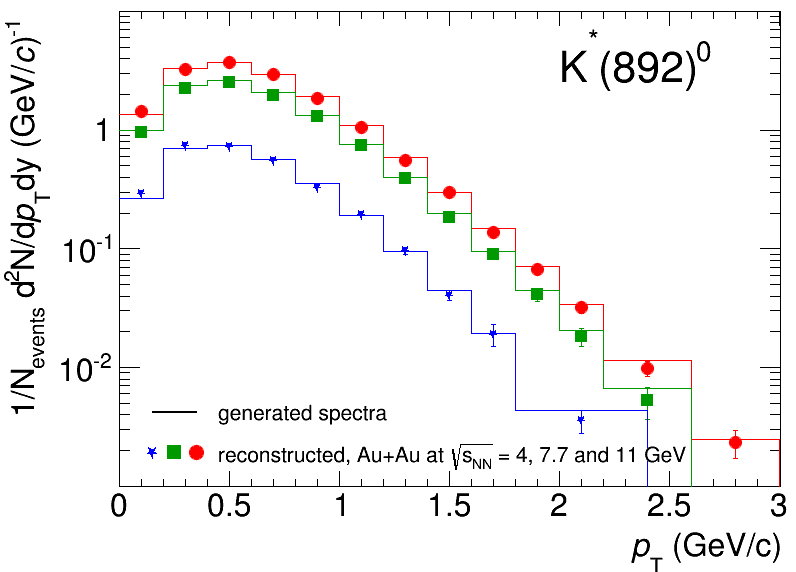}
\includegraphics[width=0.66\columnwidth]{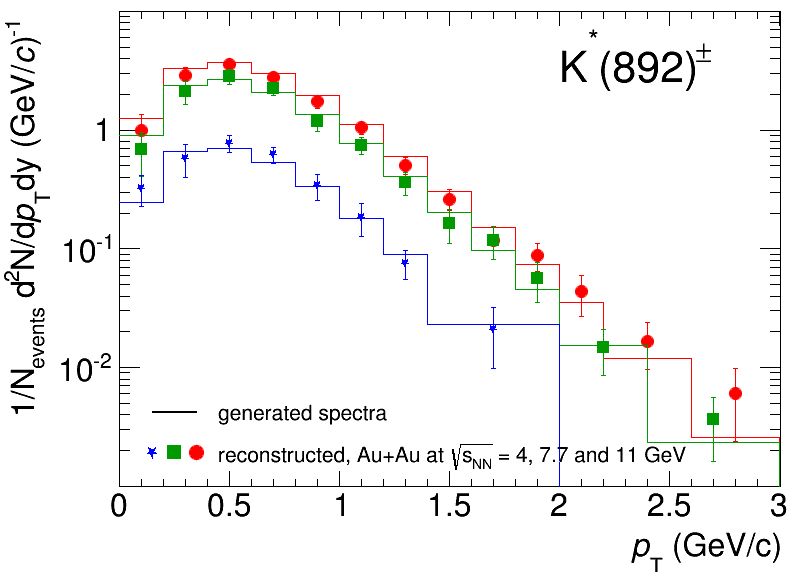}
\includegraphics[width=0.66\columnwidth]{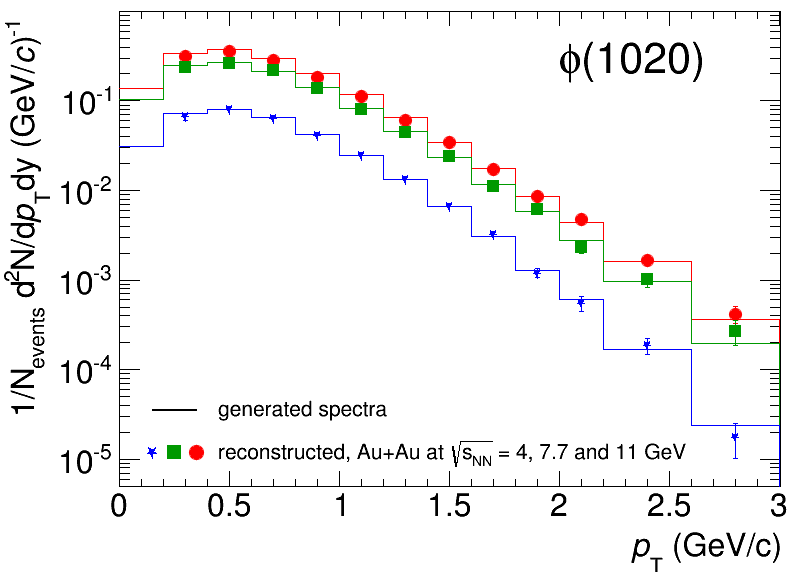}
\includegraphics[width=0.66\columnwidth]{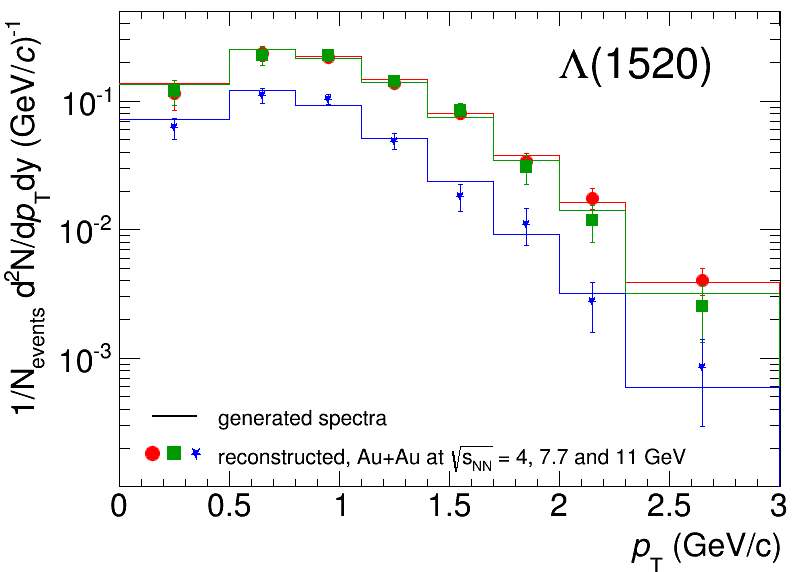}
\includegraphics[width=0.66\columnwidth]{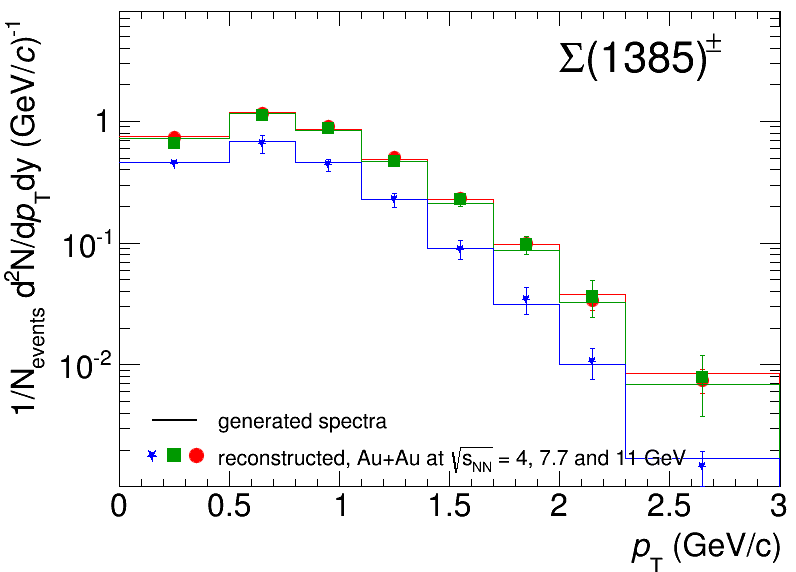}
\includegraphics[width=0.66\columnwidth]{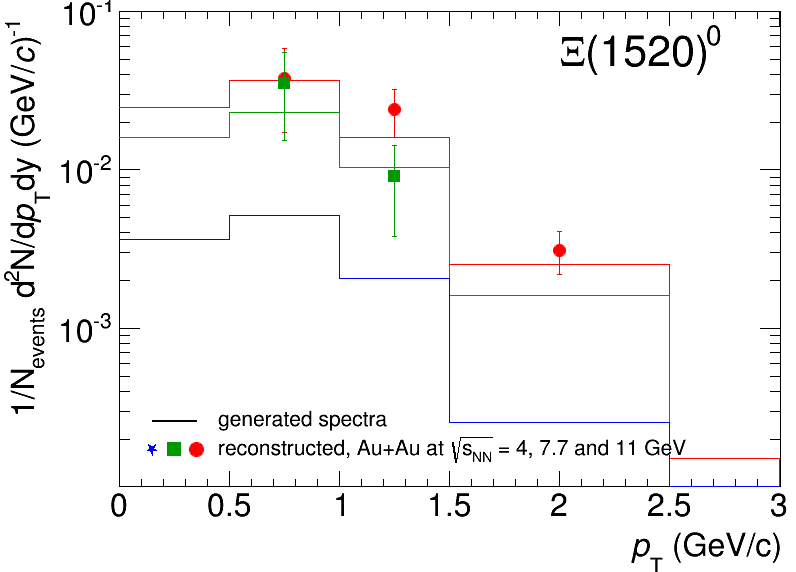}
\caption{\label{ResCT} Transverse momentum spectra reconstructed in Au+Au collisions at $\sqrt{s_{\rm NN}} = 4, 7.7$ and 11 GeV are shown with markers of different colors and compared to true generated spectra shown with solid lines of the same color for $\rho(770)$, $K^{*}(892)^{0}$, $K^{*}(892)^{\pm}$, $\phi(1020)$, $\Sigma(1385)^{\pm}$, $\Lambda(1520)$ and $\Xi(1530)^{0}$ resonances.}
\end{figure*}

Having extracted the evaluated reconstruction efficiencies and the particle raw yields from the invariant mass distributions, the fully corrected transverse momentum spectra of the resonances were obtained. They are shown with markers in Fig.~\ref{ResCT}. Markers of different color correspond to Au+Au collisions at  $\sqrt{s_{\rm NN}} = 4,7.7$ and 11 GeV. The reconstructed spectra are compared to the generated spectra shown with solid lines in the same figures. The two match within statistical uncertainties. 
This validates the reconstruction chain for its use in real data analyses.
The reconstruction of resonances is possible from very low momenta, for most of the cases from zero momentum, which is beneficial for the extraction of the integrated yields with small uncertainties. The sampled $p_{\rm T}$ ranges on the higher end depend on the accumulated statistics. First measurements of resonances will be possible with $~10^{7}$ events sampled in Au+Au collisions at $\sqrt{s_{\rm NN}} = 4-11$~GeV. The multiplicity dependent study would need ten times more collected events. A detailed study of $\Xi(1530)^{0}$ baryon production may need accumulation of up to one billion events.

\subsection{\label{sec:emprobes}Electromagnetic probes}
The measurement of electromagnetic probes such as dielectrons and direct photons has always been a challenging task in heavy-ion collisions due to large combinatorial background. The main sources of background for dielectron measurements are Dalitz decays of $\pi^{0}$ and $\eta$ mesons as well as conversion pairs produced in the beam pipe and detector material. Predictions of UrQMD, PHSD and AMPT event generators for $\pi^{0}$ and $\eta$  yields in heavy-ion collisions at top NICA energies are consistent within $\sim 20\%$, which is acceptable for the purpose of feasibility studies. However, the predicted yields of $\rho(770)^{0}$, $\omega(782)$ and $\phi(1020)$ resonances show a significant model dependence. In this study, we used a data sample of $10^{7}$ events of Au+Au collisions at $\sqrt{s_{\rm NN}} = 11$~GeV simulated with PHSD. 

The particle transport, track reconstruction and pattern recognition were performed within the MpdRoot framework using GEANT4. The following basic track selection cuts were used: tracks with a rapidity cut of $|\eta| < 1.0$ and at least 39 hits in the \acrshort{tpc}, primary particles matching to the primary vertex within $2\sigma$, transverse momentum $p_{\rm T} > 100$~MeV/c. 
To extract dielectron signals, a reliable identification of electrons (eID) is required. Electrons are identified by $\langle{\rm d}E/{\rm d}x\rangle$ measurements in the \acrshort{tpc}, time-of-flight measurements in the \acrshort{tof}, time-of-flight and $E/p$ ratio measurements in the \acrshort{ecal}, where $E$ is the energy measured for a charged track in the \acrshort{ecal} and $p$ is the particle momentum measured in the tracking system. Only electrons with $p_{\rm T} > 150 (200)$~MeV/c reach the \acrshort{tof}  (\acrshort{ecal}) thus limiting the dynamic range of the detectors. 

\begin{figure*}
\includegraphics[width=0.99\columnwidth]{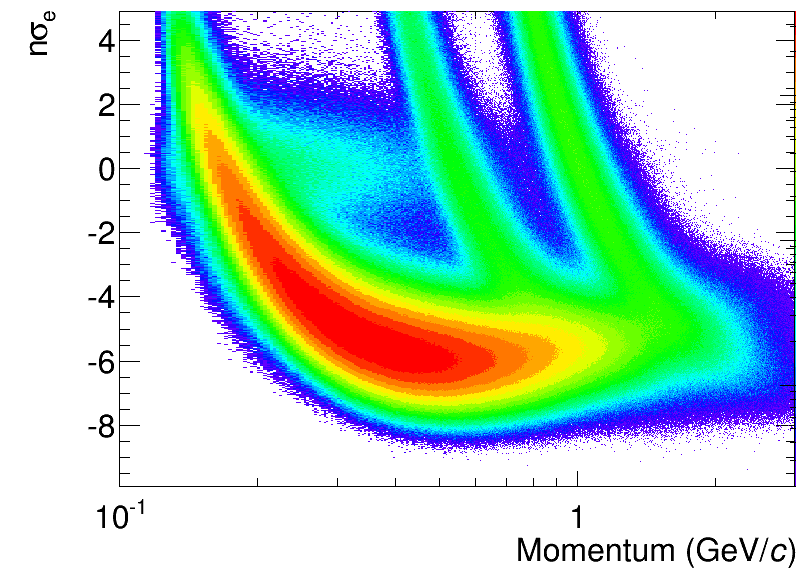}
\includegraphics[width=0.99\columnwidth]{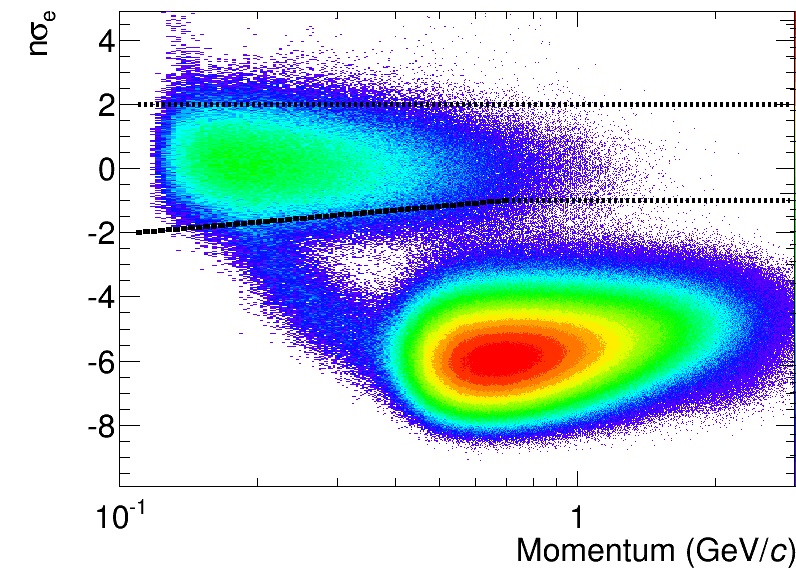}
\caption{\label{ElSel} The $n\sigma$ deviations of  $dE/dx$ signals measured in the \acrshort{tpc} for charged tracks matched to the \acrshort{tof}  from the signal expected for an electron with a given momentum.  The plot on the right was obtained with an additional requirement of eID in the \acrshort{tof}  by $2\sigma$ selection. Dashed lines in the plot show the electron selection cuts in the \acrshort{tpc}.}
\end{figure*}

Figure~\ref{ElSel} shows the $n \sigma$ deviations of $\langle{\rm d}E/{\rm d}x\rangle$ signals measured in the \acrshort{tpc} for charged particle tracks matched to the \acrshort{tof}  detector from the signal expected for an electron with a given momentum. The plot on the right shows the same distribution with the additional requirement of eID in the \acrshort{tof}, $|\frac{1}{\beta} - 1|< 2\sigma(p_{\rm T})$. With eID in the \acrshort{tof}, the electron band at $n\sigma_{e} \sim 0$ becomes much more prominent, although some background remains, mostly from wrong association of tracks and \acrshort{tof}  signals in high multiplicity events. The dashed lines in the plot on the right show the used electron selection cuts in the \acrshort{tpc}.

\begin{figure*}
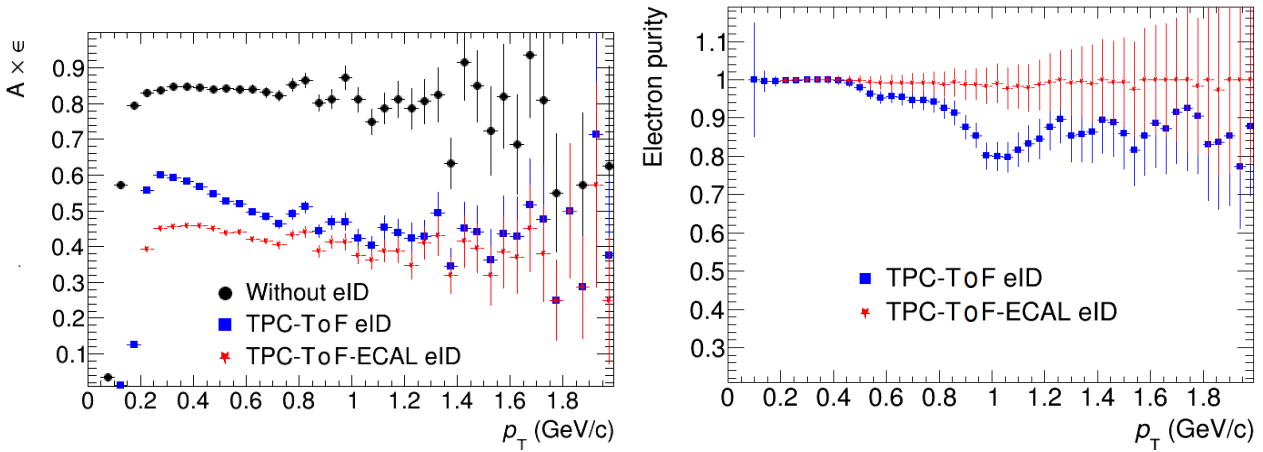

\includegraphics[width=0.99\columnwidth]{ElEff.png}
\includegraphics[width=0.99\columnwidth]{ElPur.png}
\caption{\label{ElEffPur} Reconstruction efficiency for electron tracks (left) and electron purity (right) as a function of particle transverse momentum with different eID options.}
\end{figure*}

Figure~\ref{ElEffPur} demonstrates the expected MPD performance for electron selection. The plots show the efficiency of the electron track reconstruction and the electron purity versus particle transverse momentum. The used eID selections reduce the electron reconstruction efficiency by about a factor of two at $p_{\rm T} > 300$~MeV/c (and even larger at lower momenta) but significantly improve the electron purity from a value of $\sim 5\times 10^{-3}$ obtained without particle identification to almost unity. The electron reconstruction efficiency and purity are similar to those reported by STAR~\cite{Seck:2021mti} with similar selections. The eID with the \acrshort{tpc} and \acrshort{tof}  alone is sufficient for most of the physical analyses. However, additional eID in the \acrshort{ecal} helps to clean up the electron sample at high momenta and large dielectron masses.

\begin{figure*}
\includegraphics[width=0.99\columnwidth]{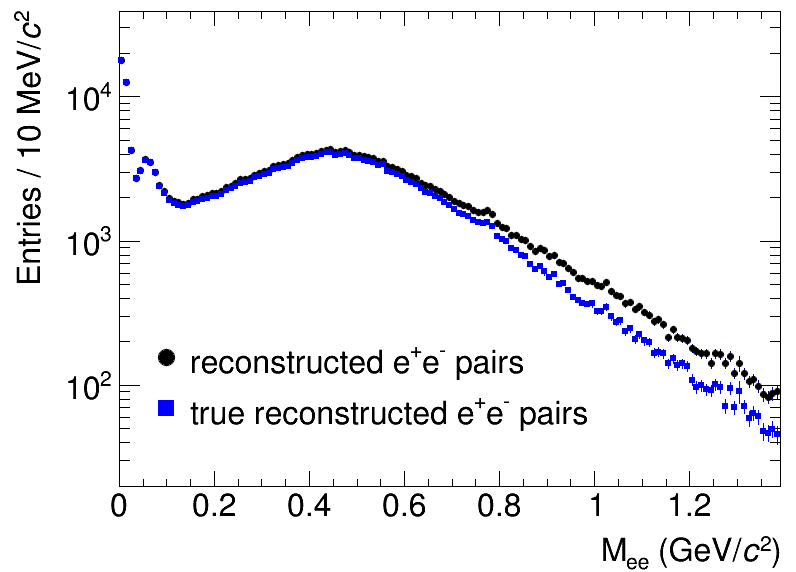}
\includegraphics[width=0.99\columnwidth]{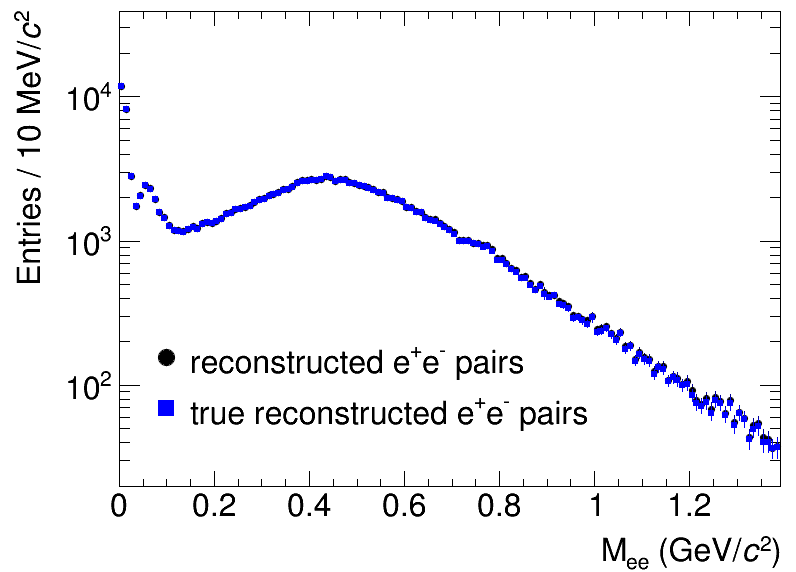}
\caption{\label{ElDiel} Dielectron invariant mass distributions measured as a function of mass for a $p_{\rm T}$-integrated bin in Au+Au collisions at $\sqrt{s_{\rm NN}} = 11$~GeV. The plot on the left was obtained with eID in the \acrshort{tpc} and \acrshort{tof}  while the plot on the right was obtained with an additional eID in the \acrshort{ecal}.}
\end{figure*}

Figure~\ref{ElDiel}  shows examples of the dielectron invariant mass distributions obtained with the eID in the \acrshort{tpc} and \acrshort{tof}  on the left, and with eID in the \acrshort{tpc}, \acrshort{tof}  and \acrshort{ecal} on the right. The distributions are shown for Au+Au collisions in a $p_{\rm T}$-integrated bin.   In the plot on the left, we observe a clear difference between the reconstructed and true $e^{+}e^{-}$ distributions, and this difference is due to hadron contamination. With the additional eID in the \acrshort{ecal}, the hadron contamination is significantly reduced. The role of the \acrshort{ecal} becomes more important with increase of the pair mass and transverse momentum. 

To summarize, electron identification in the \acrshort{tpc} and \acrshort{tof}  is sufficient for low-mass ($m_{ee} < 0.8$ GeV/c$^{2}$) $e^{+}e^{-}$ measurements. Additional eID in the \acrshort{ecal} is important at higher masses and transverse momenta. For a meaningful measurement of low-mass $e^{+}e^{-}$ and light vector mesons a sample of $\sim 10^{8}$ events is needed for Au+Au collisions at $\sqrt{s_{\rm NN}} = 11$~GeV.


\subsection{\label{sec:flowv1}Anisotropic Flow}

Directed ($v_1$) and elliptic  ($v_2$)
flows are  the dominant flow signals and have been studied very extensively both
at top  RHIC  and LHC energies \cite{Snellings:2014kwa,Lacey:2005qq,Bernhard:2019bmu}. The collision
energy dependence of flow coefficients is of particular importance. The $v_1$ and $v_2$
results from the STAR Beam Energy Scan program at RHIC \cite{Adamczyk:2017nxg,Singha:2016mna,Adamczyk:2012ku,Adamczyk:2015fum} and
NA49/NA61 program at SPS \cite{Alt:2003ab,Kashirin:2020evw}  have indicated a strong
non-monotonic behaviour at NICA energies~\cite{Taranenko:2019uyv}.
In this section, we discuss the
anticipated performance of the MPD detector for differential anisotropic flow measurements of identified hadrons
at NICA energies~\cite{Parfenov:2019pxf,Idrisov:2020yca,Parfenov:2020msw}.
We used the cascade version of the
\acrshort{urqmd} model (version 3.4) ~\cite{Bleicher:1999xi,Bass:1998ca}
to simulate the heavy-ion collisions at $\sqrt{s_{\rm NN}}$=4.5, 7.7 and 11 GeV.
Samples of $10^{8}$ minimum bias Au+Au  and Bi+Bi events  were
produced to  analyse  the directed and elliptic flow signals of identified hadrons. We denote as \lq\lq true'' the $v_1$ and $v_2$  results, obtained from the direct  analysis of the generated events.
In the next step, the sample of \acrshort{urqmd} generated events was used as input for the full chain of  realistic simulations of the MPD
detector subsystems based on the GEANT4 platform and reconstruction algorithms built in the MpdRoot.
The $v_n$ results obtained from the flow analysis of these fully reconstructed events are termed  as the \lq\lq reco'' data.
The track reconstruction in MPD is based on the Kalman filter technique \cite{kalman,kalman-vertex,reco} and the requirement of a minimum of 16 \acrshort{tpc} hits ensures a low momentum error. The primary tracks were selected by a $2\sigma$ cut on the 3D distance of closest approach  (\acrshort{dca}) to the primary vertex.
The analysis was restricted to tracks from the kinematic regions
of \acrshort{tpc} with high tracking efficiency: $0.2 < p_{\rm T} < 2.5$ GeV/$c$ and $|\eta| < 1.5$.  

The centrality classes are defined based on the uncorrected charged particle multiplicity ($N_{\rm ch}$)
distribution in the \acrshort{tpc} pseudorapidity range $|\eta| < 0.5$ and full azimuth, for details, see Subsection~\ref{sec:centrality}.
Identification of charged hadrons in the MPD experiment is
based on a combination of momentum information, the specific
energy loss ($\langle {\rm d}E/{\rm d}x\rangle$) in the \acrshort{tpc} and time-of-flight measurements from the \acrshort{tof} detector.  
Short-lived weakly-decaying particles, such as $K_{S}^{0}$ and $\Lambda$, are reconstructed
using the invariant mass technique. The combinatorial
background from uncorrelated particles is reduced by selection criteria based on the topology of the
specific decay. The topological information about the  primary and secondary  decay vertex
positions are obtained by the Kalman filtering algorithm based on the
MpdParticle tool~\cite{kalman,kalman-vertex,reco}. 

\begin{figure}[hbt]
 \centering
 \begin{minipage}[h]{0.87\linewidth}
   \includegraphics[width=70mm,angle=0] {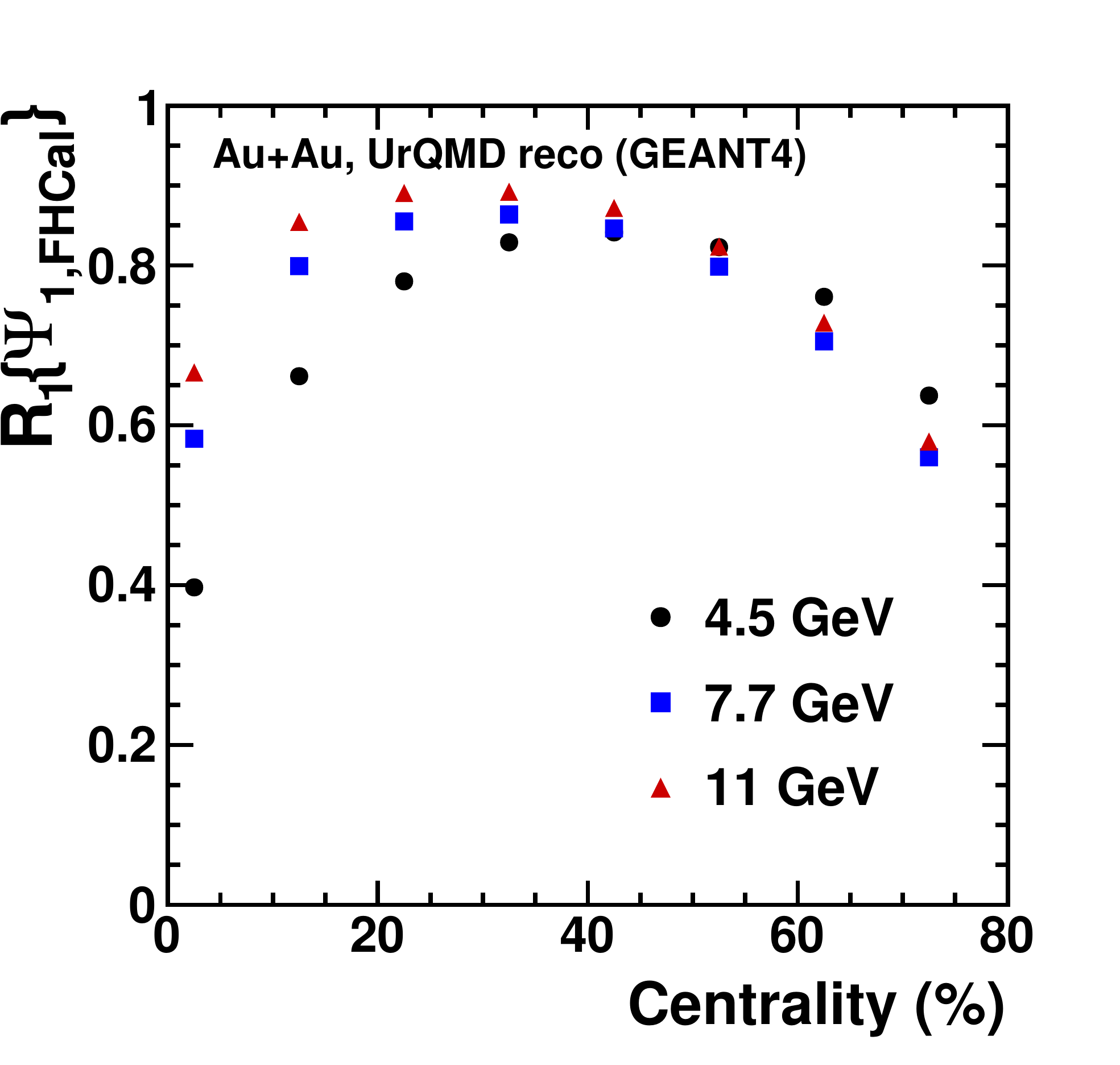}    
 \end{minipage}
 
 \begin{minipage}[h]{0.87\linewidth}
   \includegraphics[width=70mm,angle=0] {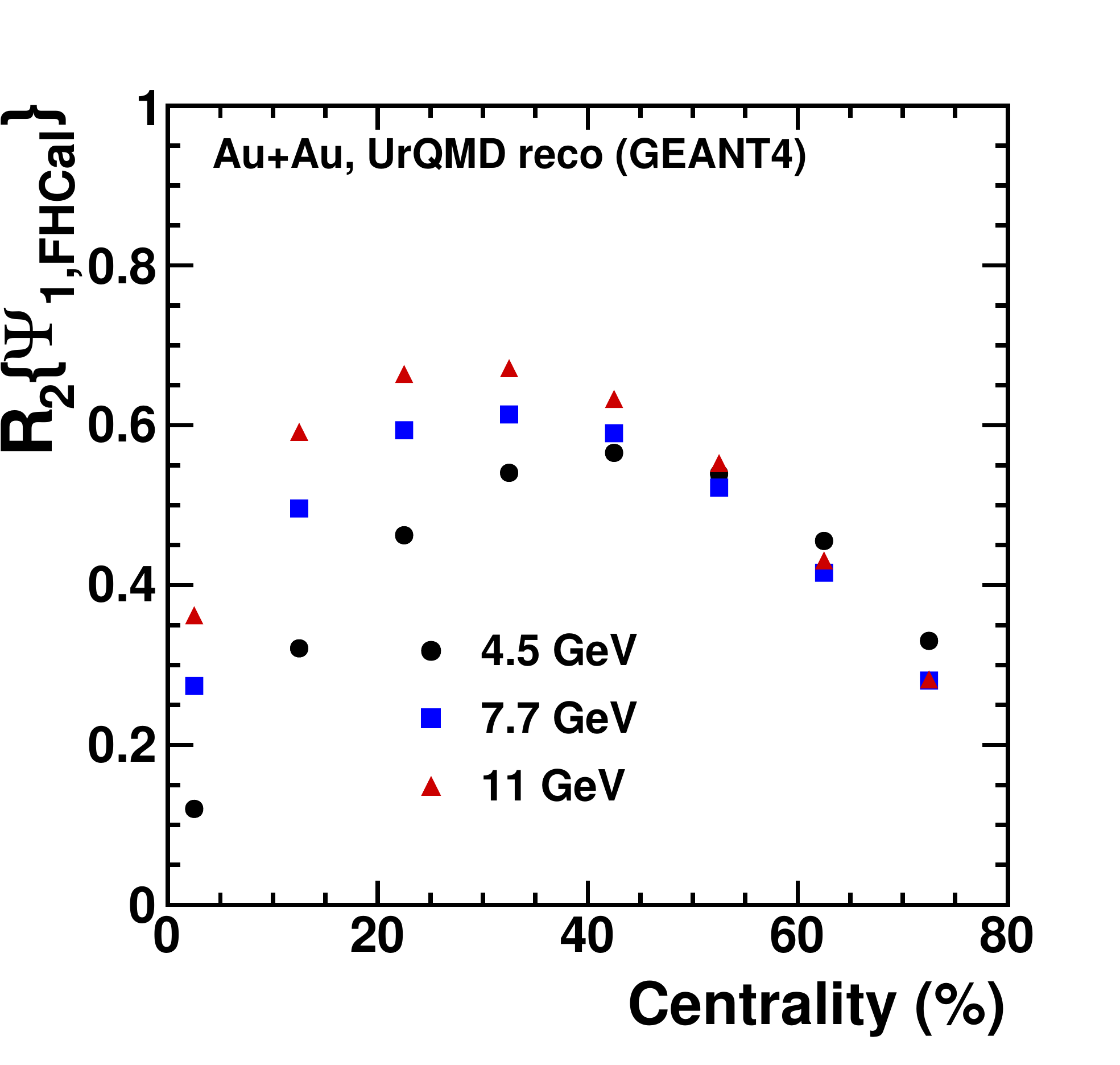}
 \end{minipage}
	\caption{Centrality dependence of event plane resolution factors
          $R_n(\Psi_{1,\textrm{FHCal}})$ for the directed $v_1$ (top) and elliptic  $v_2$ (bottom) flow measurements
          for Au+Au collisions at $\sqrt{s_{\rm NN}}$ = 4.5, 7.7 and 11 GeV.}
\label{fig:resol}
\end{figure}

The directed flow $v_1$ is large at NICA energies  compared to other flow
harmonics. It is the strongest in the forward rapidity region
(in the \acrshort{fhcal} acceptance area: 2$<|\eta|<$ 5).
For these reasons, the first harmonic event plane
$\Psi_{1,\textrm{FHCal}}$ is chosen to study the MPD flow performance.  The event plane angle $\Psi_{1,\textrm{FHCal}}$ is calculated from the energy deposition in
a given module of the \acrshort{fhcal} by constructing the so-called
flow Q-vector $Q_{1,\textrm{FHCal}}$ (a two-dimensional vector in the plane transverse to the beam axis), see
Refs.~\cite{Parfenov:2019pxf,Idrisov:2020yca,Parfenov:2020msw} for details. The reconstructed $\Psi_{1,\textrm{FHCal}}$ can be used to measure  the differential directed flow
$v_{1}^{\Psi_{1,{\rm FHCal}}}$  and elliptic  $v_{2}^{\Psi_{1,{\rm FHCal}}}$ flow coefficients
of the produced  particles detected in the \acrshort{tpc} ($|\eta|<$ 1.5),
\begin{equation}
v_{n}^{\Psi_{1,{\rm FHCal}}}=\frac{\langle\cos(n(\phi - \Psi_{1,{\rm FHCal}}))\rangle}{R_{n}(\Psi_{1,{\rm FHCal}})},    
\end{equation}
where  ${R_{n}(\Psi_{1,{\rm FHCal}})}$ represents the event plane resolution factor.
The  2-subevent method with an extrapolation algorithm is applied to estimate the event plane resolution
factors \cite{Voloshin:2008dg}. Figure~\ref{fig:resol} shows the 
centrality dependence of the event plane resolution factors $R_n(\Psi_{1,\textrm{FHCal}})$ for directed $v_1$ (top panel)
and elliptic  $v_2$ (bottom panel) flow measurements  for Au+Au collisions at $\sqrt{s_{\rm NN}}$ = 4.5, 7.7 and 11 GeV.
For the mid-central Au+Au events at $\sqrt{s_{\rm NN}}$=11 GeV the  resolution factor is as
high as 0.9 for $v_1$ and 0.65 for  $v_2$.
The event plane resolution degrades slowly with decreasing collision energy. 

\begin{figure}[hbt]
 \centering
 \begin{minipage}[h]{0.87\linewidth}
   \includegraphics[width=70mm,angle=0] {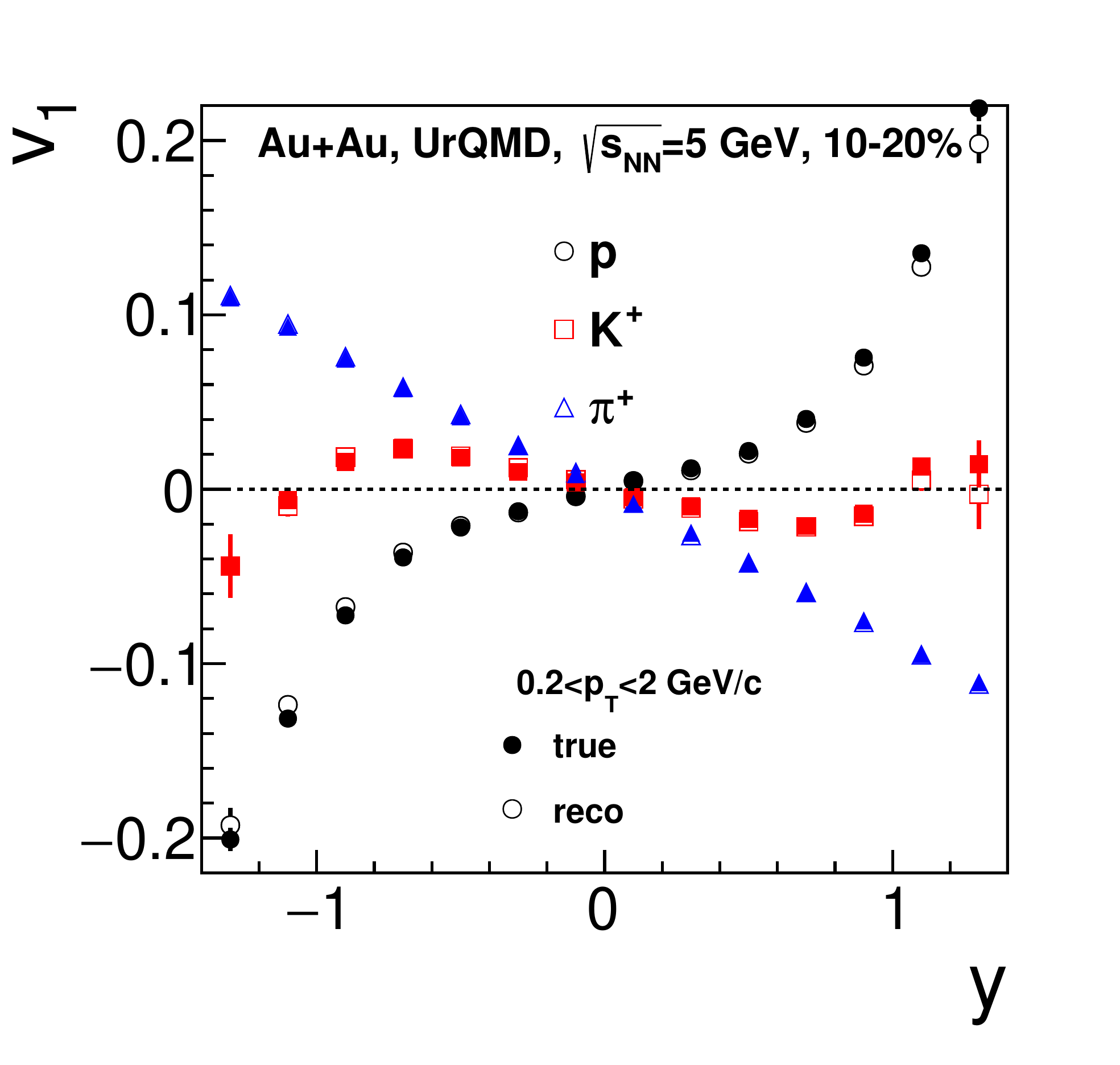}    
 \end{minipage}
 
 \begin{minipage}[h]{0.87\linewidth}
   \includegraphics[width=70mm,angle=0] {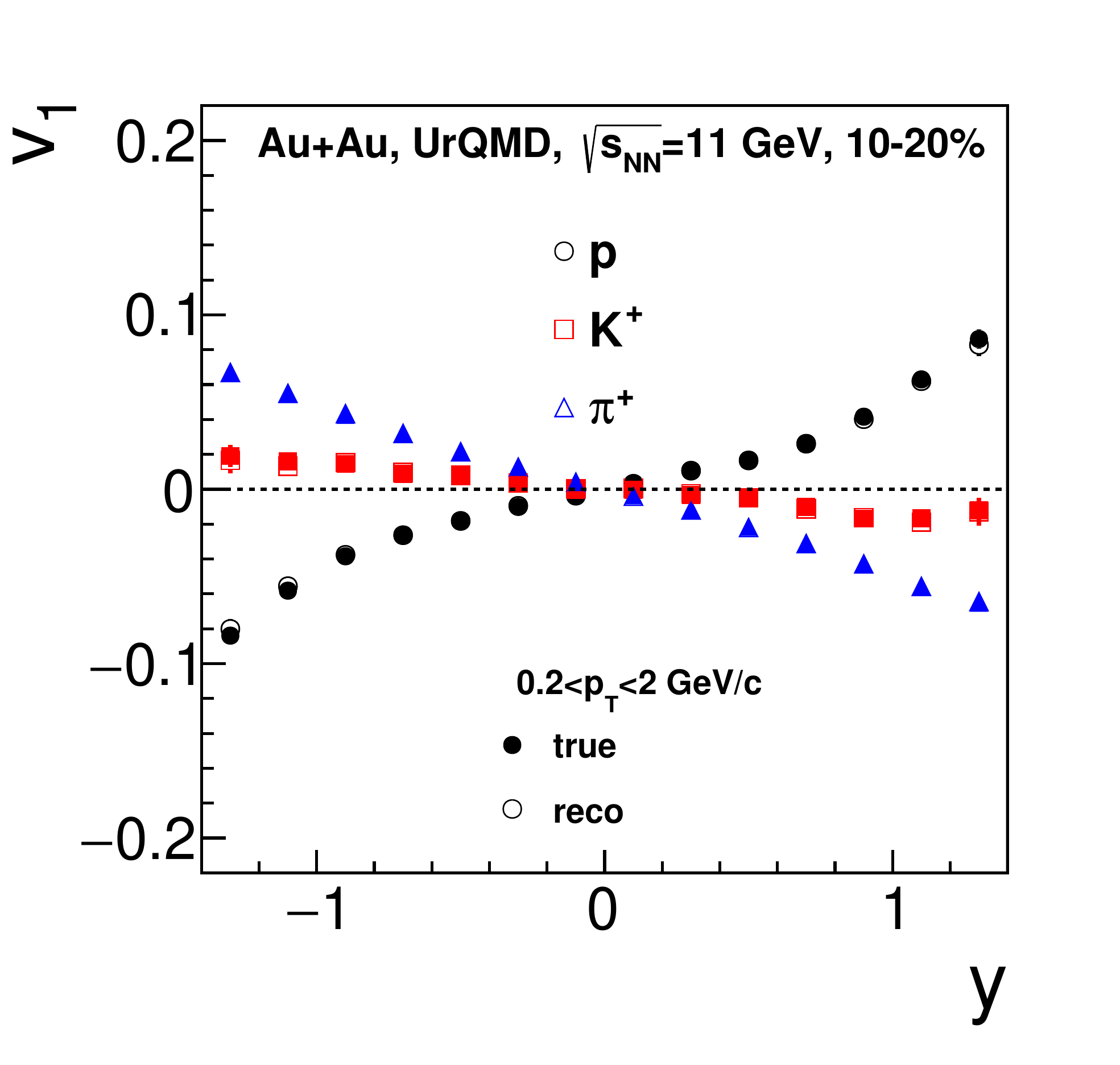}
 \end{minipage}
	\caption{Directed flow $v_1$ for charged pions, kaons and protons
          as a function of the rapidity $y$ for Au+Au collisions at $\sqrt{s_{\rm NN}} = 5$~GeV (top) and $11$~GeV (bottom).
          The results from the \acrshort{urqmd} model  are marked as \lq\lq true'', and the ones from the full reconstruction
          procedure are marked as \lq\lq reco''.}
	\label{fig-v1}       
\end{figure}

For $V^{0}$ particles, like $K_{S}^{0}$ and $\Lambda$, the $v_{n}$ of the selected sample contains
both  a signal component $v_{n}^{S}$ and a combinatorial background component $v_{n}^{B}$. The invariant mass ($M_{inv}$) fit method \cite{flowfit} is applied to extract
the  anisotropic  flow values $v_{n}^{S}$  for $V^{0}$ particles.
The $v_n$ results obtained by the event plane analysis  can be affected by non-flow
and flow fluctuations \cite{Voloshin:2008dg}.
The non-flow effects are mainly due to few particle correlations not associated
with the reaction plane: Bose-Einstein correlations, resonance decays, momentum conservation, di-jets.
In the present study, the large rapidity gap $\Delta\eta >$ 0.5 between the particles detected in \acrshort{tpc}
and the particles  in \acrshort{fhcal}
reduces the influence of possible non-flow contributions. The elliptic flow results $v_{2}^{\Psi_{1,{\rm FHCal}}}$
obtained with respect to the
spectator first-order event plane are expected to be less affected by the elliptic flow
fluctuations because of  the strong correlation between the $\Psi_{1,\textrm{FHCal}}$ and the true reaction plane
$\Psi_{RP}$. Figure~\ref{fig-v1} shows the rapidity dependence of $v_1(y)$  of charged pions, kaons
and protons from 10-20\% central Au+Au collisions at $\sqrt{s_{\rm NN}}$ = 5 GeV (left panel)  and 11 GeV (right panel).
For all particle species, the directed flow crosses 0 at midrapidity. The reconstructed values ``reco''
of  $v_1$ are fully consistent with the  generated ``true'' values  in all centrality classes and collision energies.

\begin{figure}[t]
        \includegraphics[width=0.45\textwidth,clip]{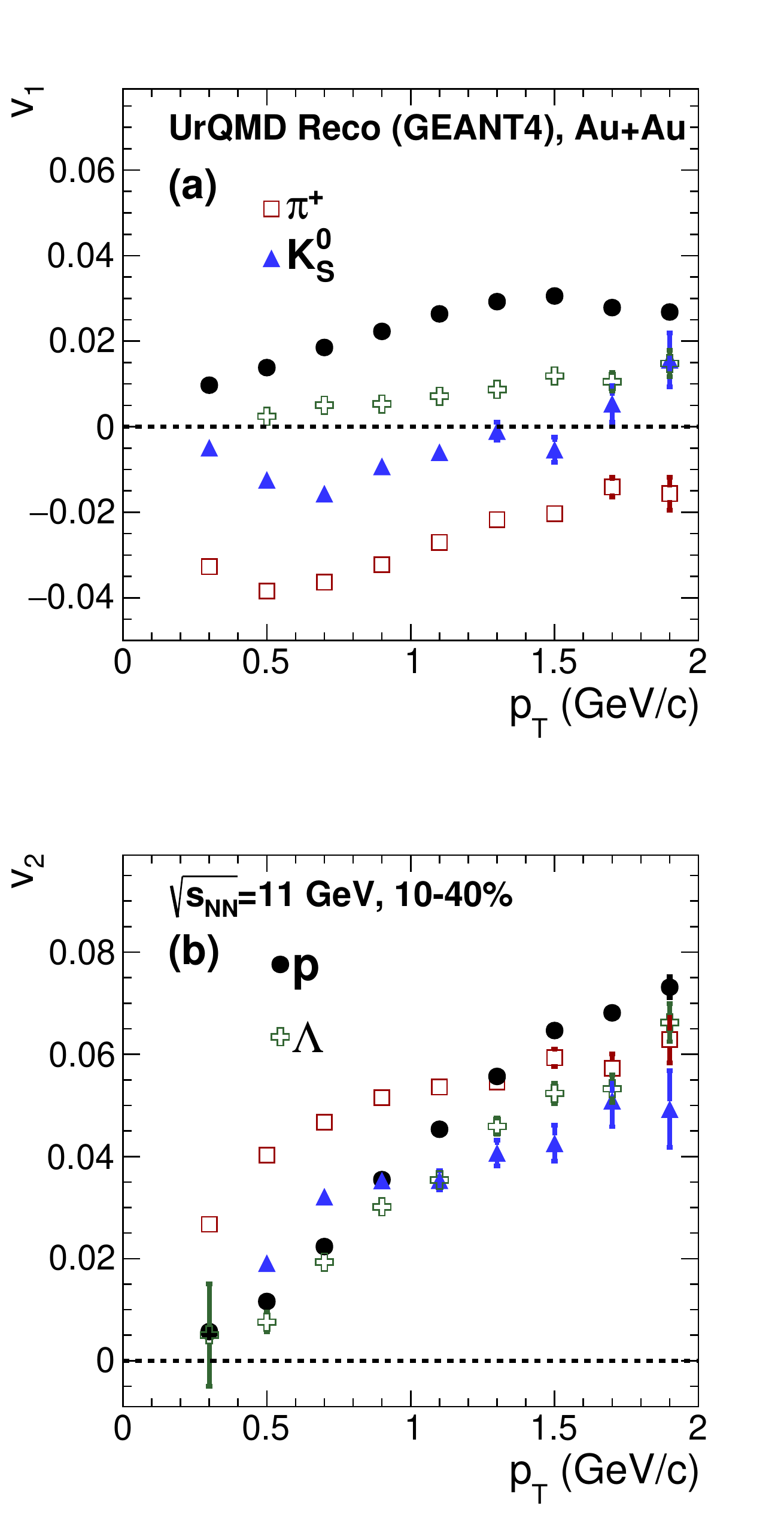}
	\caption{ $p_{\rm T}$-dependence of directed (top) and elliptic (bottom) flow of charged pions, protons,
		$K_S^0$ and $\Lambda$ particles
		from 10-40\% central Au+Au collisions at $\sqrt{s_{\rm NN}}=11$~GeV.
		Directed flow extracted for $0.2<|y|<1.2$. }
	\label{fig:v1v2kslambda}
\end{figure}

\begin{figure}[t]
	\centering
	\includegraphics[width=0.45\textwidth,clip]{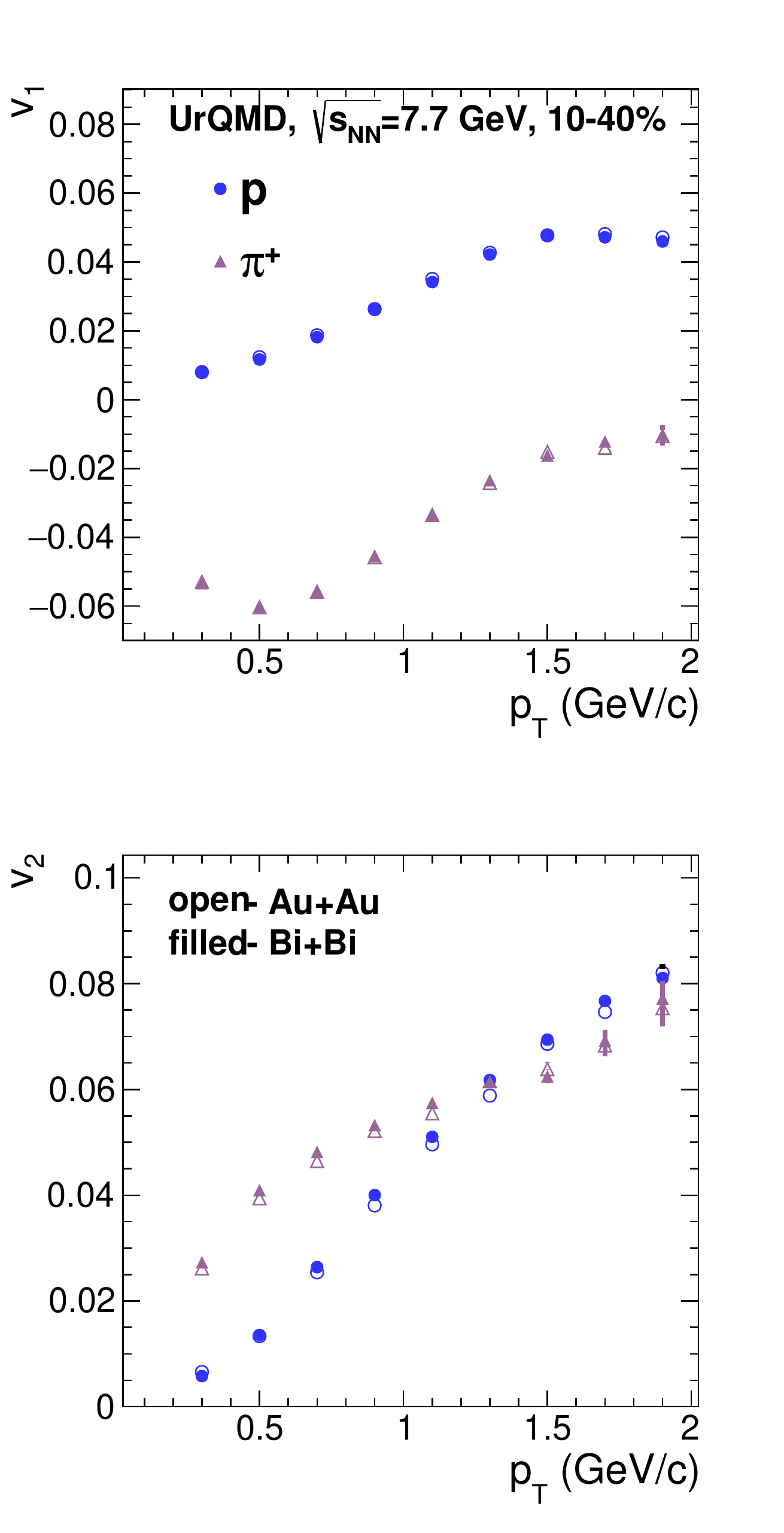}
	\caption{$p_{\rm T}$-dependence of directed $v_1$ (top) and elliptic $v_2$ (bottom) flow signals of pions and protons
		from 10-40\% central Au+Au (open symbols) and Bi+Bi (filled symbols) collisions at $\sqrt{s_{\rm NN}}=7.7$~GeV.}
	\label{fig:Sys_compare_pt}       
\end{figure}

Figure~\ref{fig:v1v2kslambda} illustrates the  MPD detector performance
for the $p_{\rm T}$ differential directed and elliptic flow measurements of charged pions, protons, $K_{S}^{0}$  and $\Lambda$ 
particles from 10-40\% central Au+Au collisions at $\sqrt{s_{\rm NN}}=11$~GeV.
The $v_n$ results for $K_{S}^{0}$  and $\Lambda$  particles are  obtained from the event plane 
analysis of 20 millions minimum bias, fully reconstructed \acrshort{urqmd} events using the
invariant-mass fit method. 
The rapidity ranges of $0.2<|y|<1.2$ and $|y|<1.2$ were used for the measured $p_{\rm T}$ dependencies of directed and elliptic flow correspondingly.
Figure \ref{fig:Sys_compare_pt}  shows the  performance for the measurements of 
the $p_{\rm T}$ dependence of  directed $v_1$ (left) and elliptic $v_2$ (right) flow
of charged pions and protons from
10-40\% midcentral Au+Au (open symbols) and Bi+Bi (filled symbols) collisions at $\sqrt{s_{\rm NN}}=7.7$~GeV.  The $v_n$ results
were obtained by the event plane method, using the first order event plane ($\Psi_{1,\textrm{FHCal}}$) from \acrshort{fhcal}.
In both cases there is a small difference between the $v_n$ results for the two colliding systems.

\begin{figure*}
	\centering
	\includegraphics[width=0.97\textwidth,clip]{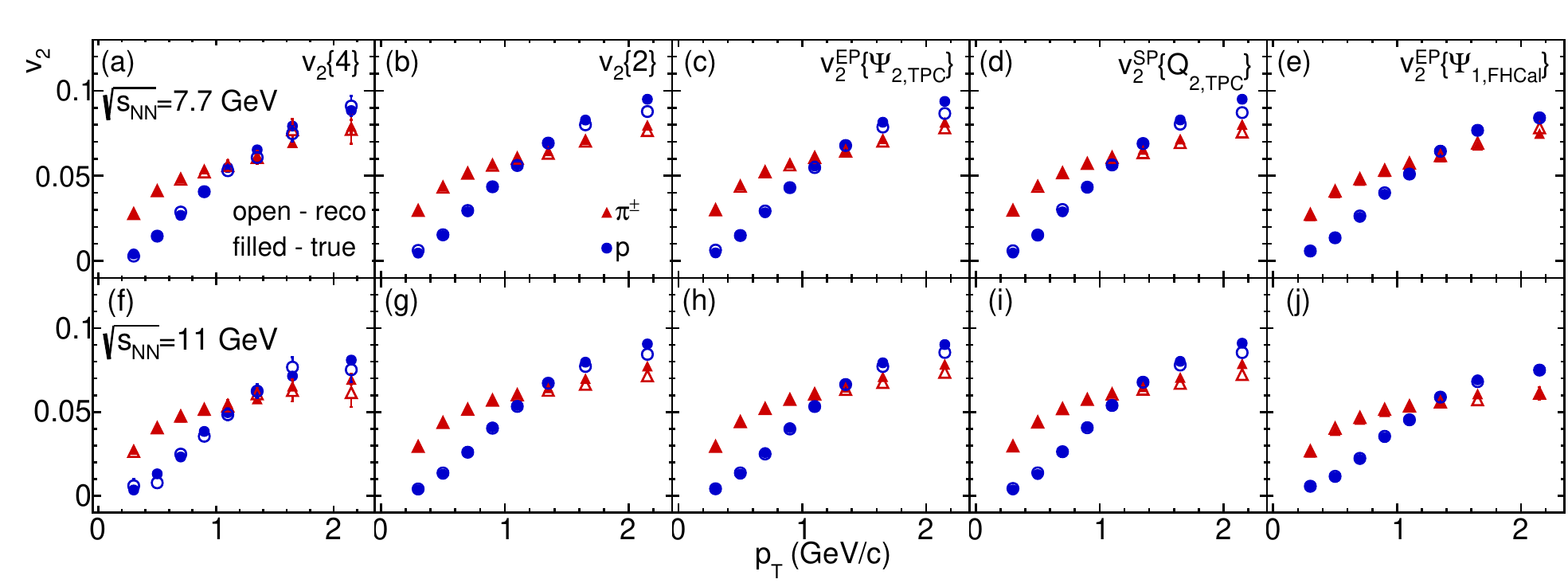}
	\caption{Comparison of $v_2(p_{\rm T})$ for charged pions (triangles) and protons (circles) from
          10-40\% central  Au+Au collisions at $\sqrt{s_{\rm NN}}=7.7$~GeV (upper panels) and
          $\sqrt{s_{\rm NN}}=11.5$~GeV (lower panels)
          obtained  by Q-cumulant, event plane and scalar product  methods of analysis.
          The solid points represent the true results from the UrQMD model and the open points are the reconstructed values.}
	\label{fig-v2}       
\end{figure*}

The large and uniform acceptance of the \acrshort{tpc} allows us to use multiparticles methods for flow measurements such as direct
cumulants, see  \cite{Idrisov:2020yca,Parfenov:2020msw} for details. As an example,
Fig.~\ref{fig-v2} shows the  $v_2(p_{\rm T})$ of charged pions and protons 
from 10-40\% central Au+Au collisions at $\sqrt{s_{\rm NN}} = 7.7$ (upper panel) and $11$ GeV (lower panel).
The $v_n$ results are based on  the analysis of  \acrshort{urqmd} events by the
event plane ($v^\textrm{EP}_2\{\Psi_{1,\textrm{FHCal}}\}$, $v^\textrm{EP}_2\{\Psi_{2,\textrm{TPC}}\}$)
and Q-cumulant ($v_2\{2\}$, $v_2\{4\}$) methods from left to right. Perfect agreement is observed between the
$v_2$ results from the analysis of fully reconstructed (\lq\lq reco'') and generated (\lq\lq true'') \acrshort{urqmd} events for all methods.

To conclude, the current studies demonstrate that the MPD detector is able to provide detailed differential measurements of directed and elliptic flows with high accuracy.

\subsection{\label{sec:fluctuations}Event-by-event net-proton and net-kaon measurements}

\begin{figure}[b]
\includegraphics[width=8cm]{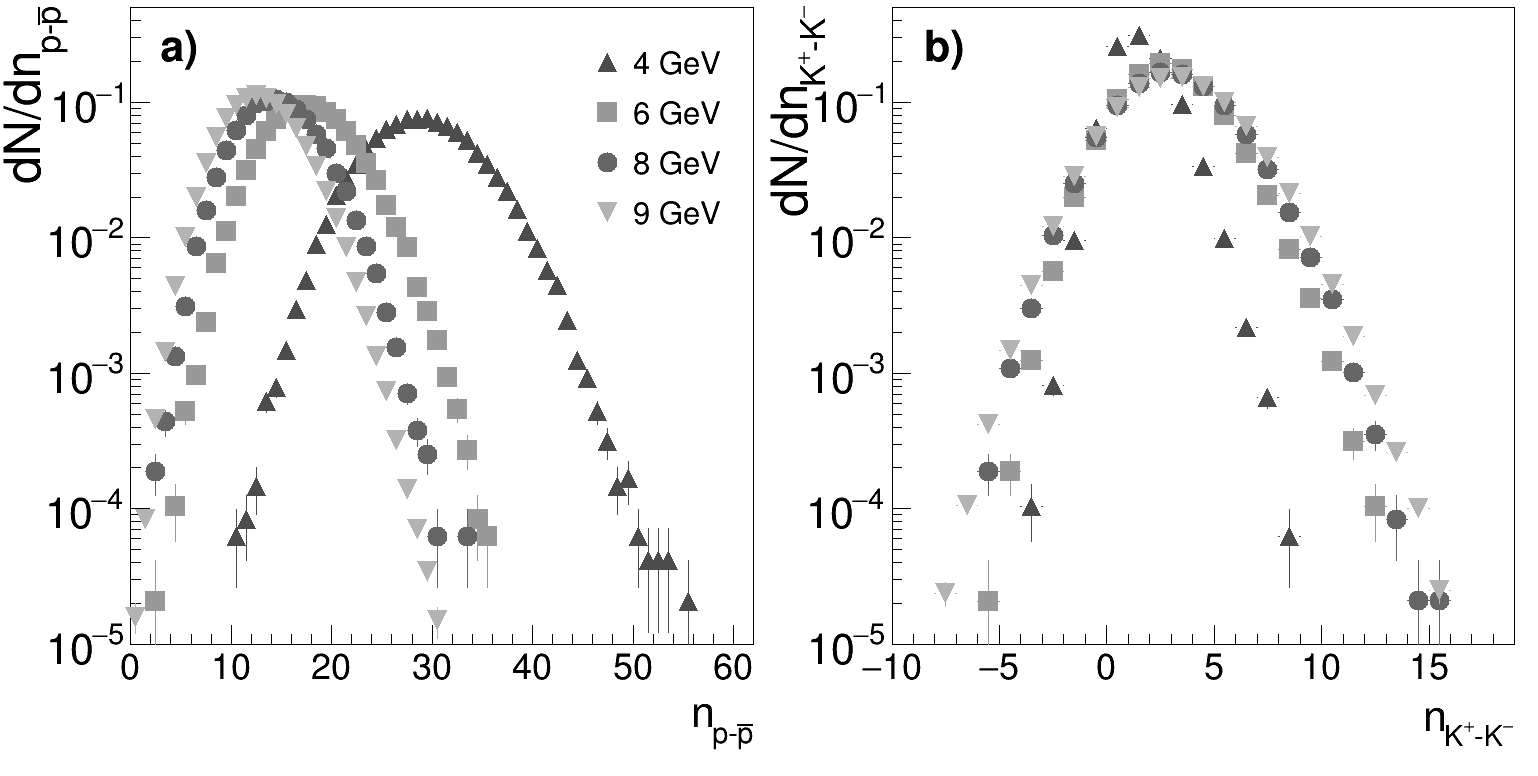}
\caption{\label{fig:netprot} a) Distributions of net-protons registered in the MPD detector in central Au+Au collision at several collision energies. b) The same for net-kaons. Results were obtained using the PHSD event generator.}
\end{figure}

\begin{figure}[b]
\includegraphics[width=8cm]{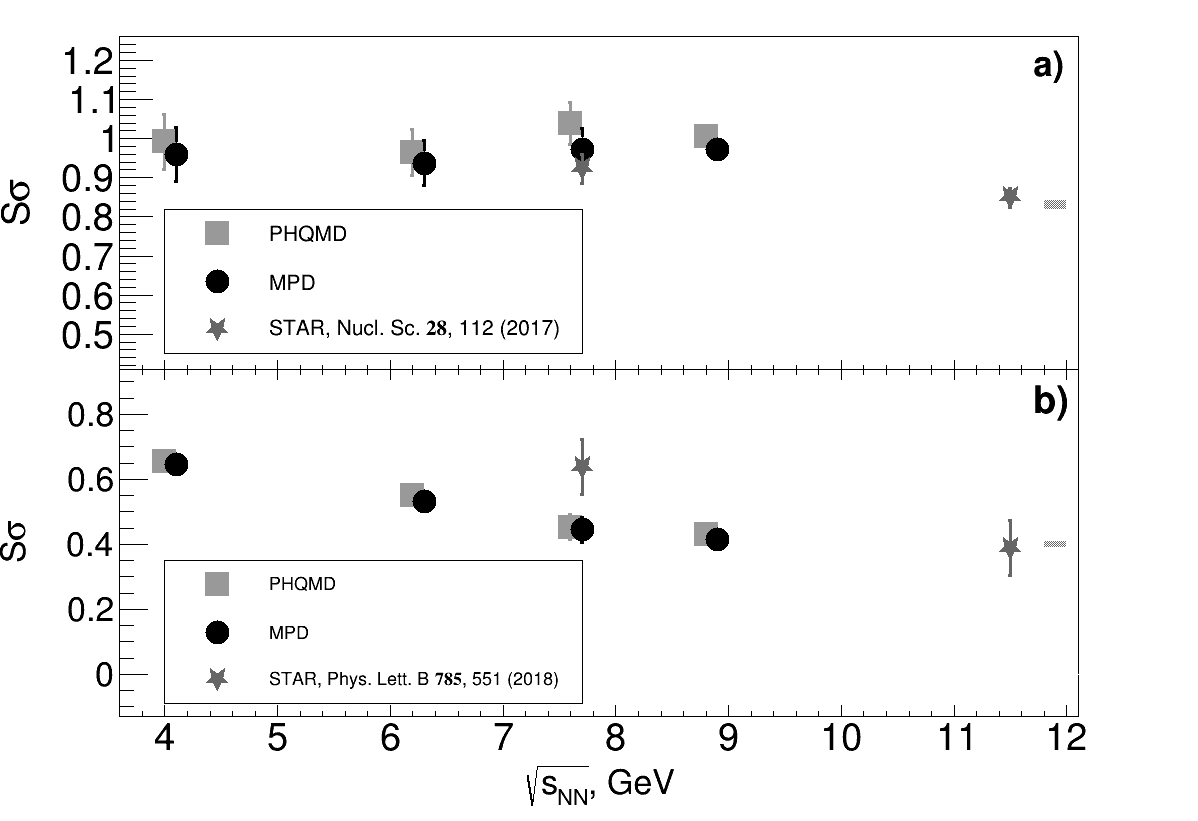}
\caption{\label{fig:moments}Effective skewness $S\sigma$ for net-protons (a) and net-kaons (b) in Au+Au interactions at several collision energies. The label PHQMD indicates the results from the model, the MPD label shows the reconstructed numbers, the STAR label is meant to represent experimental data from~\cite{Adam:2020unf,Luo:2017faz}. Squares and circles are shifted relative to each other along the horizontal axis for clarity. The expected value of the experimental uncertainties for a data set of $10^6$ central events are shown by shaded areas in the right part of the panels.}
\end{figure}

As mentioned in Sec.~2, the moments of event-by-event multiplicity distributions are sensitive to critical phenomena. To assess the sensitivity of the MPD setup to these signals, a study of the moments of the event-by-event distributions of net-protons (protons minus antiprotons) and net-kaons (positively charged kaons minus negatively charged ones) was performed. Central Au+Au collisions from the PHSD event generator at several collision energies were used. Net-protons and net-kaons are proxies for the baryon number and strangeness, respectively. 
It is important to exclude the variation in the system’s volume from event to event; for this we study various moment combinations, namely, the combination of the third and second moment (so called effective skewness, $S\sigma$) and the combination of the fourth order moment and second order one (effective kurtosis, $k\sigma^2$). 
The results of the higher moment analysis from the reconstructed data were compared to the initial data from the model. In addition, a comparison with available experimental data was performed in order to estimate the baseline of the model. The simulation, reconstruction, and particle identification procedures in this analysis are the same as described in Sec.~\ref{sec:bulkspectra}. The distributions of identified net-protons at several collision energies are shown in the left panel of Fig.~\ref{fig:netprot}. In order to minimize the sensitivity of the results for the MPD efficiency, the analysis was performed in limited rapidity and transverse momentum intervals, namely, $|y|<0.5$ and $0.4<p_{\rm T}<0.8$~GeV/c. The average number of net-protons per central collision varies from 28.7 at 4 GeV to 12.5 at 9 GeV. The distributions for net-kaons ($K^+$-$K^-$) are shown in Fig.~\ref{fig:netprot} (right panel). The average number of net-kaons per event increases with collision energy from 1.1 at 4 GeV to 2.7 at 9 GeV. One can note a significant change in the number of net-protons and net-kaons predicted by the model at the lower collision energy of $\sqrt{s_{\rm NN}}$~=~4~GeV. The MPD reconstruction efficiency for charged hadrons varies from 70 to 80\% in the selected phase space region. In order to correct the moments of the net-proton and net-kaon distributions for the detector efficiency, the method from Ref.~\cite{Bzdak:2013pha} was used.
Figure~\ref{fig:moments} shows the final results for the effective skewness $S\sigma$ of net-protons (upper panel) and net-kaons (lower panel) from Au+Au interactions at several collision energies. The initial values of the model are shown by squares (PHQMD label); the reconstructed values (plotted by circles) are indicated with the label MPD. Recent experimental data from the STAR Collaboration~\cite{Adam:2020unf,Luo:2017faz} are shown by stars. The reconstructed numbers for the moments agree with the initial values of the model within the errors. Moreover, the estimates for the measurement uncertainty for a data set of $10^6$ events are indicated by shaded areas in the right part of the figure. Finally, we can conclude that with the current MPD track reconstruction and particle identification accuracy, the study of the energy dependence of fluctuations of conserved charges at NICA is possible with good precision and small systematic uncertainty.

\section{Conclusions}\label{sec:conclusions}

The MPD experiment will be the primary 
detector at the newly constructed \acrshort{nica} Accelerator Complex at the \acrlong{jinr}. It will  fill an energy gap in the landscape of heavy-ion collision experiments, devoted to the investigation of the onset of deconfinement and chiral symmetry restoration and the exploration of the QCD phase diagram in the energy range $\sqrt{s_{\rm NN}}=4-11$ GeV. Moreover, the MPD experiment is unique since it will cover
this energy range in the collider mode with homogeneous coverage of the kinematic variables rapidity and transverse momentum.

In this review, we have highlighted a few unique physics goals which characterize the discovery potential of the \acrshort{mpd} experiment:
\begin{itemize}
    \item The energy scan covers the transition region from baryon to meson dominance in the chemical freeze-out regime. At the same time, this is the challenging domain of the transition from baryon stopping to nuclear transparency, where new experimental data would be very useful to progress in the theoretical understanding.
    \item The search for the critical end point in the QCD phase diagram is possible via a combination of observables, such as fluctuations of conserved quantities
    \item There is a minimum in the freeze-out volume related to the \lq\lq softest point'' in the EoS in the NICA energy window which is accessible to verification by femtoscopy measurements.
    \item Simulations predict an antiflow of protons, antiprotons, $\Lambda$, $\bar{\Lambda}$, $\phi$ as well as light clusters in the NICA energy range.
    \item Onset of QGP formation and chiral symmetry restoration effects through measurements of dileptons in the low and intermediate mass regions, respectively.
    \item Measurement of the yields and flow coefficiecnts for inclusive and thermal photons.
    \item Potential for pioneering measurements of charm production at (sub-)threshold energies.
\end{itemize}

The MPD, with all its major components, was presented. Construction, installation and commissioning plans are on schedule. 
Performance studies of the MPD detector components were shown and some examples of specific physics analyses, possible in the initial stages of NICA operation were given. The MPD experiment is on track to become operational and take data as soon as first beams and collisions are available in the NICA complex.

\begin{acknowledgements}
\begin{sloppypar}
We are very grateful to the following agencies and organizations for
financial support of the MPD Experiment and the MPD members: 
\end{sloppypar}

\begin{sloppypar}
China -- Ministry of Science and Technology (MOST) National Key R\&D Program of China (Grant No.~2020YFE0202000), National Natural Science Foundation of China (NSFC), Chinese Academy of Sciences (CAS), Ministry of Education (MOE);
The Czech Republic -- Grants "3+3" and the Grants of the Plenipotentiary of the
Government of the Czech Republic in JINR as well as by MEYS Grant LTT18021;
The Mexican Collaboration (MexNICA) thankfully acknowledges support from UNAM-DGAPA-PAPIIT grant number IG100219, from the Consejo Nacional de Ciencia y Tecnología (CONACyT), grant numbers A1-S-7655 and A1-S16215, the permission to use computer resources, the technical advise and the support provided by the Laboratorio Nacional de Supercómputo del Sureste de México (LNS), a member of the CONACyT national network of laboratories, with resources from grant number 201701035C and the BUAP Medical Physics and Elementary Particles Labs;
Poland -- National Science Centre (NCN) grants 2016/23/B/ST2/00692 and 2020/39/O/ST2/00277, Ministry of Science and Higher Education, The Polish Plenipotentiary for JINR, WUT ID-UB; 
The Russian Federation -- Russian Foundation for Basic Research under grant 18-02-40084/19, 18-02-40137/19, 18-02-40065/19, 18-02-40054/19, 18-02-40056/19, 18-02-40079/19, 18-02-40085/19, 18-02-40045/19, 18-02-40038/19, 18-02-40086/19, 18-02-40051/51, 18-02-40044/19, 18-02-40037/19, 18-02-40060/19. NRNU MEPhI acknowledges support from Program Priority 2030 and by the Ministry of Science and Higher Education of the Russian Federation, Project
”Fundamental properties of elementary particles and cosmology” No 0723-2020-0041.

We would like to thank Anton Andronic for preparing the figure which we show as Fig. 2,
Krzysztof Redlich for Fig. 3 and Bedanga Mohanty for preparing the figure version that we 
show as  Fig. 7.
\end{sloppypar}
\end{acknowledgements}

\bibliographystyle{spphys}       

\bibliography{apssamp-1}

\clearpage
\printglossary[type=\acronymtype]

\newpage
\onecolumn

\section*{The MPD Collaboration}
\label{app:collab}
\newcommand*{\aanlarm}{A. Alikhanyan National Lab of Armenia, Yerevan, ARMENIA}
\newcommand*{\aanlarmindex}{1}

\newcommand*{\bsuaze}{Baku State University, Baku, AZERBAIJAN}
\newcommand*{\bsuazeindex}{2}

\newcommand*{\bnrrus}{Belgorod National Research University, Belgorod, RUSSIA}
\newcommand*{\bnrrusindex}{3}

\newcommand*{\ccnuchn}{Central China Normal University, Wuhan, CHINA}
\newcommand*{\ccnuchnindex}{4}

\newcommand*{\cieamex}{Centro de Investigación y Estudios Avanzados del IPN, Mexico City, MEXICO}
\newcommand*{\cieamexindex}{5}

\newcommand*{\udsmex}{Departamento de Investigaci\'o{}n en F\'i{}sica, Universidad de Sonora, Sonora, MEXICO}
\newcommand*{\udsmexindex}{6}

\newcommand*{\udcmex}{Facultad de Ciencias - CUICBAS, Universidad de Colima, Colima, MEXICO}
\newcommand*{\udcmexindex}{7}

\newcommand*{\uapmex}{Facultad de Ciencias Físico Matemáticas, Benemérita Universidad Autónoma de Puebla, Puebla, MEXICO}
\newcommand*{\uapmexindex}{8}

\newcommand*{\uadsmex}{Facultad de Ciencias F\'i{}sico-Matem\'a{}ticas, Universidad Aut\'o{}noma de Sinaloa, Culiac\'a{}n, MEXICO}
\newcommand*{\uadsmexindex}{9}

\newcommand*{\fiuadsmex}{Facultad de Informática, Universidad Autónoma de Sinaloa, Culiac\'a{}n, MEXICO}
\newcommand*{\fiuadsmexindex}{10}

\newcommand*{\fuchn}{Fudan University, Shanghai, CHINA}
\newcommand*{\fuchnindex}{11}

\newcommand*{\tsugeo}{High Energy Physics Institute, Tbilisi State University, Tbilisi, GEORGIA}
\newcommand*{\tsugeoindex}{12}

\newcommand*{\huchn}{Huzhou University, Huzhou, CHINA}
\newcommand*{\huchnindex}{13}

\newcommand*{\inrrus}{Institute for Nuclear Research, Russian Academy of Sciences, Moscow, RUSSIA}
\newcommand*{\inrrusindex}{14}

\newcommand*{\iapmol}{Inst. of Applied Physics, Chisinev, MOLDOVA}
\newcommand*{\iapmolindex}{15}

\newcommand*{\impchn}{Institute of Modern Physics of the Chinese Academy of Science, Langzhou, CHINA}
\newcommand*{\impchnindex}{16}

\newcommand*{\icnmex}{Instituto de Ciencias Nucleares, Universidad Nacional Autónoma de México, Mexico City, MEXICO}
\newcommand*{\icnmexindex}{17}

\newcommand*{\jkupol}{Jan Kochanowski University, Kielce, POLAND}
\newcommand*{\jkupolindex}{18}

\newcommand*{\jinr}{Joint Institute for Nuclear Research, Dubna, RUSSIA}
\newcommand*{\jinrindex}{19}

\newcommand*{\miptrus}{Moscow Institute of Physics and Technology, Moscow, RUSSIA}
\newcommand*{\miptrusindex}{20}

\newcommand*{\ncnrpol}{National Center for Nuclear Reseach, Otwock, POLAND}
\newcommand*{\ncnrpolindex}{21}

\newcommand*{\nnrcaze}{National Nuclear Research Center, Baku, AZERBAIJAN}
\newcommand*{\nnrcazeindex}{22}

\newcommand*{\nrcrus}{National Research Center \lq\lq Kurchatov Institute'', Moscow, RUSSIA}
\newcommand*{\nrcrusindex}{23}

\newcommand*{\iteprus}{National Research Center \lq\lq Kurchatov Institute'' – Institute of Theoretical and Exprimental Physics, Moscow, RUSSIA}
\newcommand*{\iteprusindex}{24}

\newcommand*{\mephirus}{National Research Nuclear University MEPhI (Moscow Engineering Physics Institute), Moscow, RUSSIA}
\newcommand*{\mephirusindex}{25}

\newcommand*{\nosurus}{North Ossetian State University, Vladikavkaz, RUSSIA}
\newcommand*{\nosurusindex}{26}

\newcommand*{\npiczh}{Nuclear Physics Institute CAS, Rez, CZECH REPUBLIC}
\newcommand*{\npiczhindex}{27}

\newcommand*{\puczh}{Palacky U., Olomouc, CZECH REPUBLIC}
\newcommand*{\puczhindex}{28}

\newcommand*{\pnpirus}{Petersburg Nuclear Physics Institute, Gatchina, RUSSIA}
\newcommand*{\pnpirusindex}{29}

\newcommand*{\uopbul}{Plovdiv University Paisii Hilendarski, Plovdiv, BULGARIA}
\newcommand*{\uopbulindex}{30}

\newcommand*{\spsurus}{Saint Petersburg State University, Sankt Petersburg, RUSSIA}
\newcommand*{\spsurusindex}{31}

\newcommand*{\ihepchn}{School of Nuclear Science and Technology, University of Chinese Academy of Sciences, Beijing, CHINA}
\newcommand*{\ihepchnindex}{32}

\newcommand*{\shuchn}{Shandong University, Qingdao, CHINA}
\newcommand*{\shuchnindex}{33}

\newcommand*{\sinprus}{Skobeltsyn Institute of Nuclear Physics, Moscow, RUSSIA}
\newcommand*{\sinprusindex}{34}

\newcommand*{\tguchn}{Three Gorges University, Yichang, CHINA}
\newcommand*{\tguchnindex}{35}

\newcommand*{\tuchn}{Tsinghua University, Beijing, CHINA}
\newcommand*{\tuchnindex}{36}

\newcommand*{\utsmchl}{Universidad T\'ecnica Federico Santa Mar\'\i a, Valpara\'\i so, CHILE}
\newcommand*{\utsmchlindex}{37}

\newcommand*{\ustuchn}{University of Science and Technology of China, Hefei, CHINA}
\newcommand*{\ustuchnindex}{38}

\newcommand*{\uspol}{University of Silesia, Katowice, POLAND}
\newcommand*{\uspolindex}{39}

\newcommand*{\uscchn}{University of South China, Hengyang, CHINA}
\newcommand*{\uscchnindex}{40}

\newcommand*{\uwpol}{University of Warsaw, Warsaw, POLAND}
\newcommand*{\uwpolindex}{41}

\newcommand*{\uwrpol}{University of Wroclaw, Wroclaw, POLAND}
\newcommand*{\uwrpolindex}{42}

\newcommand*{\wutpol}{Warsaw University of Technology, Warsaw, POLAND}
\newcommand*{\wutpolindex}{43}

\begingroup
\renewcommand{\thefootnote}{\Roman{footnote}}
V.~Abgaryan$^{\jinrindex}$\textsuperscript{,}$^{\aanlarmindex}$, 
R.~Acevedo Kado$^{\icnmexindex}$, 
S.V.~Afanasyev$^{\jinrindex}$, 
G.N.~Agakishiev$^{\jinrindex}$, 
E.~Alpatov$^{\mephirusindex}$, 
G.~Altsybeev$^{\spsurusindex}$, 
M.~Alvarado Hern\'andez$^{\icnmexindex}$, 
S.V.~Andreeva$^{\jinrindex}$, 
T.V.~Andreeva$^{\jinrindex}$, 
E.V.~Andronov$^{\spsurusindex}$
% ...

\begin{description}[labelsep=0.2em,align=right,labelwidth=0.7em,labelindent=0em,leftmargin=2em,noitemsep]
\item[$^{\aanlarmindex}$] \aanlarm
\item[$^{\bsuazeindex}$] \bsuaze
\item[$^{\bnrrusindex}$] \bnrrus
\item[$^{\ccnuchnindex}$] \ccnuchn
\item[$^{\cieamexindex}$] \cieamex
\item[$^{\udsmexindex}$] \udsmex
\item[$^{\udcmexindex}$] \udcmex
\item[$^{\uapmexindex}$] \uapmex
\item[$^{\uadsmexindex}$] \uadsmex
\item[$^{\fiuadsmexindex}$] \fiuadsmex
\item[$^{\fuchnindex}$] \fuchn
\item[$^{\tsugeoindex}$] \tsugeo
\item[$^{\huchnindex}$] \huchn
\item[$^{\inrrusindex}$] \inrrus
\item[$^{\iapmolindex}$] \iapmol
\item[$^{\impchnindex}$] \impchn
\item[$^{\icnmexindex}$] \icnmex
\item[$^{\jkupolindex}$] \jkupol
\item[$^{\jinrindex}$] \jinr
\item[$^{\miptrusindex}$] \miptrus
\item[$^{\ncnrpolindex}$] \ncnrpol
\item[$^{\nnrcazeindex}$] \nnrcaze
\item[$^{\nrcrusindex}$] \nrcrus
\item[$^{\iteprusindex}$] \iteprus
\item[$^{\mephirusindex}$] \mephirus
\item[$^{\nosurusindex}$] \nosurus
\item[$^{\npiczhindex}$] \npiczh
\item[$^{\puczhindex}$] \puczh
\item[$^{\pnpirusindex}$] \pnpirus
\item[$^{\uopbulindex}$] \uopbul
\item[$^{\spsurusindex}$] \spsurus
\item[$^{\ihepchnindex}$] \ihepchn
\item[$^{\shuchnindex}$] \shuchn
\item[$^{\sinprusindex}$] \sinprus
\item[$^{\tguchnindex}$] \tguchn
\item[$^{\tuchnindex}$] \tuchn
\item[$^{\utsmchlindex}$] \utsmchl
\item[$^{\ustuchnindex}$] \ustuchn
\item[$^{\uspolindex}$] \uspol
\item[$^{\uscchnindex}$] \uscchn
\item[$^{\uwpolindex}$] \uwpol
\item[$^{\uwrpolindex}$] \uwrpol
\item[$^{\wutpolindex}$] \wutpol
\end{description}
\endgroup

\end{document}